\documentclass{aa} 
\usepackage{txfonts,graphicx,longtable,lscape,epsfig,color}

\begin{document}

   \title{A survey of stellar X-ray flares from the XMM-Newton serendipitous source catalogue: \\
Hipparcos-Tycho cool stars\thanks{Based on observations
obtained with XMM-Newton, an ESA science mission with instruments and
contributions directly funded by ESA Member States and NASA.}$^,$
\thanks{Tables C.1 and C.2 are only available in electronic form at the CDS via anonymous ftp to cdsarc.u-strasbg.fr (130.79.128.5)
or via http://cdsweb.u-strasbg.fr/cgi-bin/qcat?J/A+A/.
Figures C.1 and C.2 are only available in electronic form via http://www.edpsciences.org .} 
}

   \titlerunning{XMM-Tycho stellar flares}

   \author{J. P. Pye\inst{1} \and S. Rosen\inst{1} \and D. Fyfe\inst{1} \and A. C. Schr\"oder\inst{1,2} }

   \authorrunning{J. P. Pye et al.}

   \offprints{J. P. Pye }

   \institute{University of Leicester, Department of Physics \& Astronomy, Leicester, LE1 7RH, UK
	\and {South African Astronomical Observatory, PO Box 9, Observatory 7935, Cape Town, South Africa} }

   \date{Received 30 March 2015 / Accepted 1 June 2015 }

  \abstract
   {The X-ray emission from flares on cool (i.e.\ spectral-type F--M) stars is indicative of very energetic, transient phenomena, associated with energy release via magnetic reconnection.}
   {We present a uniform, large-scale survey of X-ray flare emission. The XMM-Newton Serendipitous Source Catalogue and its associated data products provide an excellent basis for a comprehensive and sensitive
survey of stellar flares -- both from targeted active stars and from those observed
serendipitously in the half-degree diameter field-of-view of each observation.}
   {The 2XMM Catalogue and the associated time-series (`light-curve') data products have been used as the basis for a survey of X-ray flares from cool stars
in the Hipparcos Tycho-2 catalogue. In addition, we have generated and analysed spectrally-resolved (i.e.\ hardness-ratio), X-ray light-curves. Where available, we have compared XMM OM UV/optical data with the X-ray light-curves.}
   {Our sample contains $\sim 130$ flares with well-observed profiles; they originate from $\sim 70$ stars. The flares range in
duration from $\sim 10^3$ to $\sim 10^4$~s, have
peak X-ray fluxes from 
$\sim 10^{-13}$ to  $\sim 10^{-11}$ $\rm erg\,cm^{-2}\,s^{-1}$, 
peak X-ray luminosities from 
$\sim 10^{29}$ to  $\sim 10^{32}$ $\rm erg\,s^{-1}$, and X-ray energy output from 
$\sim 10^{32}$ to $\sim 10^{35}$ erg. 
Most of the $\sim 30$ serendipitously-observed stars have little previously reported information.
The hardness-ratio plots clearly illustrate the spectral (and hence inferred temperature) variations characteristic of many flares, and provide an easily accessible overview of the data.
We present flare frequency distributions from both target and serendipitous
observations. The latter provide an unbiased (with respect to stellar activity)
study of flare energetics; in addition, they allow us to predict numbers of stellar flares that may be detected in future X-ray wide-field surveys.
The serendipitous sample demonstrates the need for care when calculating flaring rates, especially when normalising the number of flares to a total exposure time, where it is important to consider both the stars seen to flare and those 
from which variability was not detected
(i.e.\ 
measured as non-variable), since in our survey, the latter outnumber the former by more than a factor ten.}
   {}


	\maketitle

\section{Introduction } \label{intro}

The X-ray emission from flares on late-type (i.e.\ spectral-type F--M, or `cool') stars is indicative of very energetic, transient phenomena (lasting typically from minutes to hours), associated with energy release via magnetic reconnection (see e.g.\ reviews by Haisch, Strong \& Rodon\`o 1991; Favata 2002; Favata \& Micela 2003; G\"udel 2004; G\"udel \& Naz\'e 2009).
The 2XMM serendipitous source catalogue (Watson et al.\ 2009, hereafter `2XMM') and its associated database of source-specific data products (including spectra and time-series)
provide an excellent starting point for a comprehensive and uniform survey of stellar flares,
with an unprecedented combination of sensitivity, time coverage and sky coverage. In addition, simultaneous, time-resolved UV and optical-band images are available from the XMM Optical Monitor (OM) telescope for a subset of the sources.

In this paper we present a survey of physical parameters (rise time, fall time, luminosity, energy etc) of flare events identified in the 2XMM time-series (otherwise referred to as 'light-curves') of late-type stars in the (Hipparcos) Tycho catalogue (or more specifically, Tycho-2, H\o g et al.\ 2000, hereafter `Tycho'). We distinguish between those stars observed intentionally as targets of the observations and those that lie serendipitously in the XMM field of view (FOV); the latter provide, in principle, an unbiased sample of stellar flares.

The Tycho catalogue provides a large sample of stars, and has been used extensively in conjunction with previous X-ray studies, from the ROSAT mission, 
e.g.\ for galactic structure investigations (e.g.\ Guillout et al.\ 1998a,b, 1999) and for follow-up of the detailed properties of active stars (e.g.\ Frasca et al.\ 2006; Klutsch et al.\ 2008).

Our catalogue of 2XMM-Tycho flares provides a basis for various studies, including:
the identification of high-activity stars, the search for `super-flares' (c.f.\ Schaefer et al.\ 2000),
statistics of stellar flaring, comparison with large solar flares, identification of suitable datasets for more detailed analysis (e.g.\ time-resolved spectra), inputs to modelling and theoretical studies of stellar magnetic activity and flare mechanisms, and implications for the stellar-system environment including exoplanets. Some of these aspects are discussed further in Sect.~\ref{discussion}.

Section~\ref{obs} of this paper outlines the XMM observations and describes the matching with Tycho stars. 
Section~\ref{analysis} describes the selection of the `flare' lightcurves and the estimation of flare-event parameters. 
Section~\ref{analOM} describes the associated OM light-curves available for a sub-set of the observations.
Section~\ref{results} presents the results, in terms of flare parameters and distributions, together with estimates of visibility/detection thresholds and correction factors that need to be taken into account in deriving statistical properties of the flare sample. 
Section~\ref{HRcurves} presents X-ray `hardness-ratio' light-curves, useful for identifying spectral changes (and presumed temperature variations) with time, during the evolution of the flares.
Section~\ref{discussion} discusses the outcomes of the survey, and compares them to several published models and scaling relations. 
Section~\ref{conclusions} summarises the work.

\section{XMM observations and matching with Tycho counterparts} \label{obs}

\subsection{2XMM} \label{obs2XMM}

Details of the 2XMM catalogue and its construction, and an outline of the XMM EPIC X-ray instrumentation and observations, are given in 2XMM. We summarise here the key features relevant to the present work.

The 3491 pointed observations that formed the basis of 2XMM 
were distributed relatively arbitrarily over the whole sky, with a total
coverage of $\sim 1$\%. 38\,320 source detections were sufficiently bright that 2XMM time-series are available, corresponding to 30\,498 unique sources, 16\% of the total sources in the catalogue. Of these detections, 2307 were indicated as variable, from 2001 unique sources, $\sim 1$\% of the total catalogue. The median flux (in the 'Total' energy band 0.2--12 keV) of detections with time-series is $\sim 2 \times 10^{-13} \rm \,erg\,cm^{-2}\,s^{-1}$; for the `variable' subset it is $\sim 4 \times 10^{-13} \rm \,erg\,cm^{-2}\,s^{-1}$. As noted in 2XMM, the ability to detect variability falls towards lower fluxes.
Light-curves for each detection have a bin-width that is an integer multiple of 10~s (with a minimum of 10~s), and set by the requirement to have an average of $\ge 18$ ct/bin for pn and $\ge 5$ for MOS, as computed from the source-detection count rates.
The median light-curve duration is $\sim 25$\, ks; however, periods of high background can reduce the useful time, and the median exposure times excluding these high-background intervals are $\sim 12$\, ks and $\sim 16$\, ks for pn and MOS cameras respectively.
The variability indicator was based on a simple $\chi^2$-test against a null hypothesis of constancy, with a probability requirement of $\le 10^{-5}$, and used only data during `Good Time Intervals' (GTIs, see 2XMM Appendix~A). This rather stringent threshold was chosen so that the expected number of false triggers over the entire set of 2XMM time-series was less than one (2XMM Sect.~8).

The observational period covered by 2XMM was 2000.09 to 2007.25, with a corresponding epoch range for the source-detection positions in the catalogue (this is relevant for matching with high-proper-motion stars, see Sect.~\ref{obsMatch}).

\subsection{Tycho-2} \label{obsTyc}

The Tycho-2 catalogue (H\o g et al.\ 2000, hereafter `Tycho') contains positions, proper motions and two-band ($V_T$, $B_T$) photometry for the 2.5 million brightest stars in the sky (from observations with the Hipparcos satellite); it is 90\% complete to $V \sim 11.5$, but has $\sim 50$\% of its stars in the range $11.5 \la V \la 13$.
The stellar surface density on the sky varies from $\sim 150$ stars\,deg$^{-2}$ at the galactic equator, to $\sim 50$ at $ b = \pm \rm 30^o $ to $\sim 25$ at the galactic poles (H\o g et al.\ 2000).

The observational period of Hipparcos/Tycho was 1989.85 to 1993.21, with a mean observational epoch of $\sim$J1991.5; the epoch of the Tycho-2 catalogue is J2000.0 (H\o g et al.\ 2000).

As sources of further information, we have matched the Tycho-2 catalogue with the Hipparcos catalogue (ESA 1997; and used revised parallaxes from van Leeuwen 2007), the Tycho-2 spectral type catalogue (Wright et al.\ 2003) and photometric distances from the N2K project (Ammons et al.\ 2006).

\subsection{Matching of the 2XMM and Tycho-2 catalogues} \label{obsMatch}

A combined 2XMM-Tycho catalogue was generated by matching (often alternatively referred to as `cross-correlating' or `joining') the sky positions of the objects in the two input catalogues using a maximum positional offset of 5\,arcsec, a value chosen from examination of the 2XMM positional errors (c.f.\ Watson et al.\ 2009, Sect.~9.5) and from trial matching out to a distance of 10\,arcsec. (The Tycho positional errors are much less than those for 2XMM, by more than an order of magnitude.)

In performing the matching, care was required to perform proper-motion corrections to the Tycho positions before comparison with 2XMM, in order to account for the (albeit relatively small) number of high-proper-motion stars in the 2XMM fields. There were three cases where proper motion (as given in Tycho) over the time range of 2XMM ($\sim 7$ years) was $\ga 14$\,arcsec, i.e.\ $\ga 2$\,arcsec/yr, large enough to significantly affect the position matching even after correction of the Tycho positions to the nominal mean epoch of 2XMM; these were: 61~Cyg~A (\object{HD 201091}), 61~Cyg~B (\object{HD 201092}) and \object{HD 95735}, each with a proper motion of $\sim 5$\,arcsec/yr. For these three stars, there was an additional complication in that 2XMM was found to have multiple `unique' sources corresponding to the proper-motion-corrected positions of each of these stars, since clearly the algorithm used in 2XMM to match individual detections to form unique sources (2XMM, Sect.~8.1) did not have proper motions available. Each star has two corresponding unique-source entries in 2XMM\footnote{The 2XMM unique-source reference numbers (SRCID) are: 177369, 177371 for 61~Cyg~A; 177373, 177375 for 61~Cyg~B; 91087, 91088 for HD~95735. The corresponding total numbers of individual source detections (DETIDs) for each star are 6, 6 and 2 respectively (excluding a likely spurious detection [SRCID = 91090] for HD~95735). }.

To remove matches of any one 2XMM detection with more than one star of a closely-spaced set
(all were pairs of stars, with a star-star separation of $\la 6$\,arcsec),
the match was considered to be (somewhat arbitrarily) with the optically-brighter (lower $V$ magnitude) star, while the information regarding the companion object was retained for later use if needed.
There were 84 detections, from 49 sources, where this action was taken;
43 of these detections, from 23 sources, had associated time-series, with 3 being flagged as variable.

Table~\ref{tabSurveySumm} summarises the results of matching the 2XMM and Tycho catalogues.
Overall, we define `cool stars' to be those with colour $B-V \ge 0.3$ (i.e.\ spectral type F0 or later, e.g.\ Allen 1973); these comprise $\sim 75$\% of the totals, comparable with 83\% for the fraction of stars in the whole Tycho-2 catalogue with
$B-V \ge 0.3$. (Spectral types for individual stars are considered in more detail for the 2XMM variable sources, as discussed in Sections~\ref{analysis} and \ref{dis-ids}.) Hence, there were, on average $\sim 1$--2 Tycho stars detected per 2XMM field, from $\sim 10$ times that number in total.

During the matching process, no account was taken of the 2XMM source quality information (see 2XMM Sect.~7), though this was used later, when examining the individual light-curves (see Sect.~\ref{analysis} and Appendix~\ref{appQual}).

The main sources of information on the names and astrophysical properties of the stars were the SIMBAD\footnote{http://simbad.u-strasbg.fr/simbad/ } database and the XMM-Newton data products for external catalogue cross-correlations (2XMM Sect.~6).

We have estimated the expected number of chance coincidences of Tycho stars and 2XMM sources by creating `simulated' Tycho catalogues having declinations increased by (somewhat arbitrarily) between 72 arcsec and 288 arcsec relative to their actual values and matching these with 2XMM at a maximum positional offset of 5 arcsec. This procedure indicates that the false match probability is $\la 3$\% for the entire 2XMM/Tycho sample and has a similar value for the subset of (brighter) 2XMM sources with time-series (and is little affected by restricting the 2XMM source quality flag SUM\_FLAG to $\le 2$). A simple analytical calculation based on the number of Tycho stars in 2XMM fields and the sky-area searched, yielded consistent results.

The total XMM-EPIC viewing time (i.e.\ observation time per star $\times$ number of stars) of the 2XMM-Tycho survey is given in Table~\ref{tabSurveySumm}, together with values for various subsets.

\begin{table*}
\caption{Summary of the 2XMM-Tycho survey characteristics.}
\label{tabSurveySumm}
\centering
\begin{tabular}{l r r}
\hline\hline
Quantity                                         &  \multicolumn{2}{c}{2XMM-Tycho}   \\
                                                 & All           & Cool stars  \\
\hline

No.\ of Tycho stars in 2XMM fields               & $\sim 26000$  & $\sim 19000$  \\
No.\ of 2XMM sources matched with Tycho stars    & 3042          & 2357  \\
No.\ of 2XMM detections matched with Tycho stars & 4772          & 3499  \\
No.\ of 2XMM sources with light-curves           & 808           & 611  \\
No.\ of detections with light-curves             & 1393          & 933   \\
No. of X-ray variable sources / stars            & 123 / 120     & 91 / 89  \\
No.\ of variable light-curves                    & 157           & 118  \\
No.\ of variable light-curves after quality checking & 128       & 96 $a$ \\
No. of X-ray variable stars after quality checking   & 85            & 76 $a$  \\

\hline

\ \ \ {\it 2XMM summed viewing time (Ms) on Tycho stars:}  \\
for all detected stars $(b)$                     & 119           & 87 (82)  \\
for all stars with X-ray light-curves $(a)$      &               & 29 (24)  \\
for all stars with X-ray variability  $(a)$      &               & 3.9 (1.8)  \\
for all stars with flares             $(a)$      &               & 3.0 (1.4)  \\
for all stars with fully-observed flares $(a)$   &               & 2.6 (1.2)  \\
for all stars with flares, $S:N>10$   $(a)$      &               & 2.1 (0.62)  \\
for all stars with fully-observed flares, $S:N>10$ $(a)$ &       & 1.9 (0.58)  \\

\hline
\end{tabular}
\newline
{\bf Notes:} Values in parentheses (...) relate to serendipitous observations, i.e.\ the star was not the target of the XMM observation. (a) CVS sample; (b) Includes all relevant 2XMM detections not only those with light-curves.
The terms `source' and `detection' are as defined in the 2XMM catalogue (Watson et al.\ 2009, Sect.\ 8.1). 

\end{table*}

\subsection{Derived quantities} \label{obsDerived}

We summarise here the calculation of the main additional quantities that were needed for our work, and not contained in the original 2XMM or Tycho-2 catalogues.

Standard, Johnson magnitudes and colours ($V$, $B-V$) for the stars were computed from the Tycho-2 values ($V_T$, $B_T$) using the formulae given in ESA (1997; c.f.\ H\o g et al.\ 2000).

Distances to the stars were computed, in order of decreasing preference, from:
(i) Hipparcos trigonometric parallaxes (ESA 1997, and using revised values from van Leeuwen 2007);
(ii) values gleaned from the literature, for a small number (13) of stars of particular relevance to our flare study (individual references are noted in Table~\ref{tabStars}); 
(iii) `photometric parallaxes' from the N2K project (Ammons 2006);
(iv) photometric parallaxes from the apparent magnitude $V$ on the assumption of a main-sequence (i.e.\ luminosity class V) object, and utilising the Tycho colour ($B-V$) information to estimate absolute magnitude ($M_\mathrm{V}$) and hence distance modulus.

2XMM X-ray luminosity ($L_\mathrm{X}$) of each object was computed using the derived distance,
and Tycho visual luminosity ($L_\mathrm{opt}$) directly from $M_\mathrm{V}$.

\section{Analysis of the X-ray light-curves} \label{analysis}

The 2XMM light-curves of all source detections flagged as variable, and matched as above with Tycho stars, were visually examined to characterise the variability and to check for potential problems that might give rise to spurious indications of variability.
Where available, we used the EPIC-pn light-curves, due to the generally higher signal:noise relative to MOS1 or MOS2.
The inspection included the background light-curve and the plot of time intervals used in the variability test.
In addition to the light-curves themselves, we also examined the 2XMM images for any evidence of problems such as source confusion\footnote{We used the 2XMM graphical light-curve products and the 2XMM summary page for each source detection. See e.g.\ LEDAS: http://www.ledas.ac.uk/arnie5/arnie5.php?action=basic\&catname=2xmm }. Appendix~\ref{appQual} summarises the main issues that led to rejection of specific light-curves for the purposes of the present work. Of the 157 `variable' light-curves examined, we removed 29 from further consideration (and not all of these were from cool stars).

Further information on all the remaining 2XMM/Tycho stars was sought from SIMBAD and from the XMM catalogue-crosscorrelation products. We retained for further analysis those stars with spectral type F, G, K or M, and those stars lacking available spectral types which had $B-V \ga 0.3$. In addition, six earlier-type (B--A) stars\footnote{
2XMM SRCID: 26140, 38388, 41960, 130617, 50242, 99128. The first four of these displayed clear flare-like events.}
lacking detailed information were retained for completeness pending further analysis and investigation. At this stage there were 96 variable light-curves from 76 stars (and 76 unique 2XMM sources). We will refer to this subset as the {\it Cool, Variable Sample} (CVS). 41 of these stars are the intentional target of the XMM observation, leaving 35 as serendipitous measurements. 

In order to check the behaviour of the variability test and to search for flares marginally below the 2XMM variability threshold, we have visually inspected all 1143 2XMM/Tycho light-curves not indicated as variable in 2XMM and for which the Tycho $B-V \ge 0.0$. There were 19 light-curves found with apparent variability. After detailed investigation of the associated background light-curves and GTIs, 7 were not considered further due to possible contamination of the source light-curves from background variability or confusion with nearby brighter, variable sources. Of the remaining 12 light-curves (from 11 stars), two had variability probability $\la 5 \times 10^{-5}$. The remaining 10 all showed apparent variability but with much of it outside the GTIs. We retained all 12 cases for further analysis; these all had $B-V \ge 0.3$. We will refer to this subset as the {\it Cool, Low-Variable Sample} (CLVS). There was some commonality of stars between CLVS and CVS. 7 of these stars are the intentional target of the XMM observation, leaving 4 as serendipitous measurements. 

Table~\ref{tabStars} summarises the properties of each star and corresponding X-ray source, in the CVS+CLVS samples. The table contains one row per XMM observation.
Figs.~\ref{figOMlc1} -- \ref{figOMlc5} and \ref{figHR2} -- \ref{figHR7} show examples of the variable X-ray light-curves\footnote{Note that the timeseries data products (FITS files and graphical light-curve plots) which form part of the 2XMM catalogue dataset are not corrected for off-axis angle i.e.\ for vignetting. Hence, light-curves from multiple observations of the same source at different off-axis angles will, in general, need to be vignetting corrected if they are to be directly compared (or their time-resolved count rates compared with the time-integrated rates [*\_RATE values] given in the catalogue). All X-ray count rates (and derived fluxes and luminosities) in this paper are vignetting corrected to give uniform, on-axis values; this includes the X-ray light-curves in the Figures. }, 
while the complete set (CVS+CLVS, 108 X-ray light-curves) is given in Figs.~\ref{figALLlcOM} and \ref{figALLlcHR}.


\subsection{Characterisation of the X-ray variability} \label{analVar}

We have characterised the apparent form of the variability in each light-curve by visual inspection of the 2XMM time-series graphical data product and by inspecting light-curves binned at various time integrations to examine individual features such as flares, in more detail. The types of variability can be broadly placed into the following categories\footnote{
We found no examples of periodic variability in our sample.}:
{\it flare:} corresponding to a clear rise above and then fall back towards, a quiescent level;
{\it trend:} a rise or fall in the source count rate over the course of an observation;
{\it gradual:} a rise then fall, or vice versa, but without the relatively fast rise usually associated with `classic' flares;
{\it indeterminate:} there was no clear form or structure to the variability.
These definitions are somewhat subjective, and in particular it may be noted that: 
(a) the acceptance criteria for {\it flares} will need to be accounted for when considering their statistical properties later in this work; 
(b) {\it gradual} events may in some cases be `unusual' flares rather than e.g.\ due to active-region evolution or rotational modulation; 
(c) a downward {\it trend} may in some cases be the later stages of a flare whose rise and peak have not been observed; 
(d) {\it indeterminate} variability may in some cases be due to low-level flares which cannot be individually resolved;
(e) there are some {\it flares} where part of the event (one or more of the rise, peak or fall phases) has not been observed and hence the event cannot be fully characterised.
Table~\ref{tabVarSumm} summarises the results of the variability characterisation; 
flaring is the dominant form of variability, with $ > 80$\% of the X-ray variable stars displaying this behaviour.

\begin{table*}
\caption{Summary of the X-ray variability characterisation.}
\label{tabVarSumm}
\centering
\begin{tabular}{l l c c c c}
\hline\hline
Variability   & Sample & Number        & Number of    & \multicolumn{2}{c}{Number of events} \\
type          &        & of stars      & light-curves & Total     & Completely \\     
              &        &               &              &           & observed   \\
\hline
Flare         & CVS    &    63 (30)    &  76 (32)     & 133 (39)  & 116 (34) \\
              & CLVS   &     8 (4)     &   8 (4)      &  11 (4)   &  11 (4)  \\
              & CVS+CLVS &  70 (34)    &  84 (36)     & 144 (43)  & 127 (38) \\
              &        &               &              &           &      \\
Trend         & CVS    &    10 (1)     &  11 (1)      &  11 (1)   &      \\
Gradual       & CVS    &    3 (1)      &  3 (1)       & 3 (1)     &      \\
              & CLVS   &    1 (0)      &  2 (0)       & 2 (0)     &      \\
Indeterminate & CVS    &    6 (5)      &  6 (5)       & 6 (5)     &      \\
              & CLVS   &    2 (0)      &  2 (0)       & 2 (0)     &      \\ 
Poor bgd sub? {\it (a)} & CVS    &    2 (0)      &  2 (0)       & 2 (0)     &      \\
              &        &               &              &           &      \\              
All           & CVS    &    76 (36)    &  96 (39)     &           &      \\
              & CLVS   &    11 (4)     &  12 (4)      &           &      \\
              & CVS+CLVS &  84 (40)    & 108 (43)     &           &      \\
\hline
\end{tabular}
\newline
{\bf Notes:} Values in parentheses (...) relate to serendipitous observations, i.e.\ the star was not the target of the XMM observation.
(a) These occur as events within multi-event light-curves, and hence do not add to the total number of stars or light-curves. 
\end{table*}

We have characterised each X-ray flare using the parameters:
count rate\footnote{These are corrected to on-axis values. If necessary (where the original 2XMM timeseries data products were used), an approximate correction was made by multiplying the count-rates by the factor ca\_8\_RATE/AVRATE, where ca\_8\_RATE is the emldetect source count rate for the appropriate EPIC camera and AVRATE is the time-averaged count rate of the timeseries (recorded in the data-product FITS-file metadata).}
(above the quiescent level) at maximum ($c_\mathrm{max}$); 
the corresponding time ($t_\mathrm{max}$);
the times ($t_\mathrm{r}$, ($t_\mathrm{f}$), during the rise and fall of the flare, when the count rate is $c_\mathrm{max} / \mathrm{e}$; 
the quiescent count rate ($c_\mathrm{q}$) outside of flaring (usually determined from a period close but prior to the start of the flare).
Additional, derived parameters included the rise- and fall-time, $\tau_\mathrm{r} = t_\mathrm{max} - t_\mathrm{r}$ and $\tau_\mathrm{f} = t_\mathrm{f} - t_\mathrm{max}$.
Conversion of count rate to flux used the same factor as the corresponding 2XMM total-band source detection\footnote{
i.e.\ we used a single conversion factor for each light-curve, namely the ratio of the 2XMM total-band count rate and flux for the relevant EPIC camera and filter. }; 
X-ray luminosity $L_\mathrm{X}$ 
and X-ray emitted energy $E_\mathrm{X}$\footnote{$E_\mathrm{X}$ was integrated over the time interval $\tau_\mathrm{r} + \tau_\mathrm{f}$, unless otherwise stated.}
were then derived as in Sect.~\ref{obsDerived}.
In order to judge the statistical significance of a flare, 
we computed the signal:noise (S:N) ratio\footnote{
With S:N defined in the sense $ S / \sqrt{T} $, where $S$ is the
number of counts
 above the quiescent level, summed over the time interval $\tau_\mathrm{r} + \tau_\mathrm{f}$, and $T$ is the total count (i.e.\ due to flare + quiescent emission + non-source background) over the same interval.}
above the estimated quiescent emission level, over the time interval $\tau_\mathrm{r} + \tau_\mathrm{f}$.
For some of the statistical analyses presented later, we restrict the samples to flares which were `fully observed', defined  as those with measured values for both $\tau_\mathrm{r}$ and $\tau_\mathrm{f}$, and we refer to the `duration' of a flare as 
$\tau_\mathrm{r} + \tau_\mathrm{f}$.
The parameter values for each flare/event are listed in Table~\ref{tabVarDetail}.


\section{The XMM optical/ultraviolet light-curves} \label{analOM}

Optical and ultraviolet data can yield information about those regions of the stellar atmosphere at lower temperatures than are seen in X-rays, hence giving a more complete view of the flaring process (e.g.\ G\"udel 2004, Sect.~12.14; Mitra-Kraev et al.\ 2005a).

\begin{figure}
	\resizebox{\hsize}{!}{\includegraphics[angle=-90]{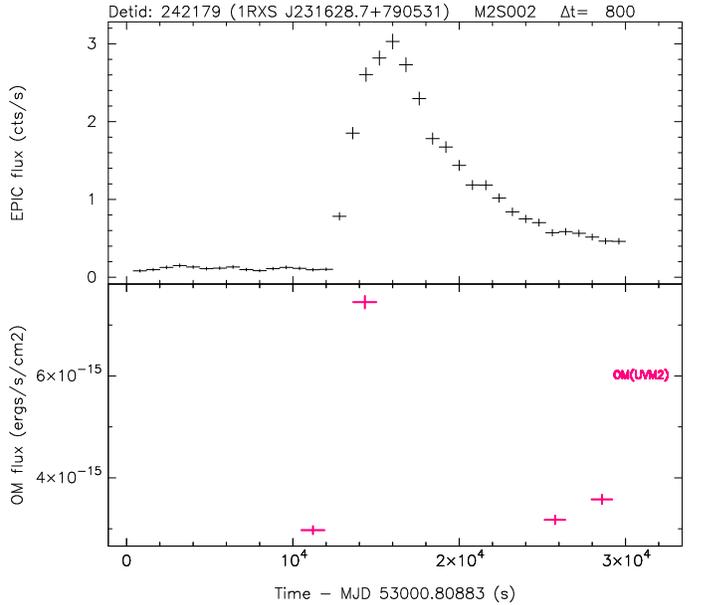}}
	\caption{{\it (From top)} {\it (a)} XMM EPIC MOS2 X-ray light-curve for the star 2MASS J23163068+7905362 (= \object{1RXS J231628.7+790531}), energy band 0.2-12 keV, 800-s time bins; {\it (b)} corresponding XMM OM UV light-curve, with filters as noted on the plots; the duration of each OM data point is indicated by the horizontal bars.}
	\label{figOMlc1}
\end{figure}

\begin{figure}
	\resizebox{\hsize}{!}{\includegraphics[angle=-90]{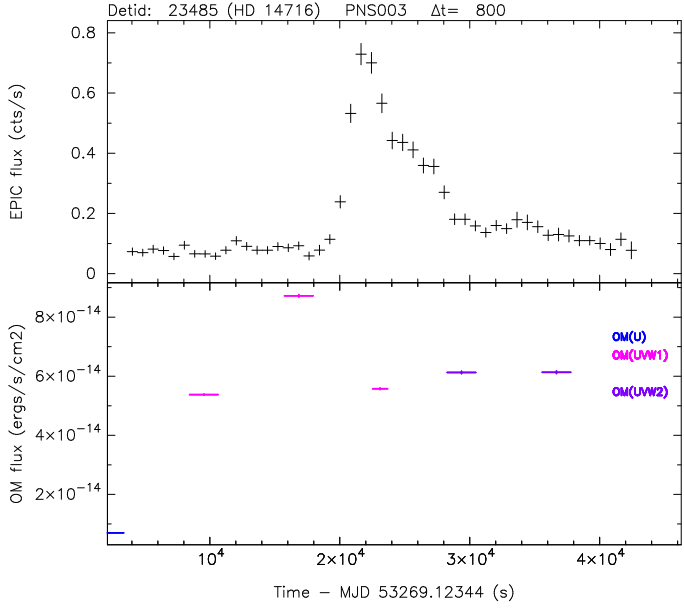}}
	\caption{As Fig.~\ref{figOMlc1}, but for the star \object{HD 14716} observed with PN.}
	\label{figOMlc2}
\end{figure}

\begin{figure}
	\resizebox{\hsize}{!}{\includegraphics[angle=-90]{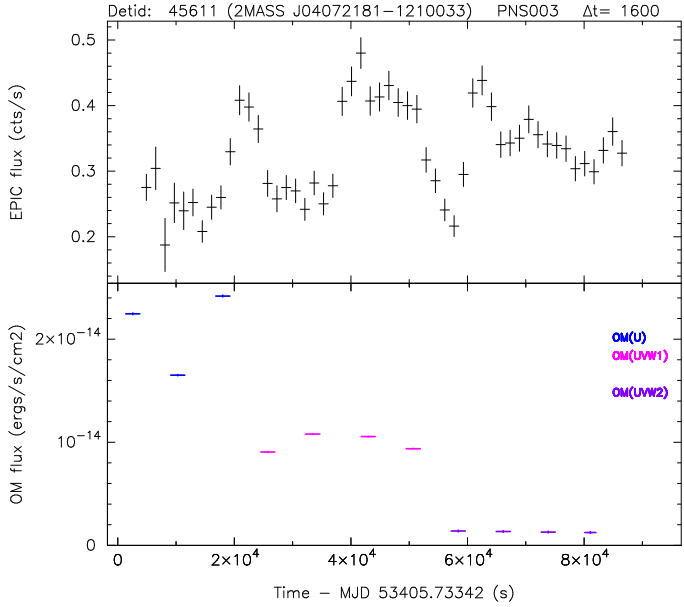}}
	\caption{As Fig.~\ref{figOMlc1}, but for the star \object{2MASS J04072181-1210033} observed with PN and with 1600-s time bins. }
	\label{figOMlc3}
\end{figure}

\begin{figure}
	\resizebox{\hsize}{!}{\includegraphics[angle=-90]{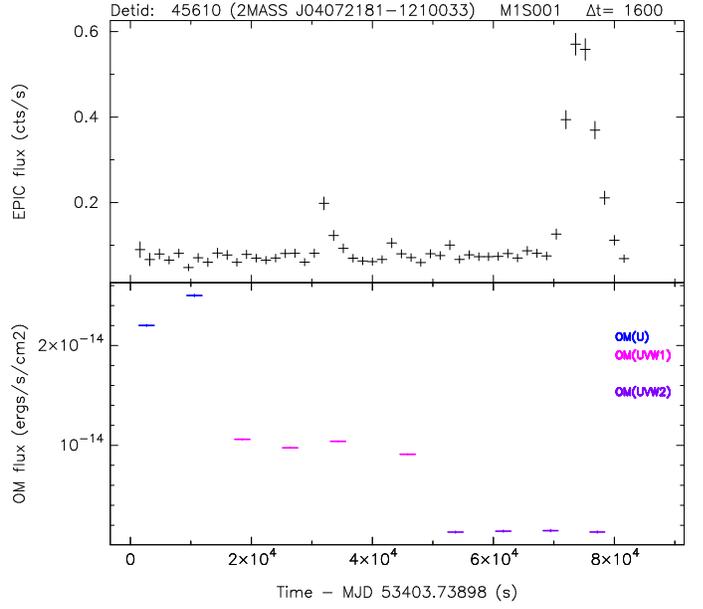}}
	\caption{As Fig.~\ref{figOMlc1}, but for the star \object{2MASS J04072181-1210033} observed with MOS1 and with 1600-s time bins. }
	\label{figOMlc4}
\end{figure}

\begin{figure}
	\resizebox{\hsize}{!}{\includegraphics[angle=-90]{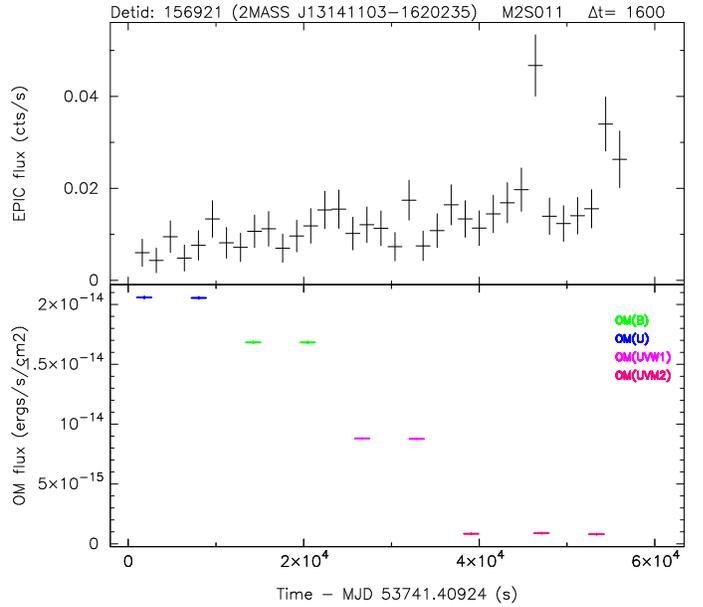}}
	\caption{As Fig.~\ref{figOMlc1}, but for the star \object{2MASS J13141103-1620235} observed with MOS2 and with 1600-s time bins. }
	\label{figOMlc5}
\end{figure}

For all CVS+CLVS stars which fell within the FOV of the XMM-Newton Optical/UV Monitor telescope (OM) (Mason et al.\ 2001) during the corresponding EPIC observations, we have extracted the optical/UV photometry from the OM source lists\footnote{
The OM full FOV is $17 \times 17$~arcmin$^2$, covering the central portion of the EPIC X-ray FOV. During a given observation, coverage of this full field may be non-uniform depending on the specification of OM observing modes.
}.
For many of the stars which were targets of the observations, OM {\it fast-mode} data were available, with a basic time resolution of 0.5~s, subsequently binned in the pipeline data products to 10~s, and then further binned at typically $\sim 300$ -- $\sim 3000$~s to match approximately the corresponding EPIC data plots. For the serendipitous sources, there were more limited data, in the form of fluxes integrated over typically $\sim 1000$~s, representing individual instrument exposures; we refer to these data as {\it imaging-mode} or {\it low-time-resolution} data. There were often more than one waveband filter used sequentially during an observation, thus complicating the identification of variability in the OM light-curves. However, there were $\sim 120$ light-curves (each corresponding to a unique 2XMM DETID) with at least two OM photometry data points in the same filter. We restrict ourselves here primarily to presentation of low-time-resolution light-curves for serendipitous sources, and with the additional constraint that the OM data were taken during at least part of the X-ray flaring period. 
This results in five OM light-curves, from four stars, as shown in
Figs.~\ref{figOMlc1} -- \ref{figOMlc5}; also plotted are the corresponding X-ray data for comparison. 

All available OM data (fast-mode where available, for target stars only, and low-time-resolution data for the serendipitous sample) are presented in Appendix~\ref{appALLlcOM} alongside the X-ray light-curves.

\section{Results } \label{results}

From this point, we will focus our studies on the set of 116 flares from the CVS sample having complete (as far as we can determine) coverage of the event profiles. Of particular interest are the 34 such flares which originated from (30) serendipitously-observed stars
(Table~\ref{tabVarSumm}), since we can use these data to estimate, or at least usefully constrain, the frequency of flare events at X-ray fluxes much lower than previous serendipitous surveys (e.g.\ Pye \& McHardy 1983).

\begin{figure}
	\resizebox{\hsize}{!}{\includegraphics{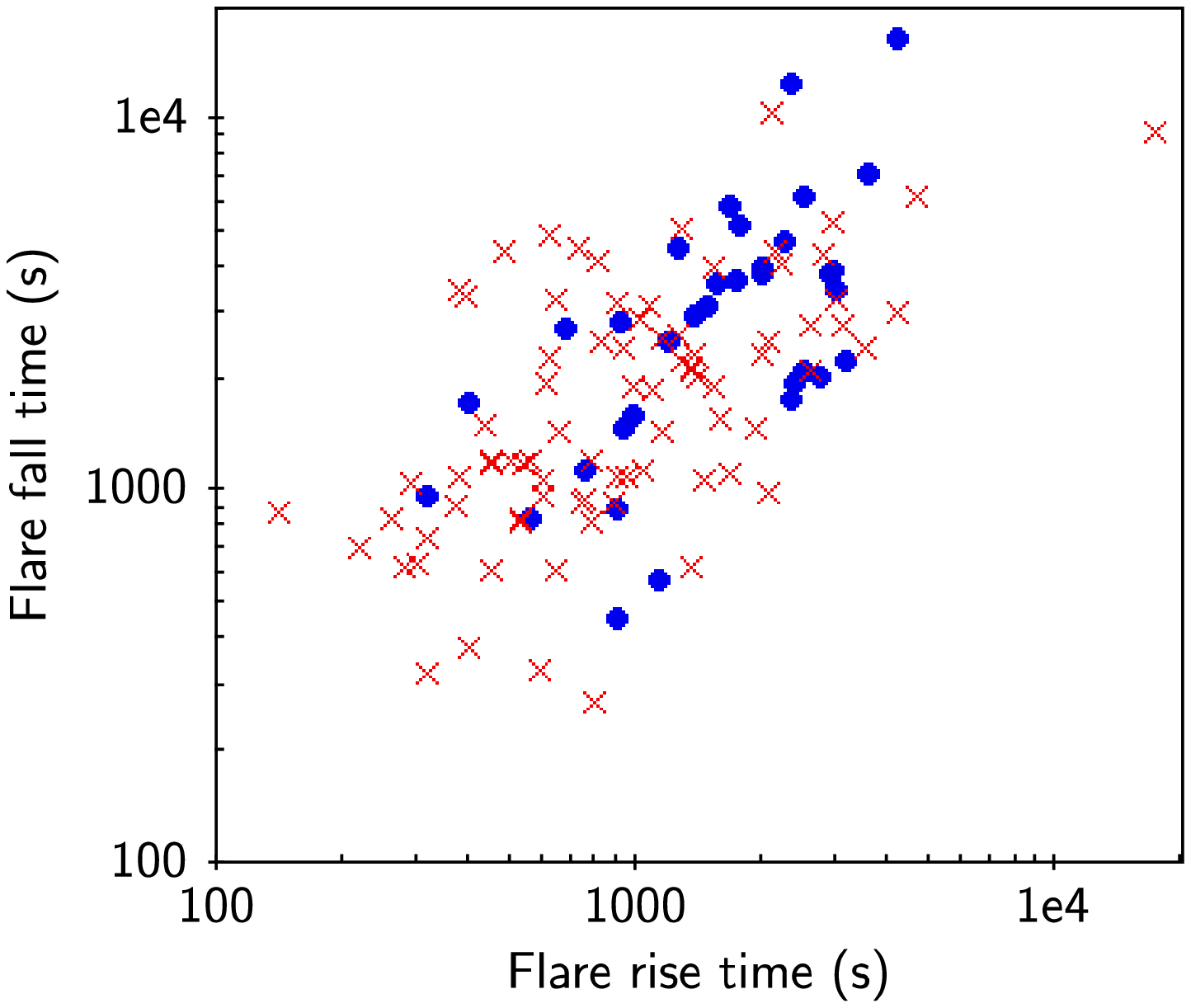}}
	\caption{Flare fall-time versus rise-time. Only `fully-observed' flares are shown
	(see text for details). Key to symbols: 
	blue circles: Serendipitously-observed stars in the CVS sample;
	red diagonal crosses: Target stars in the CVS sample.
		}
	\label{figTscatter}
\end{figure}

\begin{figure}
	\resizebox{\hsize}{!}{\includegraphics{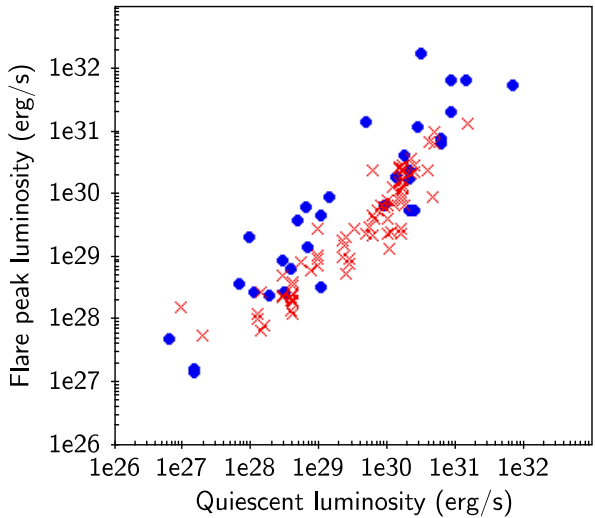}}
	\caption{Flare peak X-ray luminosity versus quiescent X-ray luminosity  (energy band 
	0.2 -- 12 keV). Symbols as in Fig.~\ref{figTscatter}. }
	\label{figLscatter}
\end{figure}

\begin{figure}
	\resizebox{\hsize}{!}{\includegraphics{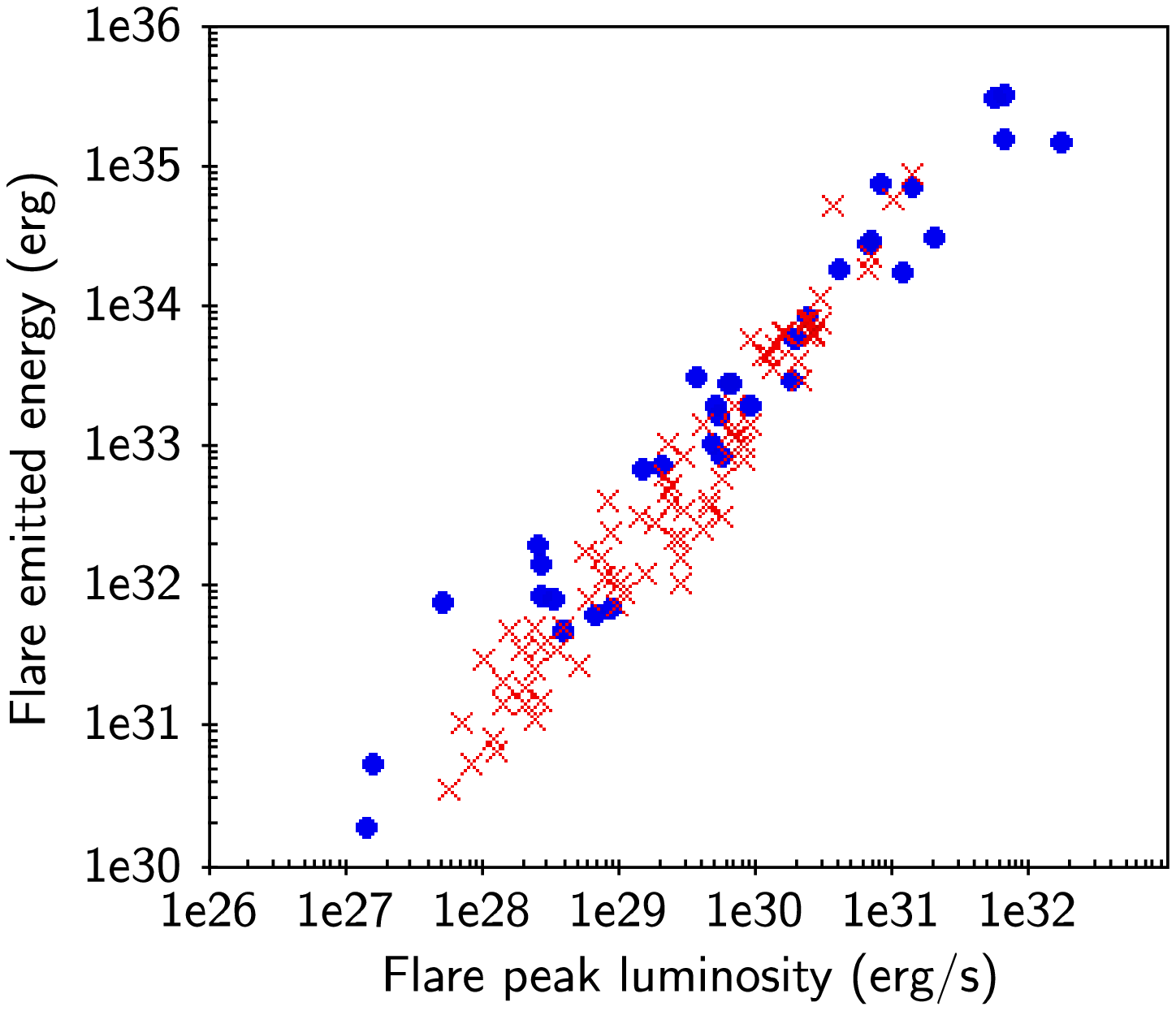}}
	\caption{Total X-ray emitted energy versus flare peak X-ray luminosity (energy band 
	0.2 -- 12 keV). Symbols as in Fig.~\ref{figTscatter}. }
	\label{figLEscatter}
\end{figure}

Figs.~\ref{figTscatter}, \ref{figLscatter} and \ref{figLEscatter} summarise respectively the distributions of 
rise time ($\tau_\mathrm{r}$) versus fall time ($\tau_\mathrm{f}$),
quiescent X-ray luminosity ($L_\mathrm{X,quies}$) versus peak X-ray luminosity ($L_\mathrm{X,peak}$), 
and $L_\mathrm{X,peak}$ versus total X-ray energy ($E_\mathrm{X}$) emitted over the time period
$\tau_\mathrm{r} + \tau_\mathrm{f}$.
Simulations have been used to estimate errors in the derived parameters, as noted below.

The observed flare rise- and fall-times each span two orders of magnitude, ranging from $\sim 200$~s to $\sim 20\, 000$~s (c.f.\ Fig.~\ref{figTscatter}). The upper and lower boundaries are set by the observational constraints: the former by the longest XMM exposure durations, the latter by the sensitivity of the instrument and data analysis method in permitting identification of short-duration events. 
Simulations indicate that the uncertainty on $\tau_\mathrm{r}$ and $\tau_\mathrm{f}$ is $\sim \pm 20$\% for S:N$\ga 10$ and $\tau \ga 2$~ks; for shorter $\tau$, the uncertainty may be larger: $\sim \pm 30$\% for $\tau \sim 1$~ks.
It is evident that although the majority of the flares have a `classsic' profile with $\tau_\mathrm{r} < \tau_\mathrm{f}$, there is apparently a significant number with $\tau_\mathrm{r} > \tau_\mathrm{f}$ (see e.g.\ G\"udel 2004, Sect.~12.11). However, taking account of the above uncertainties, low S:N in some cases, and additional uncertainties due to some flares being overlapped (`confused') in time, leaves only one potentially significant case: DETID=46637 ($\tau_\mathrm{r} \approx 17$~ks, $\tau_\mathrm{f} \approx 9$~ks), identified with the T~Tau-type star V410~Tau, and even in that case it might be argued that the apparently long rise time could be an artifact of two or more overlapping events.

The stars' quiescient X-ray luminosities range from $\sim 10^{27}$ to
$\sim 10^{31}\ \rm erg\ s^{-1}$, while the flare peak X-ray luminosities range from
$\sim 10^{28}$ to $\sim 10^{32}\ \rm erg\ s^{-1}$ (see Fig.~\ref{figLscatter}), the lower values being set by the observational sensitivity. The upper values for $L_\mathrm{X,quies}$ correspond to the maximum typically associated with cool-star coronal emission (see e.g. G\"udel 2004). 
The observed flare peak X-ray luminosities range from $\sim 0.1$ to
$\sim 50$ times the quiescent levels, i.e.\ a dynamic range of 500, with a median
$L_\mathrm{X,peak} / L_\mathrm{X,quies}$ of $\sim$1--2 depending on the precise choice of sample.
Simulations indicate that the uncertainty on $L_\mathrm{X,peak}$ (or more strictly the associated maximum count rate) is $\la \pm 20$\% for S:N$\ga 10$, with a significant bias (measured/true value) $\sim 0.7$ at S:N$\sim 10$, becoming insignificant (i.e.\ $\sim 1.0$) by S:N$\sim 15$.
The apparent strong correlation shown in Fig.~\ref{figLscatter} between $L_\mathrm{X,peak}$ and $L_\mathrm{X,quies}$, arises from a combination of observational bias and intrinsic stellar properties (c.f.\ Audard et al.\ 2000): weak flares on stars with high $L_\mathrm{X,quies}$, though likely to be occurring, are not detectable (towards lower right of plot), while very strong flares on low $L_\mathrm{X,quies}$ stars (towards upper left of plot) have a low occurrence rate (though such `super flares' are occasionally seen, e.g.\ Osten et al.\ 2010). Fig.~\ref{figLscatter} also demonstrates the intrinsically large range in $L_\mathrm{X,peak}$ at a given $L_\mathrm{X,quies}$, i.e.\ any given star produces a wide range of flare strengths. 
The distribution of flare luminosities is discussed in more detail below.

The flare total X-ray emitted energy (see Fig.~\ref{figLEscatter}) ranges from
$\sim 10^{31}$ to $\sim 10^{35}$ erg. The photon passband of the EPIC light-curves (0.2--12 keV) encompasses $>70$\% of the emitted radiation for coronal emission at the temperatures 
$kT \ga 1$~keV generally measured during stellar flares (and often in quiescence also; see e.g.\ G\"udel 2004). (Temperature values of this order are also supported by the time-averaged hardness ratios, see e.g.\ 2XMM Sect.~9.7.)
As noted later (Sect.~\ref{param-dis}(5)), Fig.~\ref{figLEscatter} indicates an approximately linear relation between $E_\mathrm{X}$ and $L_\mathrm{X,peak}$, implying that flare duration is essentially independent of amplitude.

\begin{figure}
	\resizebox{\hsize}{!}{\includegraphics{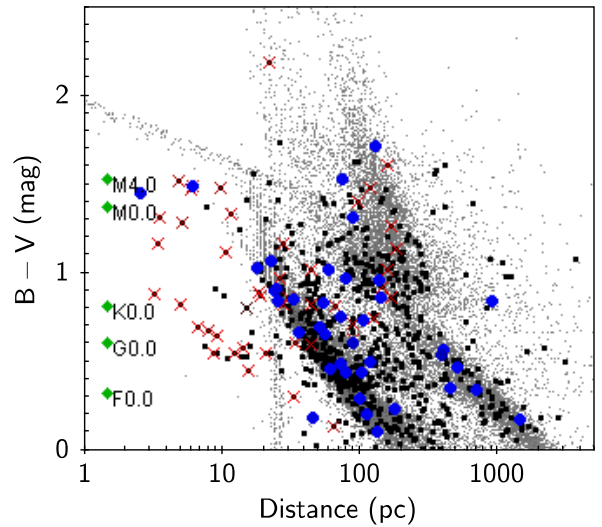}}
	\caption{Stellar distance (pc) versus colour ($B-V$, as an indicator of spectral type).
	Key to symbols: 
	blue filled circles: Serendipitously-observed stars in the CVS sample;
	red diagonal crosses: Target stars in the CVS sample;
	black dots: all Tycho stars with 2XMM light-curves; 
	grey dots: all Tycho stars falling in 2XMM fields (without regard to 2XMM detection).
	For the last two classes, some points lie outside the plot and are therefore omitted, 
	in order to improve visibility of the main distribution. }
	\label{figBVDistScatter}
\end{figure}

The maximum distance of the flaring stars in our survey is $\sim 1$~kpc, as illustrated in
Fig.~\ref{figBVDistScatter}. This figure also demonstrates that, as might be expected from the magnitude-limited nature of the Tycho catalogue, the maximum distance of the stars with 2XMM light-curves (whether detected as variable or not) is a strong function of stellar spectral type (as indicated by colour $B-V$). We can also see that for the CVS sample the serendipitously-observed stars are, overall, more distant and of earlier spectral type (smaller $B-V$) than the target stars, with the median distances being $\sim 140$ and $\sim 35$~pc respectively (though with considerable overlap in the distributions); the corresponding values for the overall Tycho `cool-star'
(i.e.\ $B-V \ge 0.3$)
light-curves set (irrespective of variability) are $\sim 145$ and $\sim 45$~pc respectively.
For all Tycho `cool' stars falling in 2XMM fields (irrespective of detection in 2XMM), the median distance is $\sim 195$~pc.
For the serendipitously-observed stars these figures give some measure of re-assurance that our sample of flares is reasonably representative, while for the target stars they reflect the obvious fact that the objects were observed as known, bright, coronally-active stars.

\begin{figure}
	\resizebox{\hsize}{!}{\includegraphics{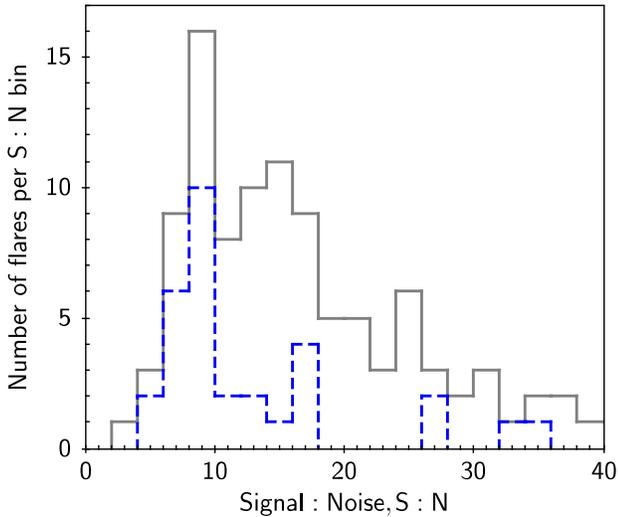}}
	\caption{Distribution of (fully observed) flares in signal:noise (S:N), 
	in S:N bins of 2. 
	Grey, solid line: all fully-observed flares (127) in the CVS+CLVS samples; 
	blue, dashed line: fully-observed flares (34) in the CVS serendipitous sample. The plotted histograms have been truncated at the right. }
	\label{figSNRhist}
\end{figure}

As mentioned earlier, the threshold for detection of the flares is essentially set by their signal:noise (S:N) relative to the quiescent emission level. Hence, the thresholds in terms of peak flux, fluence, peak luminosity or emitted energy will vary among the datasets depending on the photon-counting noise in the individual light-curves.
Fig.~\ref{figSNRhist} shows the distribution of flares in S:N, with a clear drop in the observed numbers for S:N$ \la 8$--10. Although it is not possible to give a definitive value for completeness versus S:N, simulations support the view that flares with S:N down to at least 10 should be detected at $\sim 100$\% efficiency, and we will use S:N=10.0 as threshold to define a `complete' sample where necessary, e.g.\ for estimation of flare frequency (Sect.~\ref{dis-rates}).

\section{X-ray hardness ratios and light-curves} \label{HRcurves}

As additional data products specifically for our flare survey, we have generated spectrally-resolved light-curves, utilising X-ray `hardness-ratios'.
The hardness-ratio plots clearly illustrate the spectral (and hence inferred temperature) variations characteristic of many flares, and provide an easily accessible overview of the data, prior to further analysis such as time-resolved spectral fitting.

\begin{figure}
	\resizebox{\hsize}{!}{\includegraphics{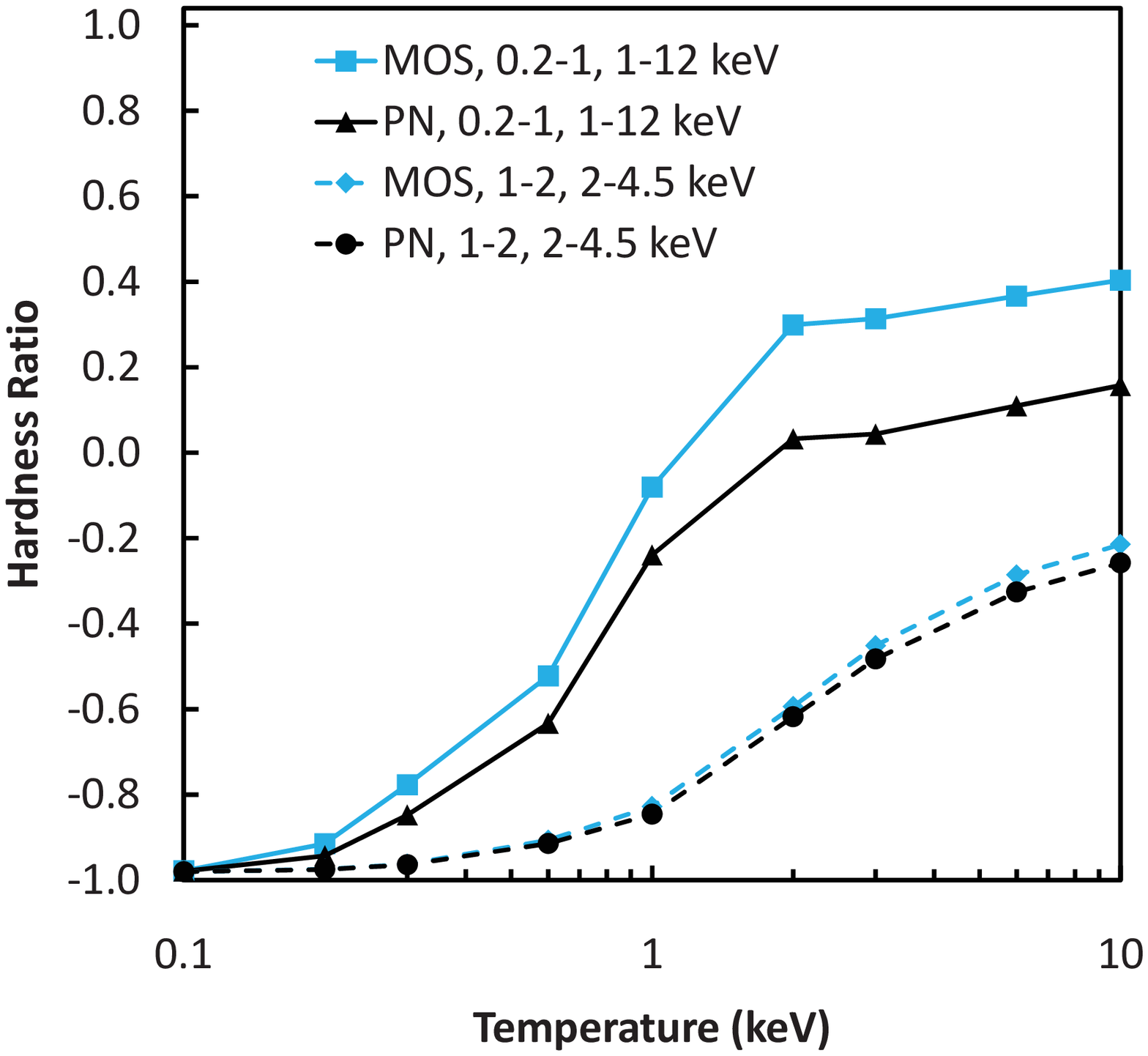}}
	\resizebox{\hsize}{!}{\includegraphics{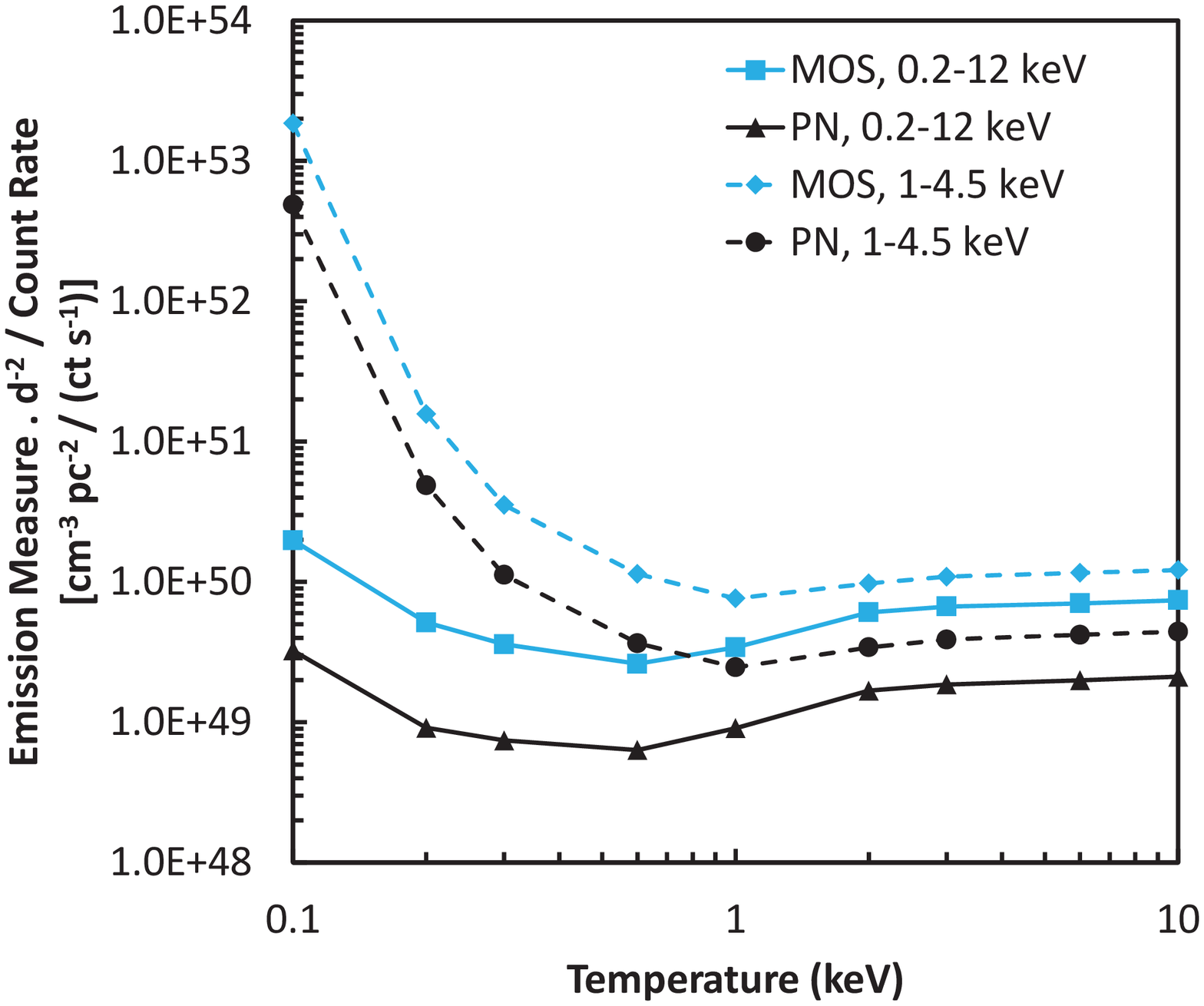}}
	\caption{{\it(From top) (a)} XMM EPIC MOS and PN hardness ratios (HR, see text for definition) versus temperature, predicted from an optically-thin, thermal spectral (MEKAL) model (using XSPEC), assuming negligible absorption column density. Solid lines -- `standard' HRs: 0.2-1, 1-12 keV; dashed lines -- `harder' HRs: 1-2, 2-4.5 keV. Symbols: MOS -- square and diamond; PN -- triangle and circle. Use of the MOS and PN medium-thickness filters was assumed in all cases, but this does not, in general, have a major effect on the derived HRs. {\it(b)} As (a), but for the predicted emission measure ($\rm cm^{-3} pc^{-2}$) per unit count rate (ct/s). }
	\label{figHRpred1}
\end{figure}

\begin{figure}
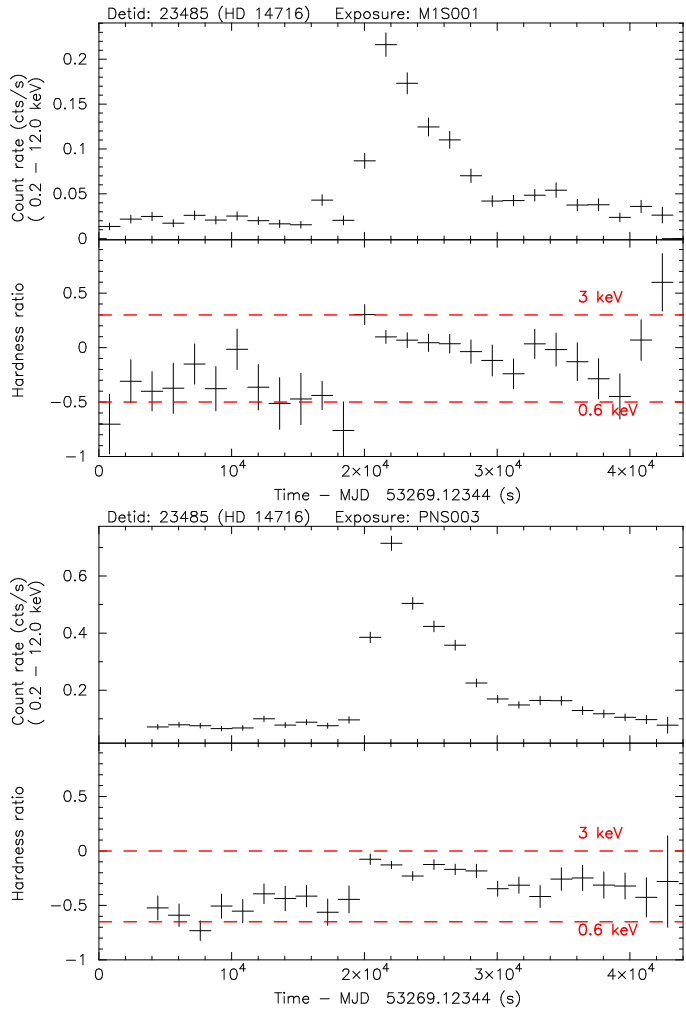

	\resizebox{\hsize}{!}{\includegraphics[angle=-90]{fig-HRlc-02a.eps}}
	\resizebox{\hsize}{!}{\includegraphics[angle=-90]{fig-HRlc-02b.eps}}
	\caption{{\it (From top)} {\it (a)} XMM EPIC MOS1 X-ray light-curve for the star \object{HD 14716}, energy band 0.2-12 keV, 1600-s time bins; {\it (b)} corresponding hardness-ratio light-curve using bands 0.2-1, 1-12 keV, approximate temperatures are indicated. {\it (c), (d)} as panels {\it (a, b)} but for EPIC PN.  }
	\label{figHR2}
\end{figure}

\begin{figure}
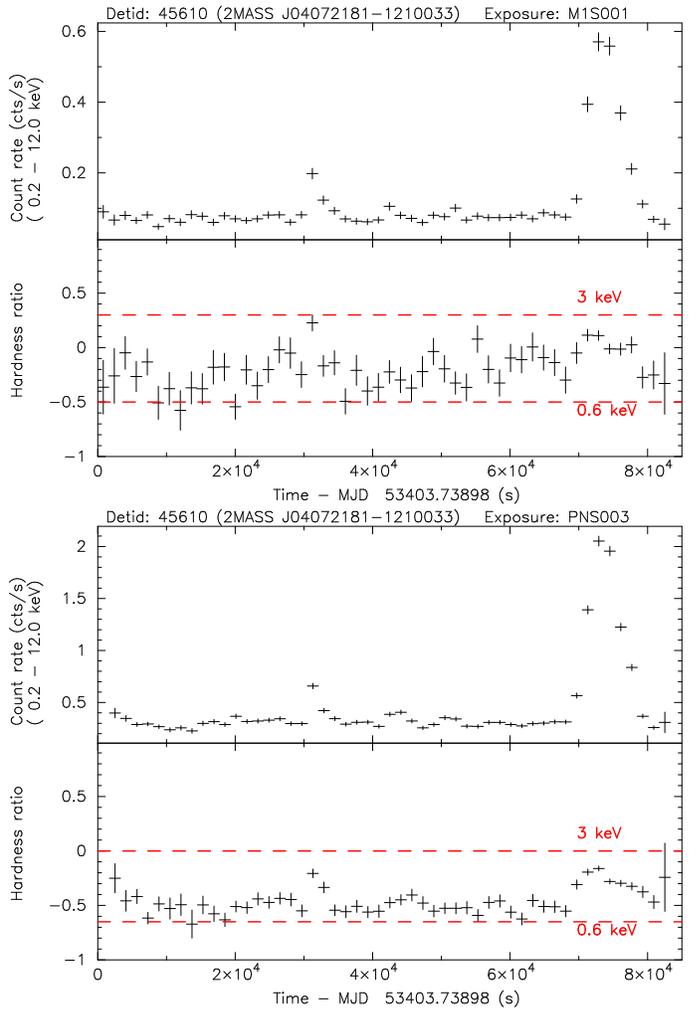

	\resizebox{\hsize}{!}{\includegraphics[angle=-90]{fig-HRlc-03a.eps}}
	\resizebox{\hsize}{!}{\includegraphics[angle=-90]{fig-HRlc-03b.eps}}
	\caption{As Fig.~\ref{figHR2}, but for the star \object{2MASS J04072181-1210033}.  }
	\label{figHR3}
\end{figure}

\begin{figure}
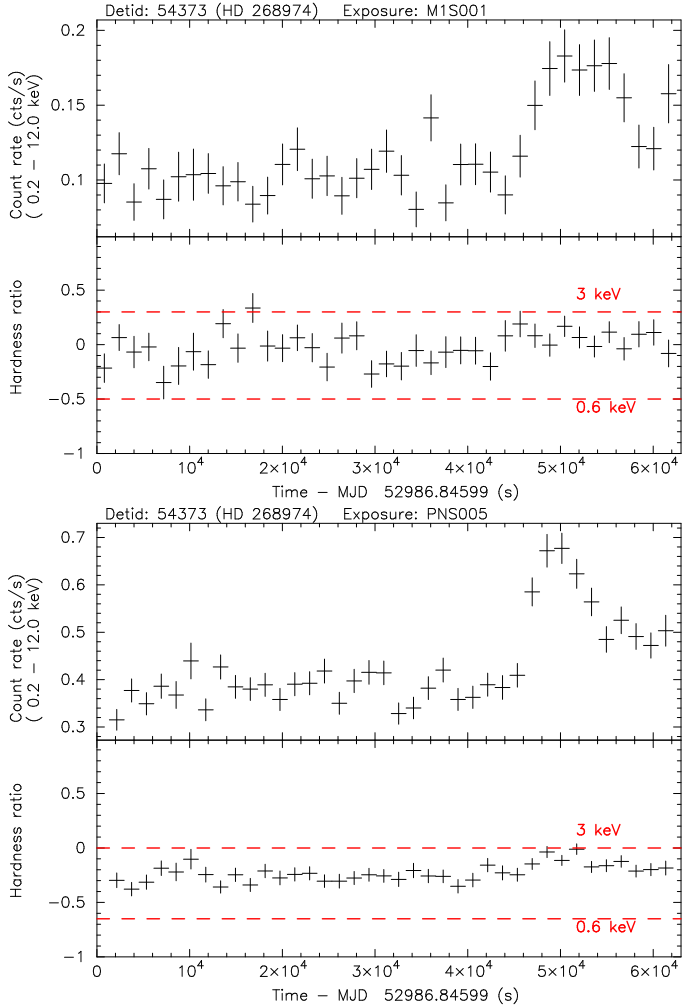

	\resizebox{\hsize}{!}{\includegraphics[angle=-90]{fig-HRlc-04a.eps}}
	\resizebox{\hsize}{!}{\includegraphics[angle=-90]{fig-HRlc-04b.eps}}
	\caption{As Fig.~\ref{figHR2}, but for the star \object{HD 268974}.  }
	\label{figHR4}
\end{figure}

\begin{figure}
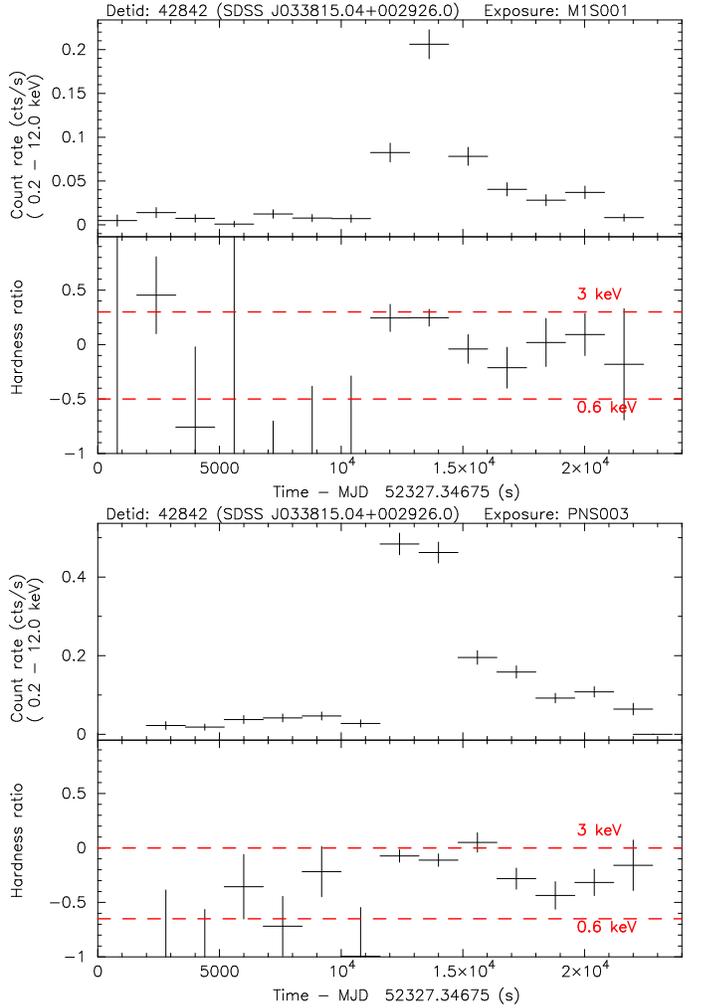

	\resizebox{\hsize}{!}{\includegraphics[angle=-90]{fig-HRlc-05a.eps}}
	\resizebox{\hsize}{!}{\includegraphics[angle=-90]{fig-HRlc-05b.eps}}
	\caption{As Fig.~\ref{figHR2}, but for the star \object{SDSS J033815.04+002926.0}.  }
	\label{figHR5}
\end{figure}

\begin{figure}
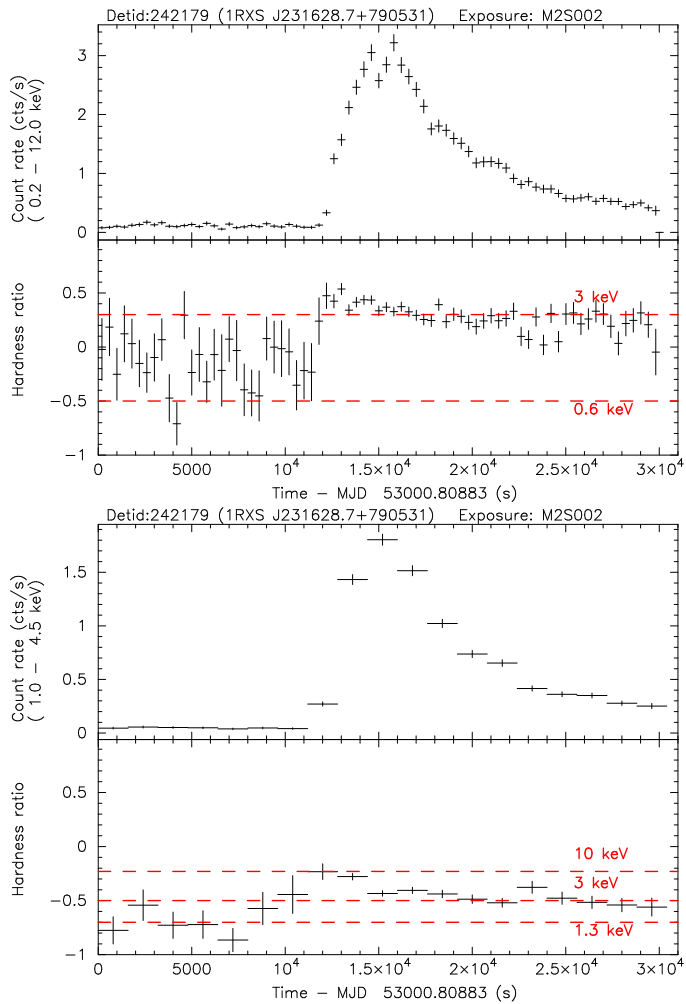

	\resizebox{\hsize}{!}{\includegraphics[angle=-90]{fig-HRlc-06a.eps}}
	\resizebox{\hsize}{!}{\includegraphics[angle=-90]{fig-HRlc-06b.eps}}
	\caption{{\it (From top)} {\it (a), (b)} As Fig.~\ref{figHR2}, but for the star 
	2MASS J23163068+7905362 (= \object{1RXS J231628.7+790531}), MOS2 only, 400-s time bins.
{\it (c), (d)} As panels {\it (a, b)}, but for the `harder' HR using the energy bands 1-2, 2-4.5 keV; 1600-s time bins.}
	\label{figHR6}
\end{figure}

\begin{figure}
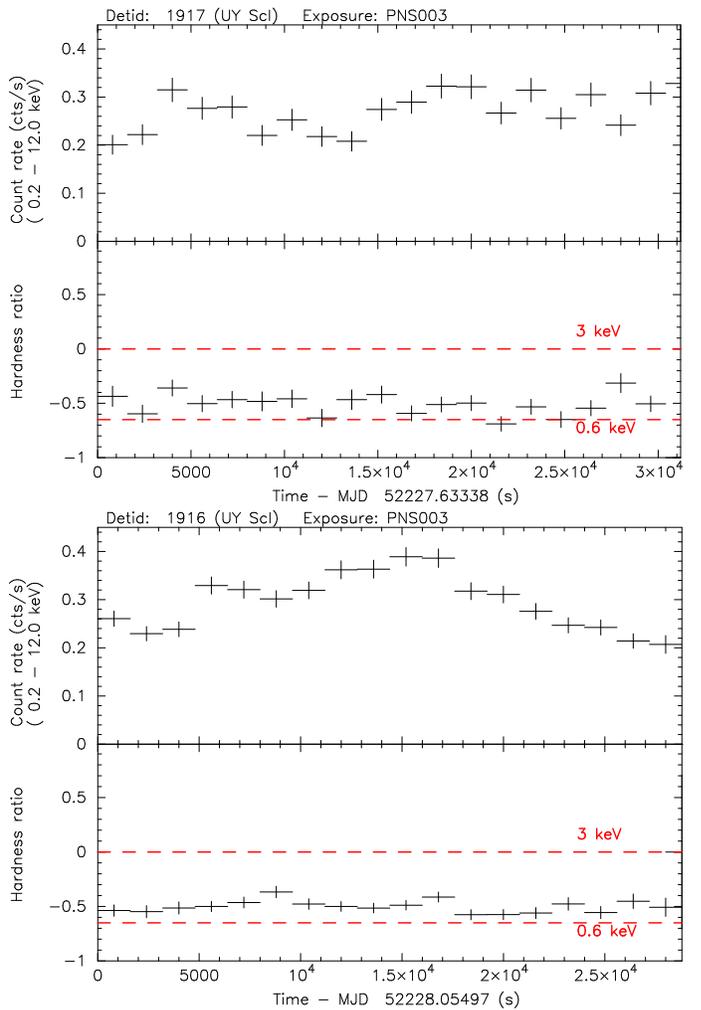

	\resizebox{\hsize}{!}{\includegraphics[angle=-90]{fig-HRlc-07a.eps}}
	\resizebox{\hsize}{!}{\includegraphics[angle=-90]{fig-HRlc-07b.eps}}
	\caption{As Fig.~\ref{figHR2}, but for the contact-binary, W-UMa-type star \object{UY Scl} (period $\approx 31$ ks, Chen et al.\ (2006)), PN only; two observations, each of duration $\sim 30$ ks, with a gap of $\sim 6$ ks from the end of the first to the start of the second observation.  }
	\label{figHR7}
\end{figure}

An X-ray spectral hardness ratio was defined as in 2XMM as:
$ {\rm HR} = (R_{j} - R_{i}) / (R_{j} + R_{i}) $,
where $R_{i}$ and $R_{j}$ are the count rates in energy bands $i$ and $j$.
Here, we choose as `standard', the bands $i$ and $j$ to be 0.2--1.0 and 1.0--12.0 keV,
in order to (a) provide a good dynamic range of sensitivity to typical temperature ranges in stellar coronae (kT $\sim 0.1$--10 keV),
(b) maximise signal-to-noise and (c) for consistency with 2XMM (i.e.\ our bands are the sum of two or more 2XMM bands).
HR light-curves were generated specifically for our flare survey.
Examples are shown in Figs.~\ref{figHR2}--\ref{figHR7}.
We have generated HR light-curves at various time resolutions; the examples presented here 
are all at a time binning of 1600~s, for good signal-to-noise whilst showing the main features of the flare events. 
In Fig.~\ref{figHR6}(c,d) we present an example for a `harder' HR, using the bands 1.0--2.0 and 2.0--4.5 keV, permitting tracking of higher temperatures than our standard band, but with reduced signal-to-noise, and hence useful only for relatively bright sources.

Fig.~\ref{figHRpred1}(a) shows predicted HRs, calculated for an optically-thin, thermal (coronal) spectrum 
and negligible line-of-sight photoelectric absorption,
 using the XSPEC software package and the MEKAL spectral model (e.g.\ Dorman \& Arnaud 2001; Arnaud, Dorman \& Gordon 2010). Fig.~\ref{figHRpred1}(b) shows the corresponding count rate to emission measure conversion factors.

All hardness-ratio timeseries plots are presented in Appendix~\ref{appALLlcHR} alongside the X-ray count-rate light-curves.

\section{Discussion } \label{discussion}

\subsection{Identifications and stellar data}  \label{dis-ids}

Most of the serendipitously-observed stars have little previously reported information (as gleaned from searching the SIMBAD database and other catalogues available via CDS) other than astrometric data and apparent magnitude, colour and a ROSAT All-Sky Survey (Voges et al.\ 1999) X-ray flux. References to stellar data are given in Table~\ref{tabStars}, where we also cite previously published XMM-based work on the target stars.

In a bid to increase our knowledge of the astrophysical properties of the serendipitously-observed stars, we searched the WASP (Pollacco et al.\ 2006) database of visible-band light-curves and found five matches with our serendipitously-observed stars. Of these, three exhibited no notable variability, but two showed a clear periodic signal (West 2009): 
\object{2MASS J04072181-1210033} (DETID = 45610 and 45611, see Figs.~\ref{figOMlc3}, \ref{figOMlc4}, \ref{figHR3}), period $\approx 6.1$~hr; \object{HD 268974} (DETID=54373, see Fig.~\ref{figHR4}), period $\approx 4.05$~dy. The WASP light-curve shapes suggest that both systems may be eclipsing binaries, with 2MASS~J04072181-1210033 being a W~UMa-type contact binary. The latter star has also been independently identified as a W~UMa system by Farrell et al.\ 2009.
We, and Farrell 2009, have also found $\approx 6.1$~hr periodicity in the XMM X-ray data of 2MASS~J04072181-1210033 (DETID = 45610 and 45611).
HD~268974 (= ASAS~J050527-6743.3 = ASAS~J050526-6743.3) was found as an eclipsing binary in the ASAS survey for periodic variable stars (Pojmanski 1998).

Other `well identified' objects in the serendipitously-observed sample include: 
\begin{itemize}
	\item \object{UY Scl}, DETID=1916: W~UMa-type contact binary (Chen et al.\ 2006), period $\approx 8.6$~hr, discussed further below);
	\item \object{78 Tau}, DETID=48976: $\delta$ Sct-type variable and member of the Hyades open star cluster;
	\item \object{V807 Tau}, DETID=50256: PMS star, Orion-type variable;
	\item \object{HD 95735} (=Gl~411), DETID=121057: dMe flare star; a possible exoplanet (GJ~411b) was reported (Gatewood 1996) but appears not to have been confirmed; it is not in the exoplanet list at http://exoplanet.eu/ (Schneider et al.\ 2011) as of 2014 October 7; 
	\item \object{IM Vir}, DETID=148010, 148011: Algol-type eclipsing binary (Pourbaix et al.\ 2004), period $\approx 1.3$~dy;
	\item \object{NU UMa} (= HD~237944), DETID=118876: eclipsing, spectroscopic binary, period $\approx 5.5$~dy (Otero \& Dubovsky 2004; Griffin 2009);
	\item \object{BN Sgr}, DETID=204777: Algol-type eclipsing binary, period $\approx 2.5$~dy (Malkov et al.\ 2006);
	\item \object{EQ Peg}, DETID=243985: dMe flare star; note that this object appears also in the target-star sample (DETID=243984).
\end{itemize}

There are at least 18 stars (i.e.\ $\sim$two-thirds) of the serendipitously-observed sample for which further optical spectroscopy and time-resolved photometry would be highly beneficial in determining the detailed nature of the objects.

We note that, as of 2014 October, there are 10 target stars out of a total of 45 in Table~\ref{tabStars} for which we can find no published reports of the XMM results in the refereed literature (from a search of the CDS SIMBAD database).

\subsection{Comparison of UV and X-ray light-curves}

Although we cannot draw any broad conclusions from such a small and incomplete sample (five UV light-curves from four stars, Sect.~\ref{analOM}), it can be seen that the UV emission in some cases clearly shows flare-like behaviour (e.g.\ Figs.~\ref{figOMlc1}, \ref{figOMlc2}) in others the UV flux does not show an obvious correlation with the X-ray flare (e.g.\ Fig.~\ref{figOMlc3}) (see e.g. review by G\"udel \& Naz\'e 2009, Sect.~2.4.2, and references therein).
Inspection of X-ray and UV (fast-mode) light-curves for those target stars where both are available, shows a wide variety of relative variability between the two wavelength regimes, even after allowing for possible time-offsets between X-ray and UV flares of up to several hundred seconds (e.g.\ Mitra-Kraev et al.\ 2005).

For the two cases where there was UV coverage both in the pre-flare quiescent phase and in proximity to the X-ray flare peak, i.e.\
DETID=242179 (star 2MASS J23163068+7905362 = \object{1RXS J231628.7+790531}, $B-V \sim 0.8$, Fig.~\ref{figOMlc1}) and DETID=23485 (star \object{HD~14716}, spectral type F5~V, Fig.~\ref{figOMlc2}), we can, at least tentatively and in a limited manner, apply the analysis of Mitra-Kraev et al.\ (2005a) in comparing the X-ray and UV luminosity and flare energy. 
(It may be noted in passing that the Mitra-Kraev et al.\ sample was composed of [five] dMe stars, while the two stars from our survey are of earlier spectral types [F--G].)
The two flares in our survey have large X-ray amplitudes (from quiescent to peak luminosity), of a factor ten or greater, compared with those discussed in Mitra-Kraev et al. (typically amplitudes of a factor 2--3), and are of much longer duration ($\sim 10^4$~s versus $\la 10^2$~s).
In contrast, the amplitude of the UV flare emission for our two 2XMM-Tycho stars is of order a factor two, similar to that in the Mitra-Kraev et al.\ sample.
Mitra-Kraev et al. refer to a similar effect, i.e.\ an apparent deficit in UV emission or excess in X-rays, for the `flat-top' flares in their sample.
Following the discussion of Mitra-Kraev et al., comparison of the X-ray-to-UV flare-emitted-energy to the X-ray-to-UV quiescent (or mean) luminosity for the two 2XMM-Tycho stars may suggest for these large-amplitude, long-duration events, the presence of an additional energy source or mechanism, boosting the `hot' X-ray-emitting plasma relative to the cooler UV-emitting region.
As a further investigation into this effect, we have selected a good example of an apparently isolated, large amplitude, long duration flare from one of the target stars in our sample: the well-known flare star 61~Cyg~B (DETID=229736, spectral type K7~V), observed in the OM UVW2 filter (see Sect.~\ref{appALLlcOM}). The conclusions are the same as those noted above for HD~14716 and 2MASS~J23163068+7905362.

\subsubsection{The eclipsing RS~CVn-type binary \object{SV Cam} }

This is one of our target-star sample.
Sanz-Forcada et al.\ (2006) have presented the XMM measurements from 2001 (DETID=75409) and 2003 (DETID=75410), and reported an eclipsed X-ray flare during the 2001 observation. They discuss only the X-ray data. We now present the corresponding OM UV timeseries (see Fig.~\ref{figALLlcOM}, DETID=75409) which clearly shows the eclipse, and its registration with the X-ray light-curve. An eclipse is also seen in the OM UV data for 2003 (see Fig.~\ref{figALLlcOM}, DETID=75410), though with no obvious X-ray counterpart.

\subsection{Hardness-ratio light-curves}

The HR light-curves demonstrate\footnote{We make the plausible assumption that other physical parameters (e.g.\ absorption column density) that might cause changes in spectral shape, are not time variable for these sources.}
 the ability to track temperature\footnote{Since stellar coronae are, in general, multi-temperature sources, the single `temperature' derived from the HR is a `mean' value, weighted by the distribution of emission measure with temperature.}
(and emission measure) changes with time.
As indicated by the predictions in Fig.~\ref{figHRpred1}, and borne-out by the observed flare light-curves in Figs.~\ref{figHR2}--\ref{figHR6}, our `standard' HR is sensitive over the temperature range kT $\sim 0.3$--3 keV, whilst the `harder' HR is useful over a somewhat higher range of kT $\sim 1$--10 keV, albeit with reduced signal-to-noise.
The somewhat different effective area as a function of energy for the XMM EPIC MOS and PN instruments leads to distinct HR(T) curves in each case, with MOS having a rather `harder' response and greater rate of change with temperature than PN (c.f.\ Fig.~\ref{figHRpred1}).

In general, as can be seen in Figs.~\ref{figHR2}--\ref{figHR6}, there is a clear, rapid temperature increase at flare onset; subsequent behaviour varies from flare-to-flare. In contrast, the flux variability in the contact-binary star UY~Scl (DETID=1916) in Fig.~\ref{figHR7}(c,d) shows no significant changes in HR. However, the same feature was not apparent on the previous rotation of the system (Fig.~\ref{figHR7}(a,b), DETID=1917). Hence it is unclear whether this variability is due to flare-like activity or e.g.\ rapid active-region evolution. An analysis of the XMM X-ray spectrum of UY~Scl has been reported by Stobbart et al.\ (2006).

\subsection{Flare parameter distributions and application to flare models} \label{param-dis}

We present here some examples of comparing the 2XMM-Tycho results with a selection of previously published flare models (or more directly, the associated diagnostic plots), scaling laws and observational flare surveys. This is not intended to be exhaustive but rather an illustrative comparison. Advantages of the current survey over many previously reported ones include uniformity of measurements and analysis, and sensitivity.

The physical parameters of temperature and emission measure have been derived from the observational quantities: hardness ratio and count rate, using the conversion factors presented in Section~\ref{HRcurves} (Fig.~\ref{figHRpred1}). The count-rate to emission-measure conversions used were those for the corresponding hardness-ratio/temperature.

\begin{figure}
	\resizebox{\hsize}{!}{\includegraphics{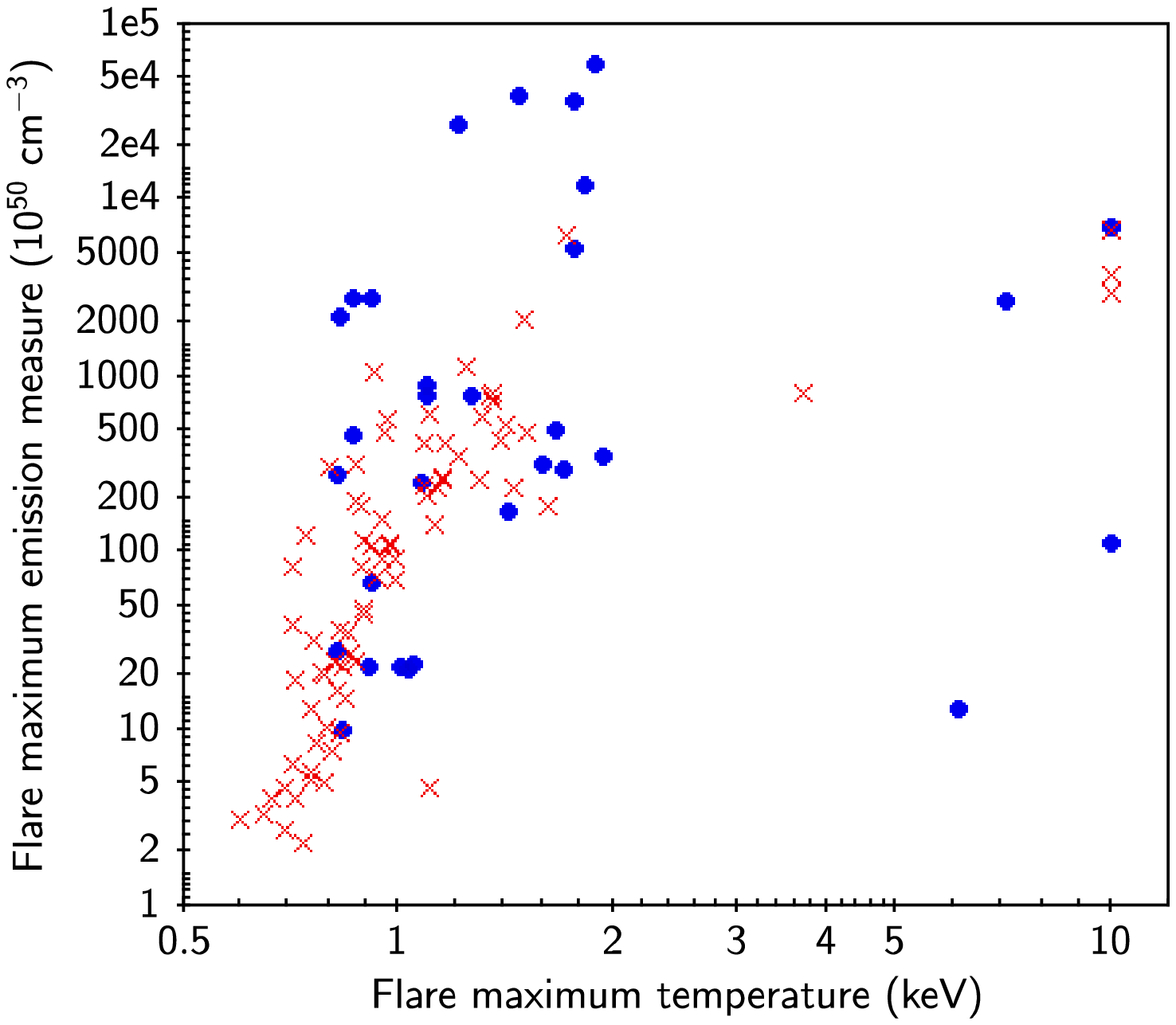}}
	\resizebox{\hsize}{!}{\includegraphics{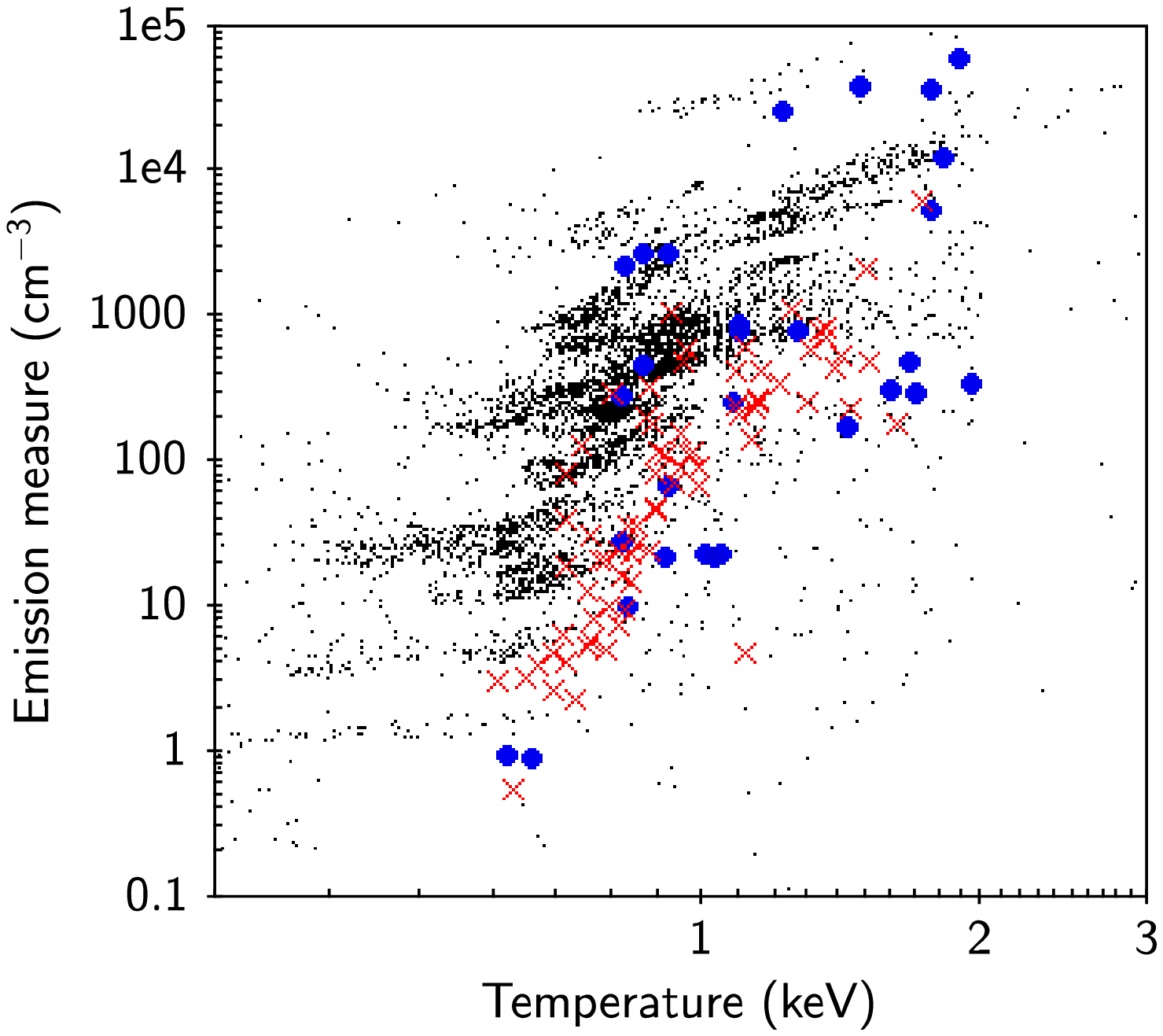}}
	\caption{{\it (a)} Flare maximum emission measure versus maximum temperature (energy band 
	0.2 -- 12 keV). Symbols as in Fig.~\ref{figTscatter}. {\it (b)} As (a) but also showing, as black dots, essentially all data points from the 2XMM-Tycho light-curves, i.e.\ each time bin (1 bin = 400 or 1600 s) in each light-curve is represented on the plot, irrespective of the presence of a flare; the temperature range has been restricted for clarity in showing the bulk of the distribution.  }
	\label{figkTEMscatter}
\end{figure}

We summarise here the main findings.
\begin{enumerate}

	\item {\it Temperatures from hardness ratios.} 
	These provide a uniformly distributed (in time) and relatively high time resolution, useful for identifying and characterising e.g.\ flare on-set (e.g.\ Fig.~\ref{figHR2} -- Fig.~\ref{figHR6})  or multiple, overlapping flares (e.g.\ Fig.~\ref{figALLlcHR}, DETID=53358, 61224, 243984). The HRs are temperature-sensitive over a limited range, from $\sim 0.3$ to $\sim 3$ keV for the `standard' ones (Fig.~\ref{figHRpred1}), and hence the `measured' temperatures will be restricted, and in particular, will tend to underestimate the flare-peak values (since, in general, the plasma may be multi-temperature), and hence care must be exercised in using them quantitatively.

We have investigated this effect by comparing our HR-derived flare-peak temperatures with those from spectral fits reported in the literature for the same XMM observations. Our comparison set, though rather inhomogeneous and incomplete, comprised 19 flares from 12 target stars, and clearly confirmed that our HR temperatures consistently fall below the spectral-fit values, with the HR temperatures $\sim 0.3$--$0.5$ of the spectral-fit ones.
	
	\item {\it Emission measures from count rates.} 
	The derived emission measures (and luminosities and emitted energies) are relatively insensitive to the precise choice of temperature; as shown by Fig.~\ref{figHRpred1}(b), the conversion factor, for the `standard' 0.2--12~keV band, varies by at most a factor $\sim 2$ over the range $\sim$1--10 keV, or over the range $\sim$0.2--1 keV.

	\item {\it Peak temperature vs peak luminosity, peak emission measure, and emitted energy.} 
	Our results (see Fig.~\ref{figkTEMscatter}(a)) are generally similar to those of Aschwanden et al.\ 
(2008) as summarised in their Fig.~1 (see also G\"udel 2004, Sect.~12.12). For example, our 
fitted\footnote{We have used the same method as Aschwanden et al.\ (2008), i.e.\ linear ordinary least-squares bisector to calculate the power-law slope. In performing the fits we have discarded some of the extreme, `outlier' points.}
power-law slopes of 7.7 and 8.1 for peak emission measure as a function of temperature, for the target and serendipitous samples respectively, are comparable with $4.5 \pm 0.4$ of Aschwanden et al., noting that (a) the angular difference in inclination for slopes of 4 and 8 is only 7~deg.; (b) the `compression' of high temperatures, previously mentioned, will tend to increase the slope. The ranges of temperatures and emission measures covered by our sample and Aschwanden et al.\ are similar, though (again probably due to the `compression' effects) the 2XMM-Tycho temperatures are, in general, rather lower than those reported in Aschwanden et al.
We have a larger (by factor $\sim 2$) and more uniform and coherent dataset (i.e.\ all from 2XMM), and clearly confirm that larger flares are hotter (G\"udel 2004).

Our fits ignored ten `outliers' at $ T > 3$~keV; these flares have relatively low peak emission measure for the peak temperature, or equivalently, relatively high peak  temperature for the peak emission measure. They are: 4 flares from the classical T~Tau-type pre-main-sequence (PMS) stars \object{SU Aur} (DETID~53358; 3 flares; Robrade et al.\ 2006, Franciosini et al.\ 2007b) and \object{CR Cha} (120401; Robrade et al.\ 2006), both XMM target objects), and 6 flares from the serendipitous sample -- \object{HD 31305} (A0 type; 53325; Arzner et al.\ 2007; Franciosini et al.\ 2007b), \object{TYC 9275-01654-1} (194425), \object{TYC 1082-02107-1} (224075), \object{1RXS J231628.7+790531} (242179), \object{BN Sgr} (F3~V Algol-type;  204777; Malkov et al.\ 2006), \object{2MASS J05350341-0505402} (possibly a PMS star in Orion Molecular Cloud 2/3; 64214)\footnote{
The flares from BN~Sgr and 2MASS~J05350341-0505402 are only in the CLVS sample and hence do not appear in Fig.~\ref{figkTEMscatter}(a).
}. 
An alternative, or possibly additional, explanation for the high hardness ratios and hence high derived temperatures, could be the presence of significant line-of-sight absorption, preferentially removing low-energy photons from the observed flux. Indeed, Robrade et al.\ 2006 report significant X-ray absorption with column density $ n_\mathrm{H} \sim 3 \times 10^{21}\ \rm cm^{-2}$ for both SU~Aur and CR~Cha. They also find a hot component with $\rm kT \sim 5\ keV$ in SU~Aur. Franciosini et al.\ (2007b) report significant, but lower absorption, $ n_\mathrm{H} \sim 6 \times 10^{20}\ \rm cm^{-2}$, for HD~31305, together with a high-temperature flaring component with $ kT \sim 9\ \rm keV$. 
For the remaining five stars, we have performed spectral 
fits\footnote{We used the standard XMM data products, fitting with XSPEC and a two-temperature MEKAL + $n_\mathrm{H}$ model.}
to the time-averaged spectra, yielding significant absorption for BN~Sgr ($ n_\mathrm{H} \sim 1 \times 10^{21}\ \rm cm^{-2}$) and 2MASS J05350341-0505402 ($ n_\mathrm{H} \sim 2 \times 10^{21}\ \rm cm^{-2}$), and low absorption ($ n_\mathrm{H} \la 4 \times 10^{20}\ \rm cm^{-2}$) in the other three cases.

G\"udel (2004, Sect.~12.12) has noted that `non-flaring' emission from a G-star sample of `solar-analogues' follows a similar X-ray luminosity / temperature trend to the flare-peak properties of his flare dataset, suggesting that flares may contribute systematically to the hotter components of the plasma. As an extension of this scheme, and somewhat speculatively, we have plotted essentially {\it all} timeseries data in our 2XMM-Tycho variable, cool-star sample in Fig.~\ref{figkTEMscatter}(b), where each point represents a single time bin (usually of $\approx 400$ or $\approx 1600$ duration); there are $\sim 5500$ data points. 
Although there is a wide spread, there appears to be a clear trend, similar to that exhibited by the flare-peak values.
For ease of cross-comparison, we also show the flare information from Fig.~\ref{figkTEMscatter}(a).

	\item {\it Peak temperature vs duration.}
	Our results are broadly similar to those of Aschwanden et al.\ (2008), who obtain a power-law slope of $1.8 \pm 0.2$ with a relatively low degree of correlation (correlation coefficient = 0.39). We find slopes of $\sim 2.1$ and $\sim 1.0$ for our target and serendipitous samples respectively, with correlation coefficients of 0.36 and 0.02.
	
	\item {\it Peak luminosity vs emitted energy} (see Fig.~\ref{figLEscatter}).
	We obtain power-law slopes of $\sim 1.2$, $\sim 1.2$ and $\sim 1.0$ for our full survey, target and serendipitous samples respectively, in good agreement with the slope of 1.16 reported by Wolk et al.\ (2005) from their {\it Chandra} survey of 41 K-type PMS stars in Orion.  As Wolk et al.\ note, this near-linear relationship implies that duration is essentially independent of peak luminosity. We have also directly confirmed this in our survey dataset by comparing these two parameters.

	\item {\it Time difference between maximum emission measure and maximum temperature,}
	$ \Delta t = t_{EM_{max}} - t_{T_{max}} $. 
	As discussed by Reale (2007), if $ \Delta t > 0 $, the heat pulse (which originates the flare event) is relatively short compared to the characteristic cooling time of the emitting loop volume, i.e.\ it indicates that the loop does not reach equilibrium conditions; conversely, $ \Delta t = 0 $ indicates that loop has reached equilibrium. 
In the majority ($\sim 70$\%) of cases in our survey 
we are unable to confirm 
evidence of non-equilibrium (i.e.\ $ \Delta t $ is not significantly greater than zero), 
though this may be due to limited time resolution and S:N; 
for $\sim 30$\% there is some evidence, but we caution that in most cases the delay corresponds to only 1 time bin ($\approx 400$ or $\approx 1600$~s); for $\la 10$\% the delays are several thousand seconds ($\sim 3$--8~ks). 
(These percentages apply to the survey as a whole and individually to the target and serendipitous samples.)
Examples of delays occur in the light-curves shown in Figs.~\ref{figHR2}, \ref{figHR3}, \ref{figHR5} and \ref{figHR6}.
Reale (2007) cites 3 example measurements, with $ \Delta t $ (his $ \Delta t_{0-M} $) $\sim \rm 20,\ 0.4,\ 0.2$~ks.  

	\item {\it Decay-phase emission measure vs temperature.}
	We have computed the power-law slope of the $\sqrt{emission~measure}$--temperature distribution during the flare decay phase, a diagnostic of the presence of continued heating during the decay phase (e.g.\ Reale et al.\ 1997; Reale 2007). However, we have not been able to achieve reliable results, probably due to the limitations of the single-temperature / hardness-ratio analysis. Hence we defer further consideration, but note that in principle, our survey could yield a relatively large sample of slope values.
	
\end{enumerate}

In summary, the diagnostic plots based on hardness-ratio temperatures and associated emission measures etc, provide excellent, general indicators for the investigation of large numbers of flares, but individual, quantitative studies may require more detailed analysis e.g.\ via time-resolved spectral fitting.

\subsubsection{Flare luminosity, energy and emission-measure distributions}

It is evident from Fig.~\ref{figLscatter} that, although flares from the serendipitously-observed stars and target stars cover broadly similar ranges of both $L_\mathrm{X,quies}$ and $L_\mathrm{X,peak}$, the $L_\mathrm{X,peak} / L_\mathrm{X,quies}$ values at a given $L_\mathrm{X,quies}$ tend to be higher, e.g.\ $\sim 50$\% of flares from the serendipitously-observed stars have $L_\mathrm{X,peak} / L_\mathrm{X,quies} \ga 2$, while the corresponding fraction for target stars is $\la 10$\%. This may be largely or wholly an observational selection effect arising from the generally smaller distances and hence higher X-ray fluxes of the target stars, enabling the detection of lower levels of variability.

Large, but not infrequent solar flares have soft X-ray peak luminosities and emission measures $\sim 10^{28}\ \rm erg\ s^{-1}$ and $\sim 10^{50}\ \rm cm^{-3}$ respectively, and X-ray emitted energies $\sim 10^{32}\ \rm erg$ (see e.g.\ G\"udel 2004, Table~4; Aschwanden et al.\ 2008, Fig.~3; Schrijver et al.\ 2012), with occurrence rates of $\sim 0.1$--1/year (see e.g.\ Thomson et al.\ 2010, Table~1; Schrijver et al.\ 2012). Thus, there is overlap between the lower end of the stellar $L_\mathrm{X,peak}$ and $E_\mathrm{X}$ distributions (Figs.~\ref{figLscatter}, \ref{figLEscatter}) and solar flares. 
Schrijver et al.\ (2012) conclude that the largest solar flares for which good evidence exists are up to an order of magnitude more energetic than those discussed above, placing them well within the 2XMM-Tycho distribution.

\subsubsection{Flare rates and duty cycles} \label{dis-rates}

\begin{table*}
\caption{Summary of the flare frequency statistics and distributions, 
for the CVS serendipitous-stars sample.
}
\label{tabFreqSumm}
\centering
\begin{tabular}{l l c c c c}
\hline\hline
\multicolumn{2}{l}{Quantity}  & \multicolumn{2}{c}{A} & \multicolumn{2}{c}{B} \\ 
               &              & Raw,     & Coverage  & Raw,     & Coverage   \\
               &              & measured & corrected & measured & corrected  \\
\hline
Min.\ selected signal:noise & 10.0 &               &        \\
\hline
Min.\ selected peak flux, $f_{\rm X,peak,min}$ & $\rm (erg\,cm^{-2}\,s^{-1})$ & $2.5 \times 10^{-13}$ & & $1.0 \times 10^{-12}$ &  \\

\ \ \ {\it Above $f_{\rm X,peak,min}$:}  \\
No.\ of flares in survey   &   & 15 & 16.3  & 10  & 10.2     \\
No. of flares predicted, all-sky/year & & & $4.4 \times 10^6$ &  & $2.8 \times 10^6$  \\
No. of flares predicted per star/year & {\it (a)} & & $4.3 \times 10^{-2}$ & & $2.7 \times 10^{-2}$  \\

Power-law index, $\alpha_\mathrm{f}$  & {\it (b,c)} & $0.55 \pm 0.14$ & $0.58 \pm 0.15$ &  $0.94 \pm 0.30$ &  $0.94 \pm 0.30$ \\
Normalisation, $N_\mathrm{ref,f}$     & {\it (b)}   & 7.0 & 7.3 & 10.0 & 10.2 \\
\ \ with $f_{\rm X,ref} = 1 \times 10^{-12} $ $\rm erg\,cm^{-2}\,s^{-1}$  \\ 

\hline
Flare duty cycle (fraction)           & {\it (a)} & & $4.3 \times 10^{-3}$ & &  $2.0 \times 10^{-3}$  \\
                                      & {\it (d)} & & $7.3 \times 10^{-2}$ & &  $3.4 \times 10^{-2}$  \\
                                      
\hline
Min.\ selected peak luminosity, $L_{\rm X,peak,min}$ & $\rm (erg\,s^{-1})$ & $2.0 \times 10^{29}$ & & $1.0 \times 10^{30}$ &  \\

\ \ \ {\it Above $L_{\rm X,peak,min}$:}  \\
No.\ of flares in survey   &   & 14 & 16.5  & 11  & 13.3     \\
No. of flares predicted per star/year & {\it (a)} & & $4.3 \times 10^{-2}$ & & $3.5 \times 10^{-2}$  \\

Power-law index, $\alpha_\mathrm{L}$ & {\it (b,c)} & $0.29 \pm 0.08$ & $0.28 \pm 0.07$ & $0.38 \pm 0.12$ & $0.37 \pm 0.11$ \\
Normalisation, $N_\mathrm{ref,L}$  & {\it (b)} & 8.8 & 10.5 & 11.0 & 13.3 \\
\ \ with $L_{\rm X,ref} = 1 \times 10^{30} $ $\rm erg\,s^{-1}$  \\

\hline
Min.\ selected emitted energy, $E_{\rm X,min}$ & (erg) & $1.0 \times 10^{33}$ & & $1.0 \times 10^{34}$ &  \\

\ \ \ {\it Above $E_{\rm X,min}$:}  \\ 
No.\ of flares in survey   &   & 13 & 15.4  & 9  & 11.3     \\
No. of flares predicted per star/year & {\it (a)} & & $4.03 \times 10^{-2}$ & & $2.9 \times 10^{-2}$   \\
                                      & {\it (d)} & &   11.6 & & 8.5  \\
                                      & {\it (a,e)} & & $4.9 \times 10^{-4}$ & & $4.4 \times 10^{-5}$  \\

Power-law index, $\alpha_\mathrm{E}$ & {\it (b,c)} & $0.29 \pm 0.08$ & $0.28 \pm 0.08$ & $0.47 \pm 0.16$ & $0.47 \pm 0.16$ \\
                                              & {\it (e)} &                 & $0.72 \pm 0.20$ &                 & $0.56 \pm 0.19$ \\
                                              
Normalisation, $N_\mathrm{ref,E}$  & {\it (b)} & 13.0 & 15.4 & 26.6 & 33.1 \\
\ \ with $E_{\rm X,ref} = 1 \times 10^{33} $ erg  \\

\hline
\end{tabular}
\newline

{\bf Notes:} 
columns A: results for thresholds ($f_{\rm X,peak,min}$, $L_{\rm X,peak,min}$, $E_{\rm X,min}$) based on lowest value in the set, with minimum S:N $\ge 10.0$;
columns B: as A, but for substantially higher thresholds. 
(a) considering all serendipitous-sample stars (504) with 2XMM light-curves, irrespective of detected variability; associated survey on-time $t_{\rm total} \sim 2.4 \times 10^7$~s; 
(b) from ML fit; 
(c) no correction has been made to these values, e.g.\ the bias correction $(N-1)/N$ suggested by Crawford et al.\ (1970), which could reduce the values by up to $\sim 10$\%; 
(d) considering only serendipitous-sample stars (30) observed to flare; associated survey on-time $t_{\rm total} \sim 1.4 \times 10^6$~s; 
(e) applying scaling factor based on each star's quiescent X-ray luminosity and normalised to a quiescent solar X-ray luminosity, see text Sect.~\ref{dis-rates} for details.

\end{table*}

\begin{figure}
	\resizebox{\hsize}{!}{\includegraphics{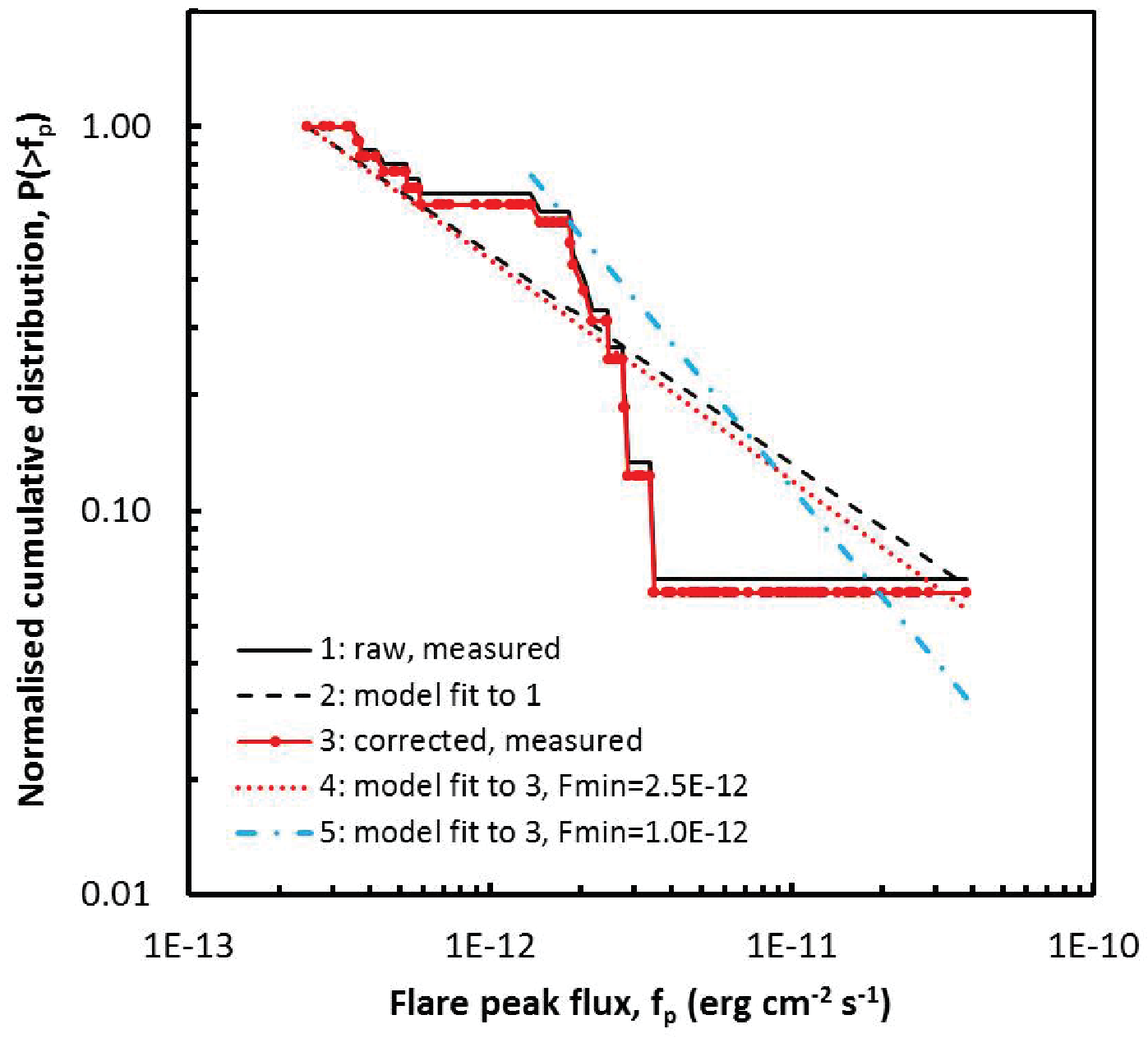}}
	\resizebox{\hsize}{!}{\includegraphics{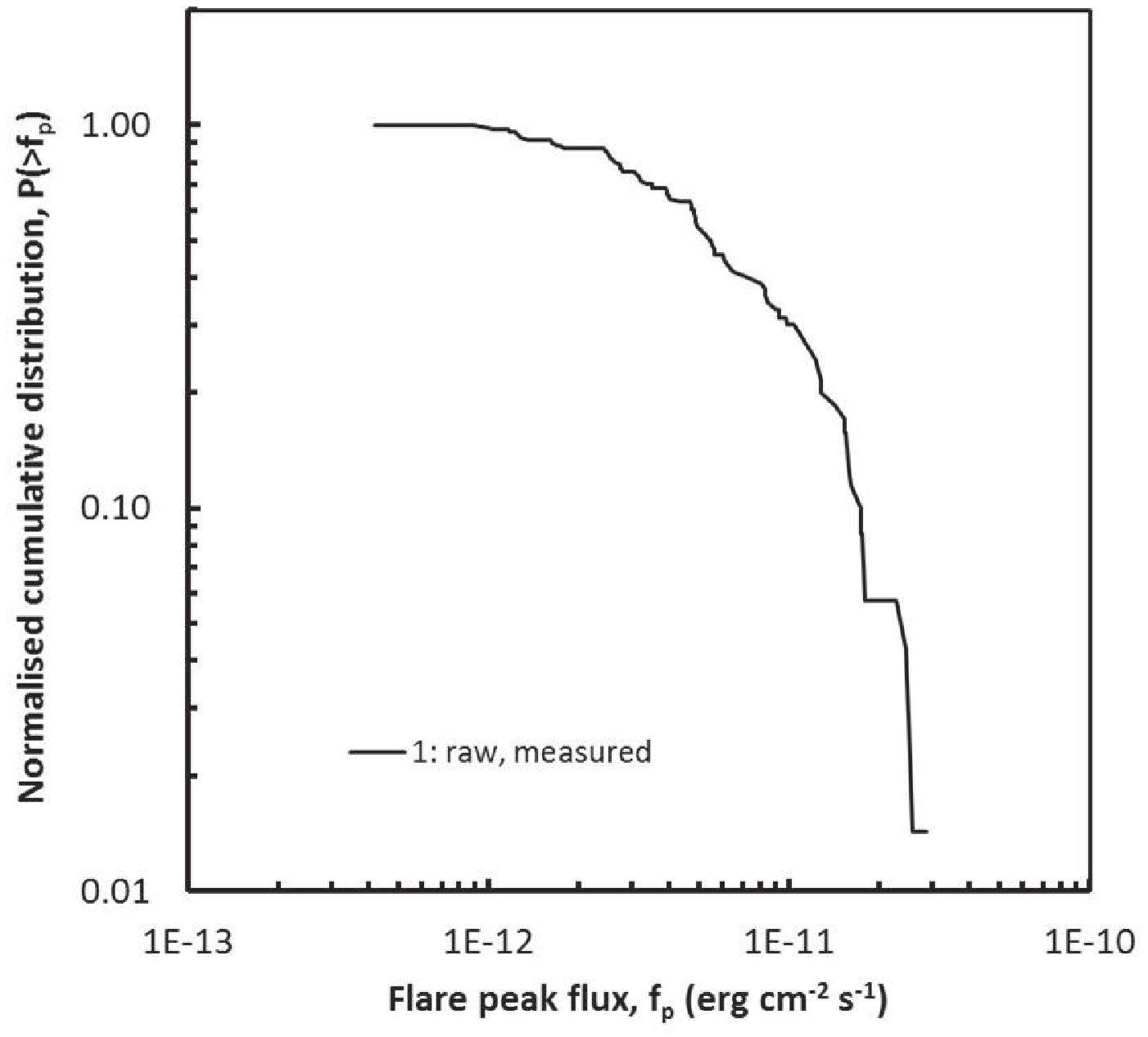}}
	\caption{{\it(From top) (a)} Cumulative frequency distribution of flare peak X-ray flux, $f_\mathrm{X,peak}$, for the CVS serendipitous-stars sample: black solid line -- `raw', measured distribution; black dashed line -- model distribution from ML power-law fit to raw, measured distribution; red solid line with dots -- coverage-corrected `measured' distribution; red dotted line -- model distribution from ML power-law fit to corrected distribution;
blue dash-dot line -- model distribution from ML power-law fit to corrected distribution, with higher minimum flux, $f_{\rm X,min}$. 
Each model distribution has been derived from the data over the flux range indicated by the model (power-law) line.
	{\it(b)} As (a), but for the CVS target-stars sample, raw, measured distribution. }
	\label{figFlPkFreq}
\end{figure}

\begin{figure}
	\resizebox{\hsize}{!}{\includegraphics{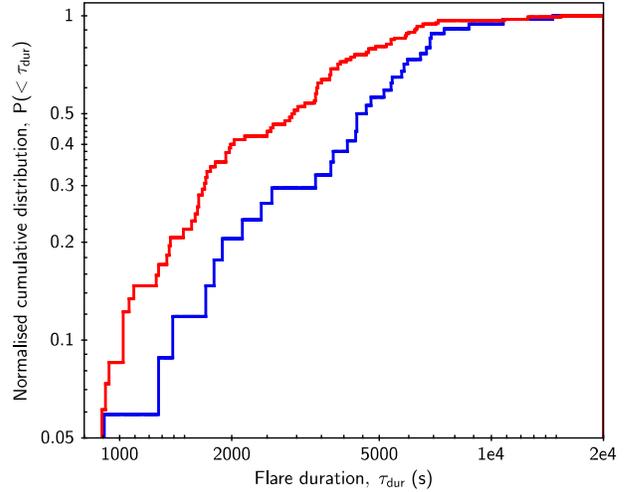}}
	\caption{Cumulative frequency distribution of flare duration: blue (lower) line -- the CVS serendipitous-stars sample; red (upper) line -- the CVS target-stars sample. The distributions are the `raw', measured ones, i.e.\ not coverage corrected. }
	\label{figFlDurFreq}
\end{figure}

\begin{figure}
	\resizebox{\hsize}{!}{\includegraphics{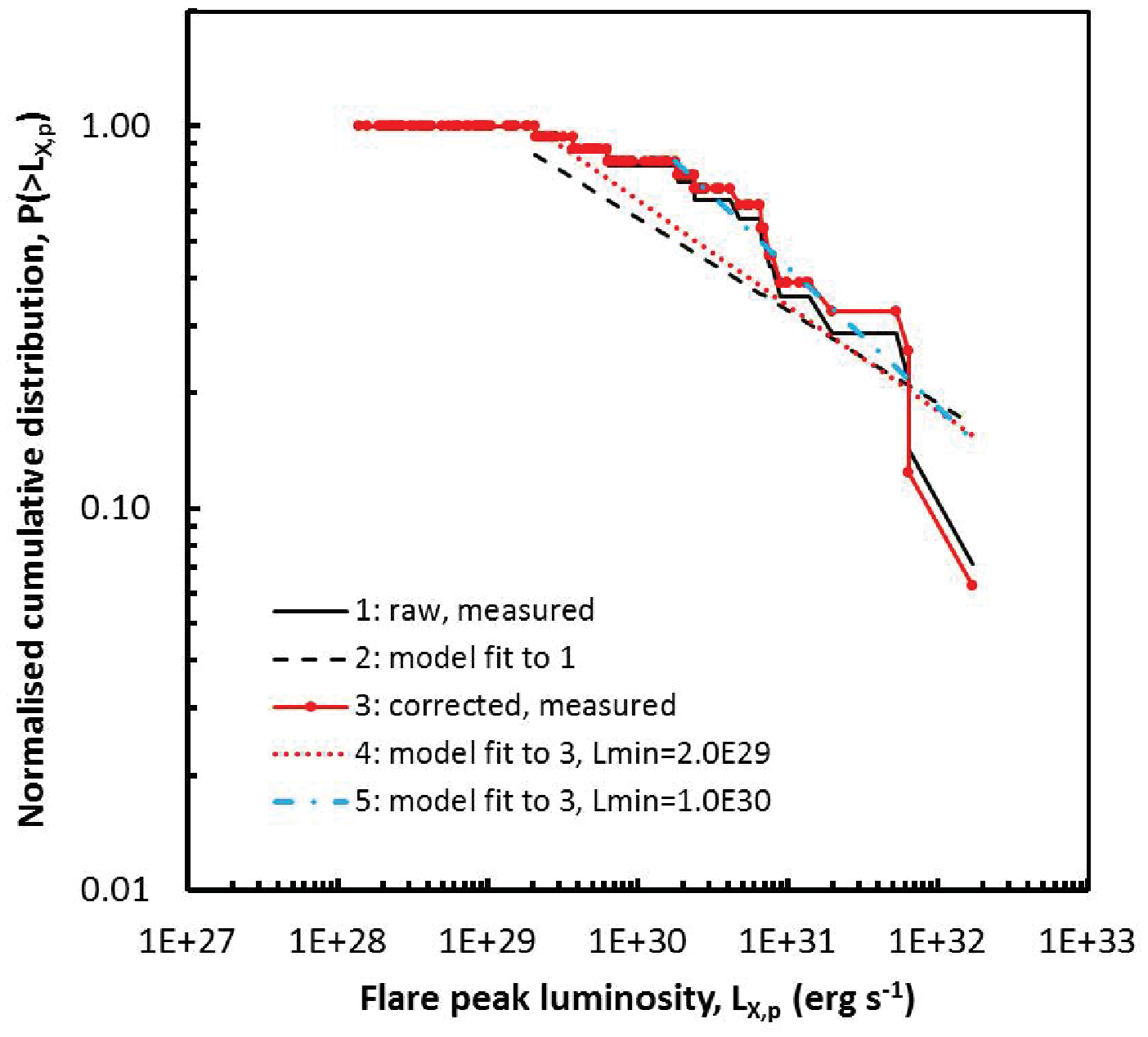}}
	\resizebox{\hsize}{!}{\includegraphics{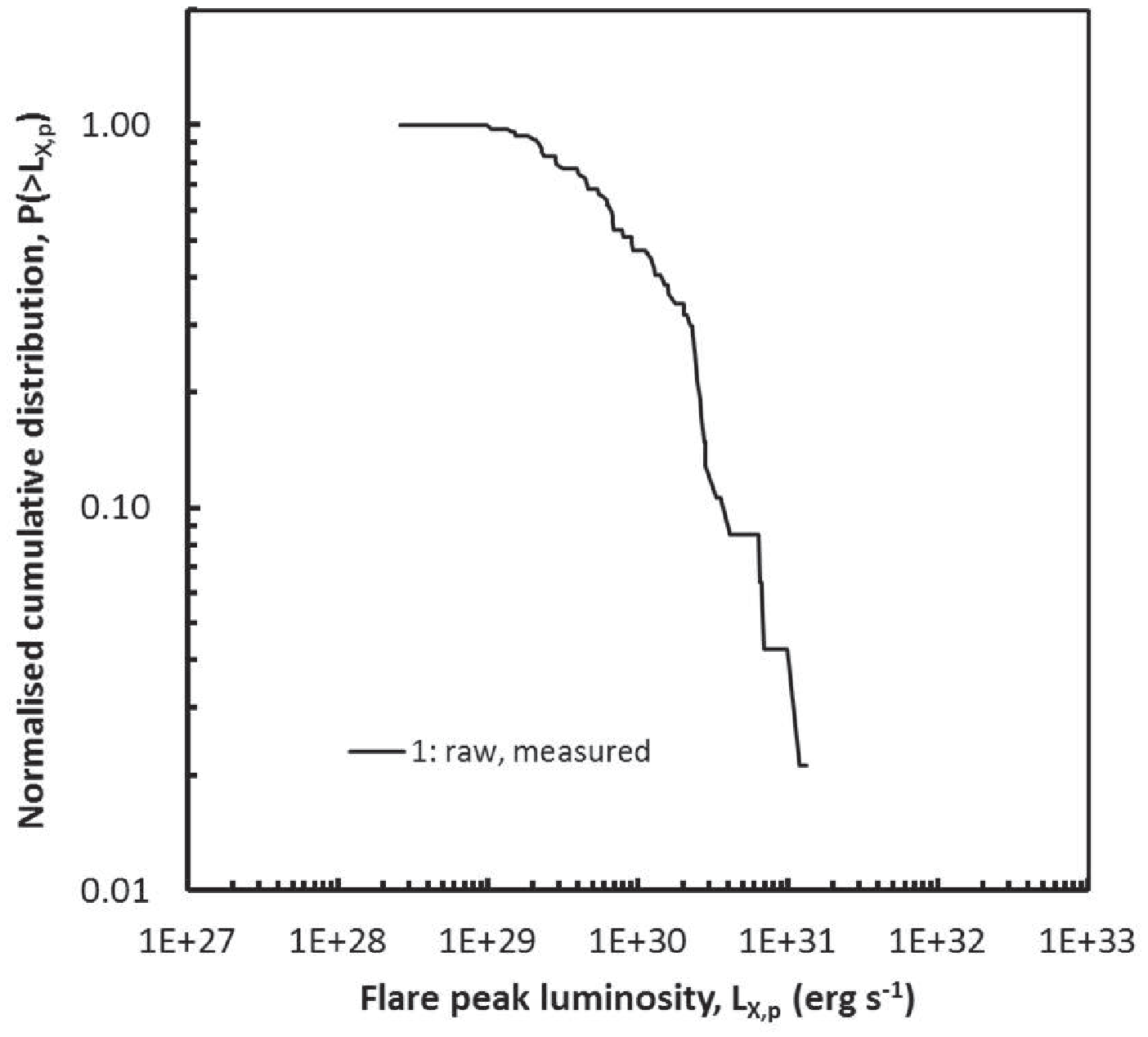}}
	\caption{As Fig.~\ref{figFlPkFreq}, but for flare peak X-ray luminosity, $L_\mathrm{X,peak}$. 
	Note that this is an {\it observed luminosity distribution}, not a (volume-normalised) 			    {\it luminosity function}. }
	\label{figFlLxFreq}
\end{figure}

\begin{figure}
	\resizebox{\hsize}{!}{\includegraphics{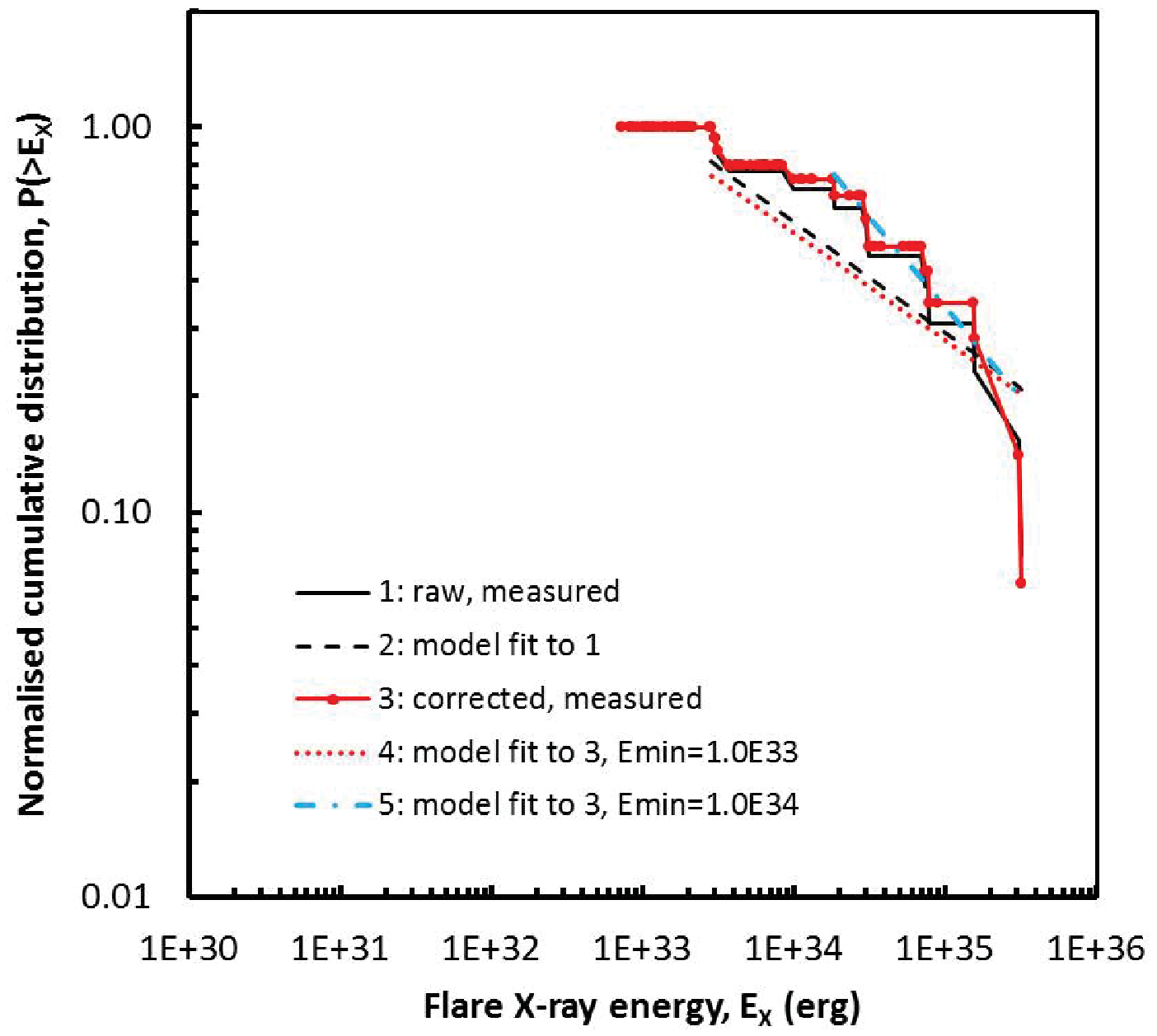}}
	\resizebox{\hsize}{!}{\includegraphics{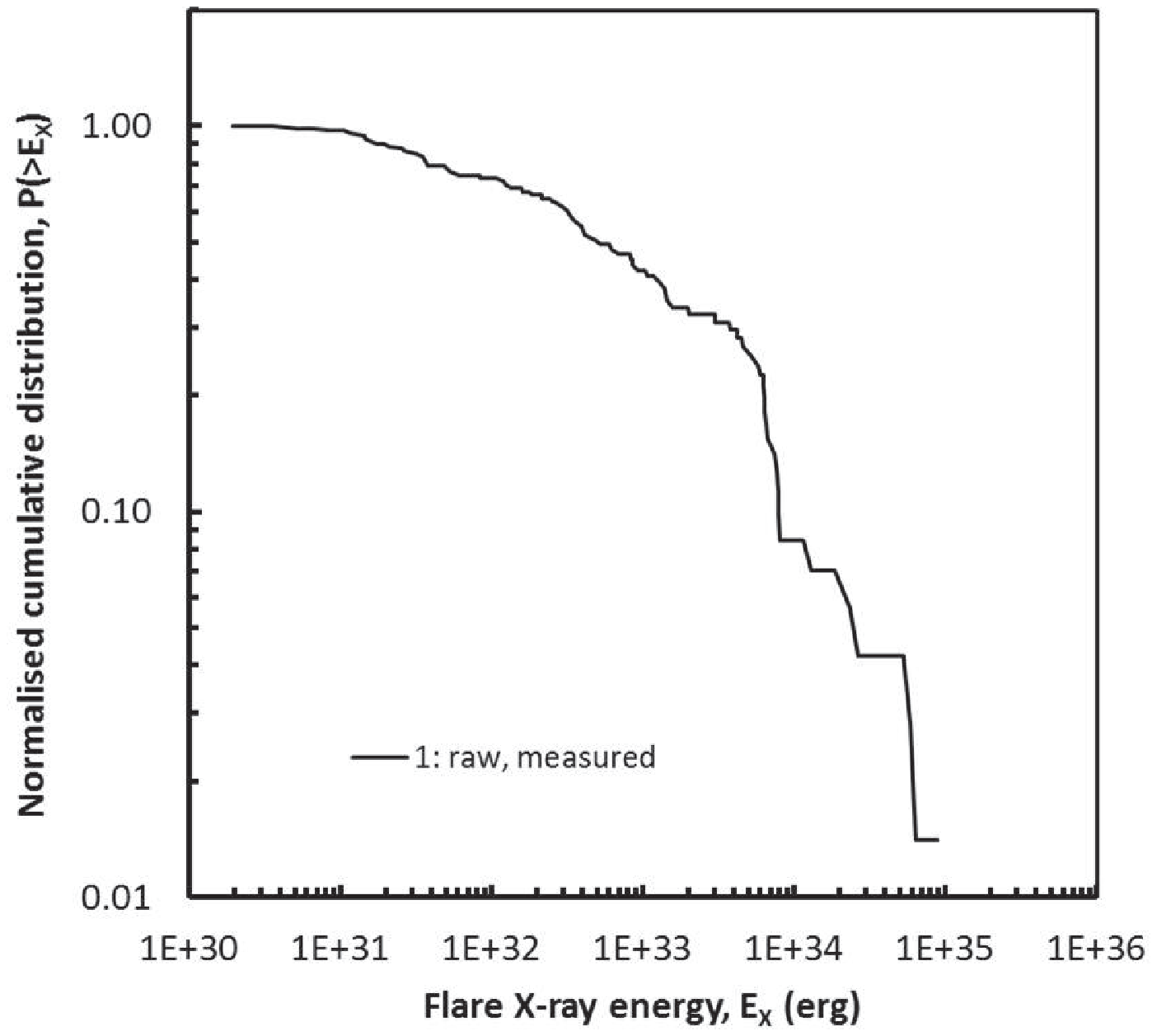}}
	\caption{As Fig.~\ref{figFlLxFreq}, but for flare X-ray emitted energy, $\rm E$. }
	\label{figFlExFreq}
\end{figure}

In a sense, frequency distributions of various flare properties represent the final distillation of the survey.
Measured rates of flaring, or in more general terms, frequency distributions of intrinsic physical properties such as peak luminosity and emitted energy, allow comparison with models for flare production. The frequency distribution of peak flux is of practical use in predicting numbers of flares observable in future surveys and missions. We discuss the peak-flux distribution first, together with the topic of survey completeness. 

In order to estimate the rate of flaring and the flare `duty cycle'\footnote{
Duty cycle = the sum of the flares' duration ($ \tau_{\rm r} + \tau_{\rm f} $) as a fraction of the sum of the light-curves' duration (the total `on-time'). }
we had to take account of possible incompleteness in the survey, due to variations in the minimum detectable flare strength arising mainly from the source quiescent flux against which the flare had to be viewed. Hence, for each flare we computed a `survey completeness' factor, C (range 0.0 -- 1.0), being the fraction of the survey time in which the flare could have been detected.
Each flare then contributed 1/C to the corrected distribution.
For the CVS serendipitous sample and our chosen S:N threshold of 10.0 (Sect.~\ref{results}), the maximum correction factor 1/C was $\sim 2$ with most flares requiring a correction of $\la 1.3$. Thus, the corrections applied were relatively small.
The calculation of C is described in Appendix~\ref{appCfactor}, and example distributions are shown in Fig.~\ref{figEminCover}.

In order to provide a simple parametrisation of the flare-peak frequency distribution, we have considered a (cumulative) power-law:

$N(>f_\mathrm{X}/f_\mathrm{X,ref}) = N_\mathrm{ref}\, (f_\mathrm{X}/f_\mathrm{X,ref})^{-\alpha_\mathrm{f}} $

and determined the power-law index $\alpha_\mathrm{f}$ from the measurements using a maximum-likelihood (ML) method (Jauncey 1967; Wall \& Jenkins 2003). Consistency between measurements and model was checked with the Kolmogorov-Smirnov test (e.g.\ Wall \& Jenkins 2003), yielding $D_{\rm max} \sim 0.25$, corresponding to formal acceptance of the fits at probabilities $\sim 20$\% for the serendipitous sample. 
We have used the power-law form 
purely
as a convenient,
empirical 
characterisation rather than to demonstrate any fundamental,
physically-based
shape for the measured distributions.

Table~\ref{tabFreqSumm} and Figs.~\ref{figFlPkFreq}, \ref{figFlDurFreq} summarise the flare-frequency statistics, in relation to peak flux ($f_\mathrm{X,peak}$) and duration/duty-cycle. 
The values presented are based on only `fully-observed' flares; this results in an under-estimate, but this amounts to $\la 20$\%.
The selected threshold flux $f_\mathrm{X,peak,min} = 2.5 \times 10^{-13}$ (Table~\ref{tabFreqSumm}, columns A) corresponded to the lowest flux in the sample; we also show the results for a substantially higher value, $f_\mathrm{X,peak,min} = 1.0 \times 10^{-12}$ (Table~\ref{tabFreqSumm}, columns B), resulting in an increase in the best-fit $\alpha$ but with much larger uncertainties. (Increasing $f_\mathrm{X,peak,min}$ still further would result in a rather small number of flares available for fitting, and hence even larger uncertainties.)
Due to the small numbers of stars, and the lack of detailed information for most of the serendipitous sample, we have not attempted to  divide them into different categories; hence the distributions and statistics refer to a rather heterogeneous collection of stellar types.
As the target sample comprises a rather arbitrary and ill-defined set of objects (other than all being well-established active stars), we have not attempted to correct and fit their frequency distributions.

We have examined the sensitivity of the derived values to changes in the selection criteria and correction factors for the serendipitous sample, as follows.
As expected, reducing the S:N threshold increased the correction factors to be applied. However, S:N thresholds of 8.0, 5.0 resulted in 
only relatively modest increases in corrected flare rates and duty cycles, by factors $\sim 1.4$ and 1.8 respectively.
Conversely, increasing the S:N threshold to 15.0 yielded a reduction by a factor $\sim 0.85$.
The effect of varying S:N threshold on the power-law index was also relatively small, with the largest change being $\alpha_\mathrm{corrected} = 0.77 \pm 0.15 $ for a S:N threshold of 5.0.
We have also examined the effect of errors in the estimation of the coverage-correction factor C; shifting the C distribution by a factor 2, in the sense of  worsening the coverage (i.e.\ ${\rm C}(f_\mathrm{X,peak,i}) \rightarrow {\rm C}(2f_\mathrm{X,peak,i}) $), increased the corrected numbers by $\la 20$\%
and increased $\alpha$ by a factor $\sim 1.3$--1.4, but otherwise had little effect.

We are reporting the activity levels of serendipitously-observed stars for a sample defined as being generally `active' (i.e.\ X-ray emission detected, but not necessarily flaring or otherwise variable; Sect.~\ref{obs2XMM}), and relatively optically bright, and incomplete in respect of M-type stars (Sect.~\ref{obsTyc}). This is important to bear in mind if comparing these results to other samples or surveys.
In addition, care must be taken in recognising the set of stars to which the various statistics are applicable, as indicated by the Notes in Table~\ref{tabFreqSumm}, e.g.\ the estimated flare frequency  when considered only within the set of observed flaring stars or only those stars with detected X-ray emission, is obviously higher than if considered over the whole set of Tycho stars falling within the 2XMM survey area. 

We note that the flare duty-cycle values from our survey are of the same order, i.e.\ typically a few percent, as those reported by Walkowicz et al.\ (2011) from a {\it Kepler} white-light sample, when considering only those stars observed to flare (and hence biasing the result upwards). 
Our flare-frequency distribution power-law indices $\alpha_\mathrm{f}$ are comparable to that of Pye \& McHardy (1983), who derive a value of 0.8 ($+0.4,\, -0.5$) from an all-sky survey of fast-transient X-ray sources, at least 60\% of which were likely cool-star flares (Pye \& McHardy 1983; Rao \& Vahia 1987).
However, we caution against too-detailed comparisons given the differing selection criteria and sample types.

Using the principles outlined above, we have also estimated the flare rates in terms of intrinsic properties of the flares: peak X-ray luminosity ($L_\mathrm{X,peak}$) and X-ray emitted energy ($E_\mathrm{X}$).
We recognise, as noted by G\"udel (2004, section 13.5), that there are likely to be significant biases and incompleteness in such an analysis.
In particular, the distributions we present are observed distributions, not volume-normalised luminosity and emitted-energy functions; the coverage correction C takes explicit account only of Tycho stars with 2XMM time-series,
and we see the probable effects of incompleteness in the flattening of the distributions towards low $L_\mathrm{X,peak}$ and $E_\mathrm{X}$ due to failure to detect intrinsically faint flares towards larger distances (see Figs.~\ref{figFlLxFreq}, \ref{figFlExFreq}).
The results are given in Table~\ref{tabFreqSumm} and Figs.~\ref{figFlLxFreq}, \ref{figFlExFreq},
in a similar way to the peak-flux values described earlier.
For $E_{\rm X,min} = 1.0 \times 10^{33}$ erg, the power-law index $\alpha_\mathrm{E}$ for the 
serendipitous sample is somewhat below the lower-end 
values ($\sim 0.5$) in the compilation of G\"udel (2004), possibly due to incompleteness towards lower $E_\mathrm{X}$. 
If we raise the serendipitous-sample threshold $E_{\rm X,min}$ to $1.0 \times 10^{34}$ erg, the best-fit slope steepens to $\sim 0.5$
(Table~\ref{tabFreqSumm}, columns B).
In order to mitigate the possible incompleteness effects, while recognising the limited number of stars and flares available, we also attempted to define an approximately volume-limited sample by lowering the S:N threshold to 5.0 and setting a maximum distance $d_\mathrm{max}$ (c.f.\ Fig.~\ref{figEminCover}(d)) of 100 pc. This gives 11 flares with $L_{\rm X,peak} > 10^{29}\ \rm erg\,s^{-1}$ and 8 flares with $E_{\rm X} > 10^{33}\ \rm erg $, with an estimated completeness $\ga 50$\% (though the maximum applied correction factor 1/C $\sim 10$), resulting in power-law indices of 
$\alpha_\mathrm{L} = 0.52 \pm 0.16$ and $\alpha_\mathrm{E} = 0.75 \pm 0.26$.

Audard et al.\ (1999, 2000) have reported EUV flare rates from a sample of 10 active cool stars.
The estimated rate of flaring for our serendipitous-stars sample, even when restricted to the stars observed to flare (i.e.\ Table~\ref{tabFreqSumm}, note (d)), is much lower, by a factor $\sim 10$ -- $\sim 100$, than those reported by Audard et al.\ for the seven stars in their sample with measured distributions extending beyond $E_{\rm X} \sim 10^{33}$ erg.
Audard et al.\ (2000) noted a roughly linear relation between flare rate (above a defined $E_{\rm X}$ threshold) and stellar quiescent coronal luminosity, $L_\mathrm{X,quies}$.
We do not have sufficient detected flares to yield useful frequency distributions for individual stars; however, we have formed a `scaled' frequency distribution for the serendipitous-stars sample by weighting the contribution of each flare inversely according to the quiescent X-ray luminosity of its star, i.e.\ by $1/L_\mathrm{X,quies}$. The resulting $\alpha_\mathrm{E} \sim 0.7$ (i.e.\ Table~\ref{tabFreqSumm} note (e)), is somewhat steeper than the unweighted value, and still consistent with the range reported by Audard et al.\ (2000) and the literature reviewed by G\"udel (2004).
Schrijver et al.\ (2012) have compared solar flare frequency distributions with those for the five G--K-type stars from Audard et al.\ (2000), scaling the latter (in frequency) by $L_\mathrm{\odot,X,quies} / L_\mathrm{\star,X,quies} $, using a nominal solar  quiescient X-ray luminosity $L_\mathrm{\odot,X,quies} = 4.3 \times 10^{27} $ $\rm erg\,s^{-1}$. They show that even with the scaling, the stellar rates are substantially higher, by a factor $\sim 100$, than the solar distribution.
For $E_{\rm X,min} \sim 10^{33}$ erg, their scaled rates are $\sim 10$ events/star/year, compared with $\sim 0.01$ solar events/year\footnote{Following Schrijver et al.\ (2012), we have assumed $ E_\mathrm{bol} / E_\mathrm{X} \sim 3$--5.  }.
The corresponding scaled rate for our serendipitous sample was $\sim 5 \times 10^{-4}$ events/star/year, somewhat below the nominal power-law indicated by Schrijver et al.\ (their figure~3), but arguably within their overall error limits.
In summary, our serendipitous-sample results can span the solar rate estimates, depending on the set of stars (and associated total on-time) within which the detected flares are considered, i.e.\ normalising  to only those  stars (30, $\sim 1.4 \times 10^6$~s) observed to flare will obviously yield a higher rate than normalising to all stars (504, $\sim 2.4 \times 10^7$~s) with 2XMM light-curves irrespective of detected variability.
In this context, we now consider in more detail those stars which did not distinguish themselves by detected flaring within the observations, and compare their general properties with those that were observed to flare.

\begin{figure*}
	\resizebox{0.35 \hsize}{!}{\includegraphics{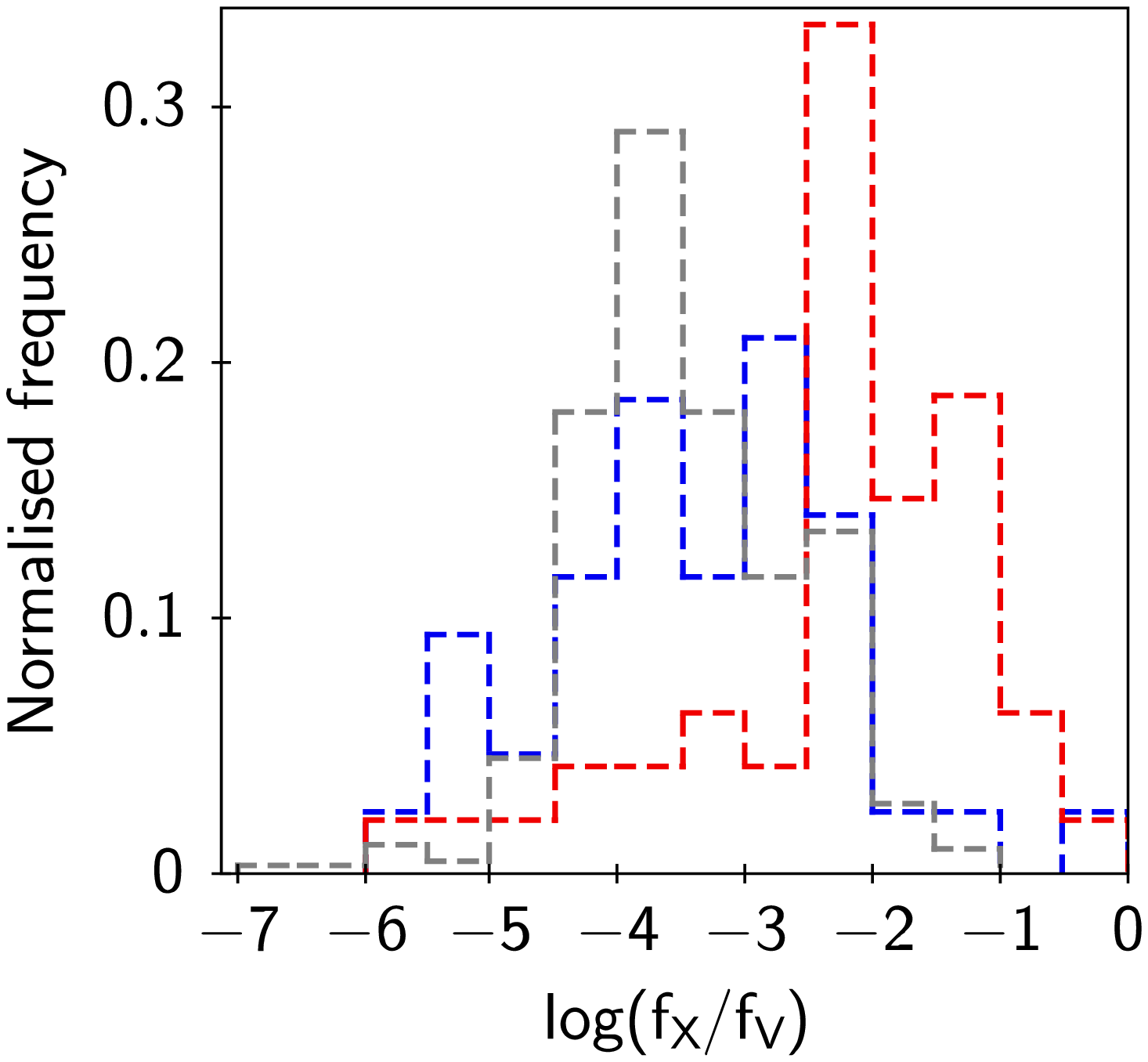}}
	\resizebox{0.35 \hsize}{!}{\includegraphics{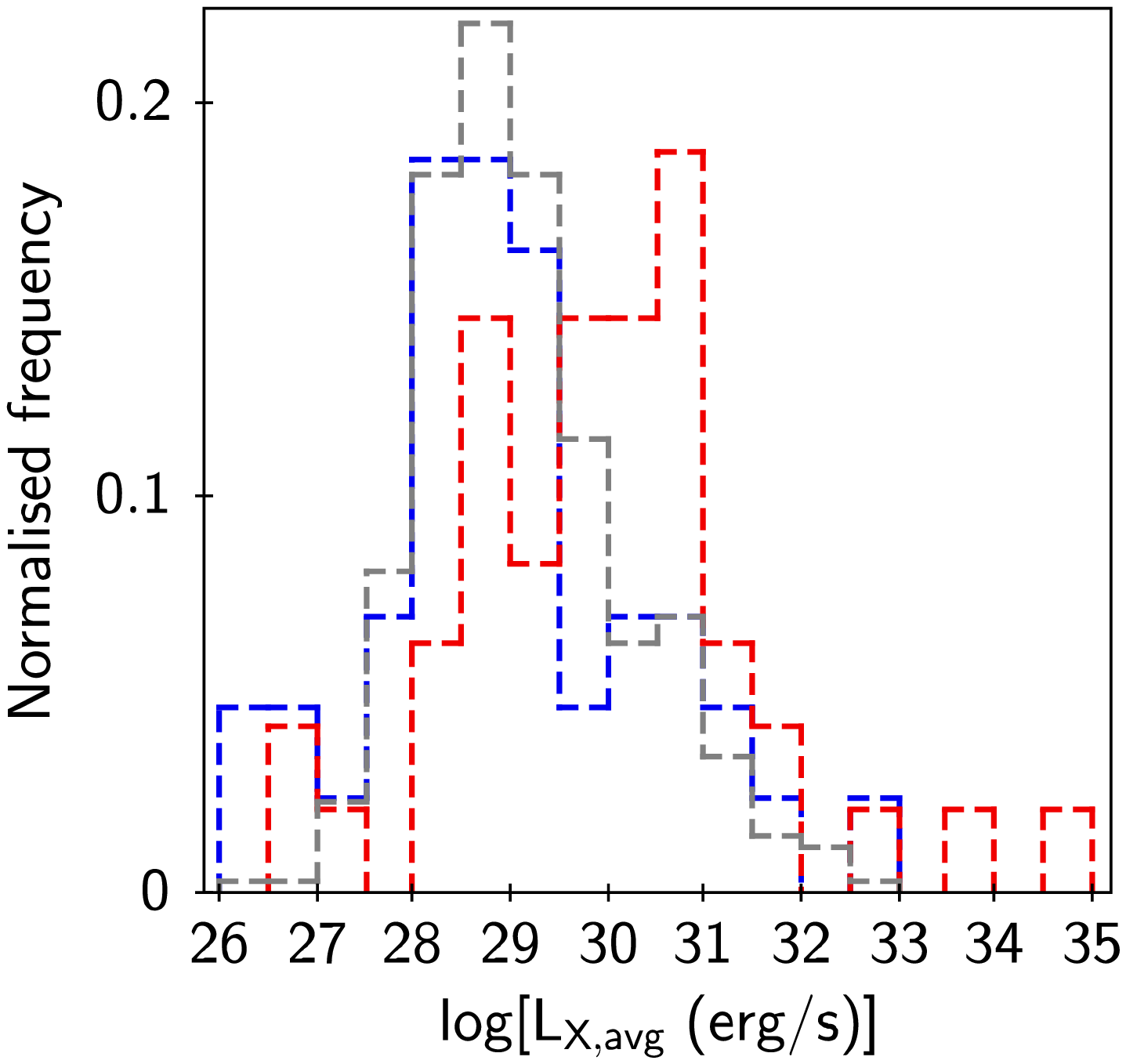}}
	\resizebox{0.35 \hsize}{!}{\includegraphics{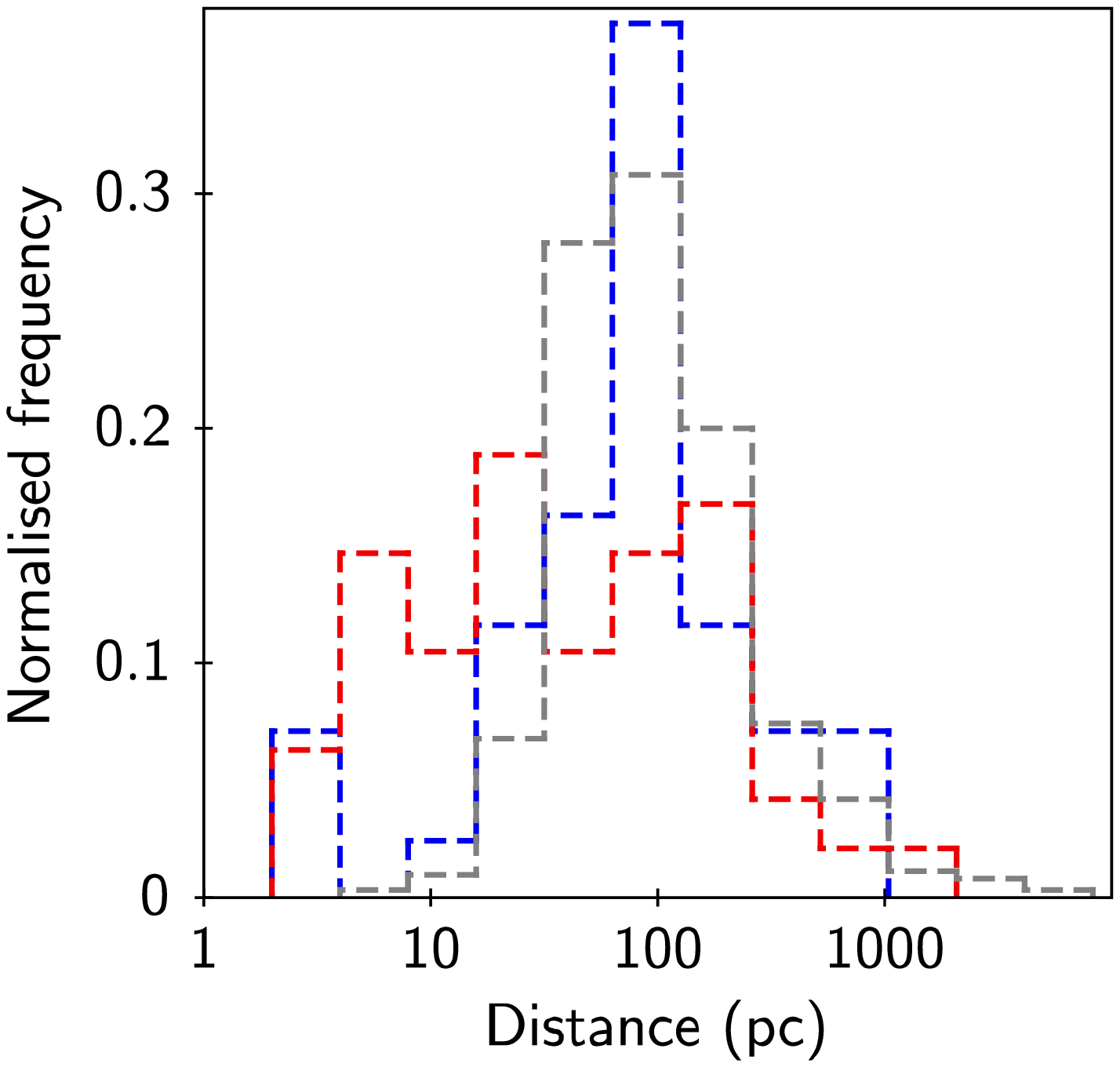}}
	\resizebox{0.35 \hsize}{!}{\includegraphics{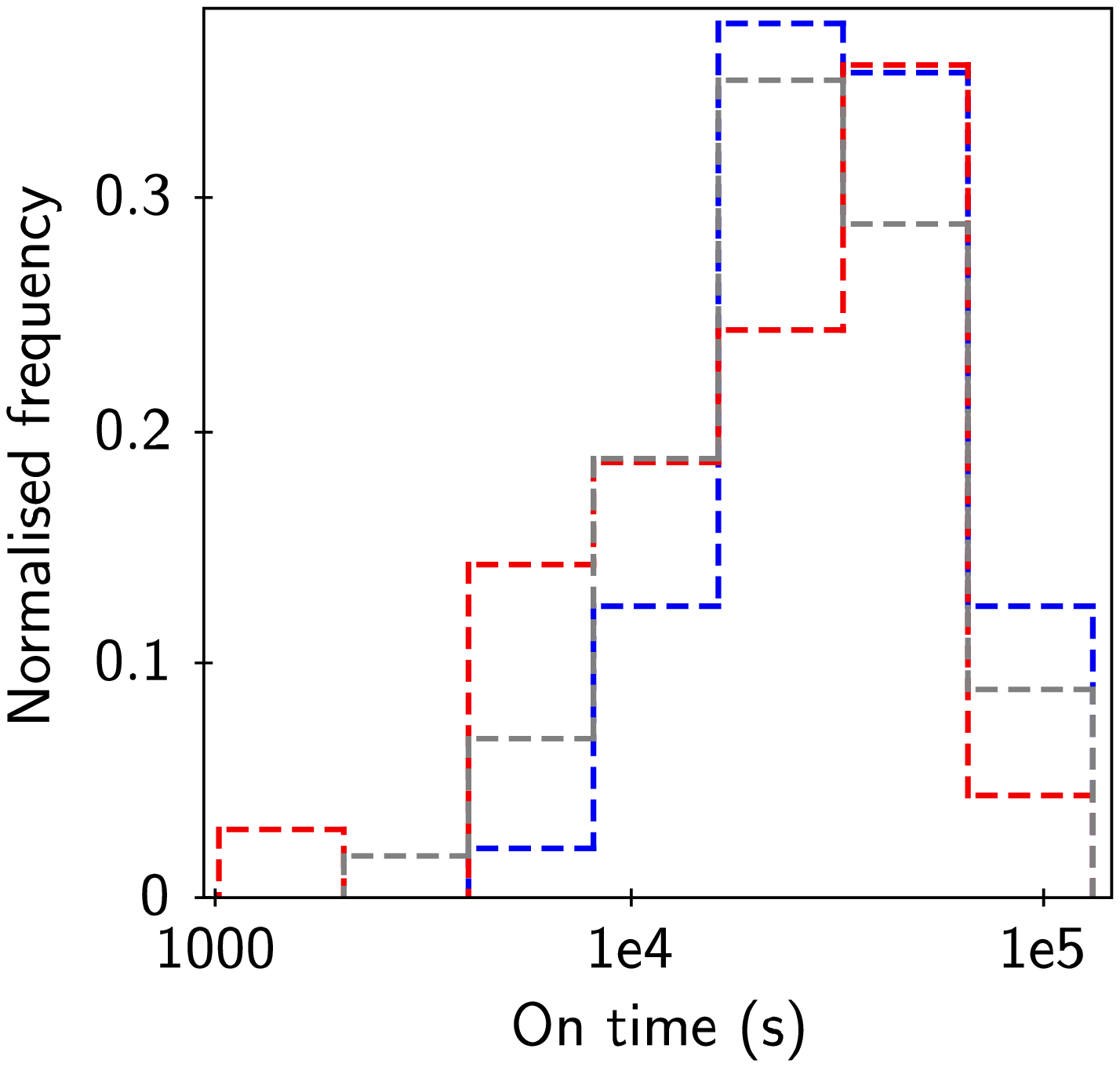}}
	\resizebox{0.35 \hsize}{!}{\includegraphics{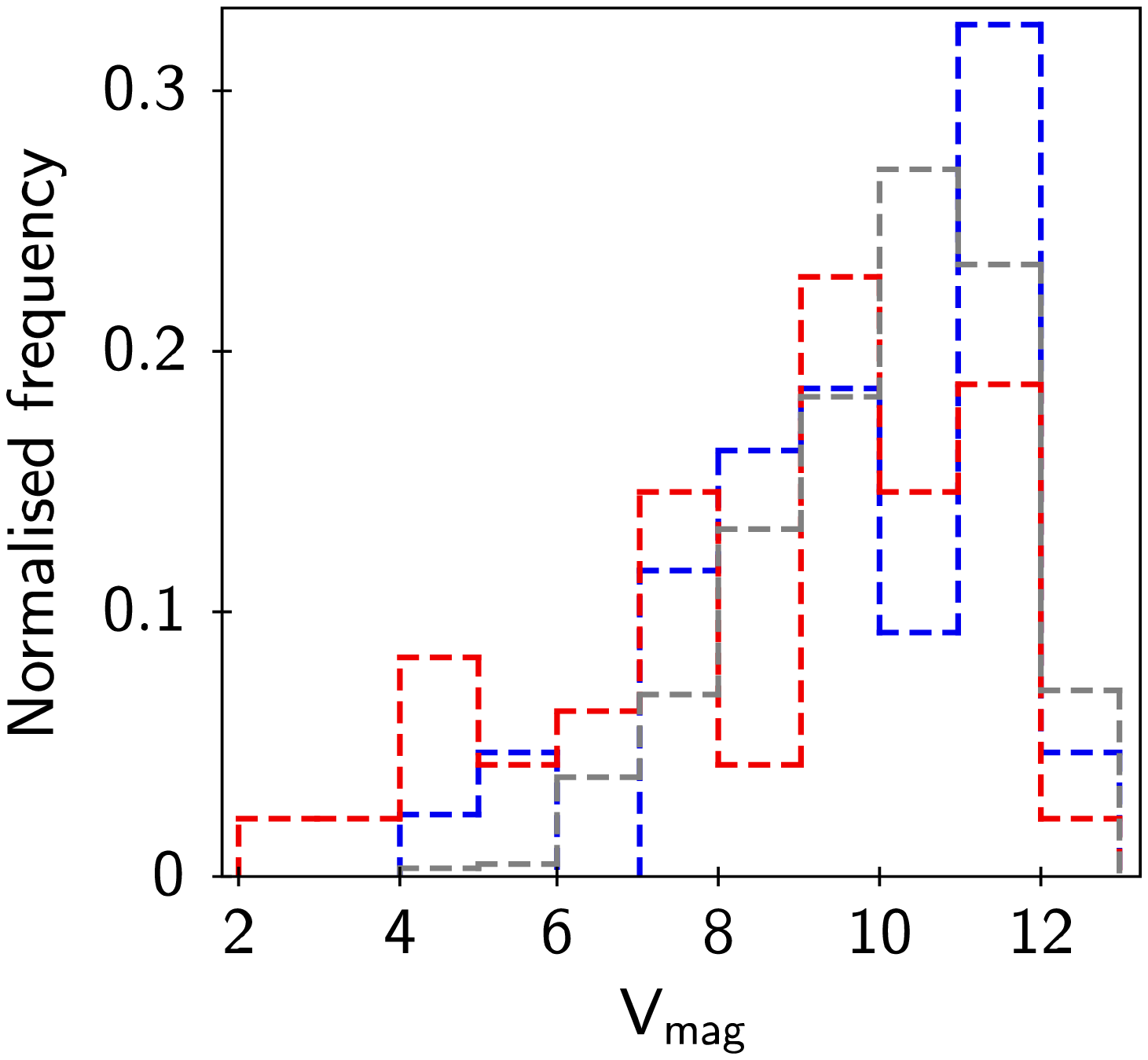}}
	\resizebox{0.35 \hsize}{!}{\includegraphics{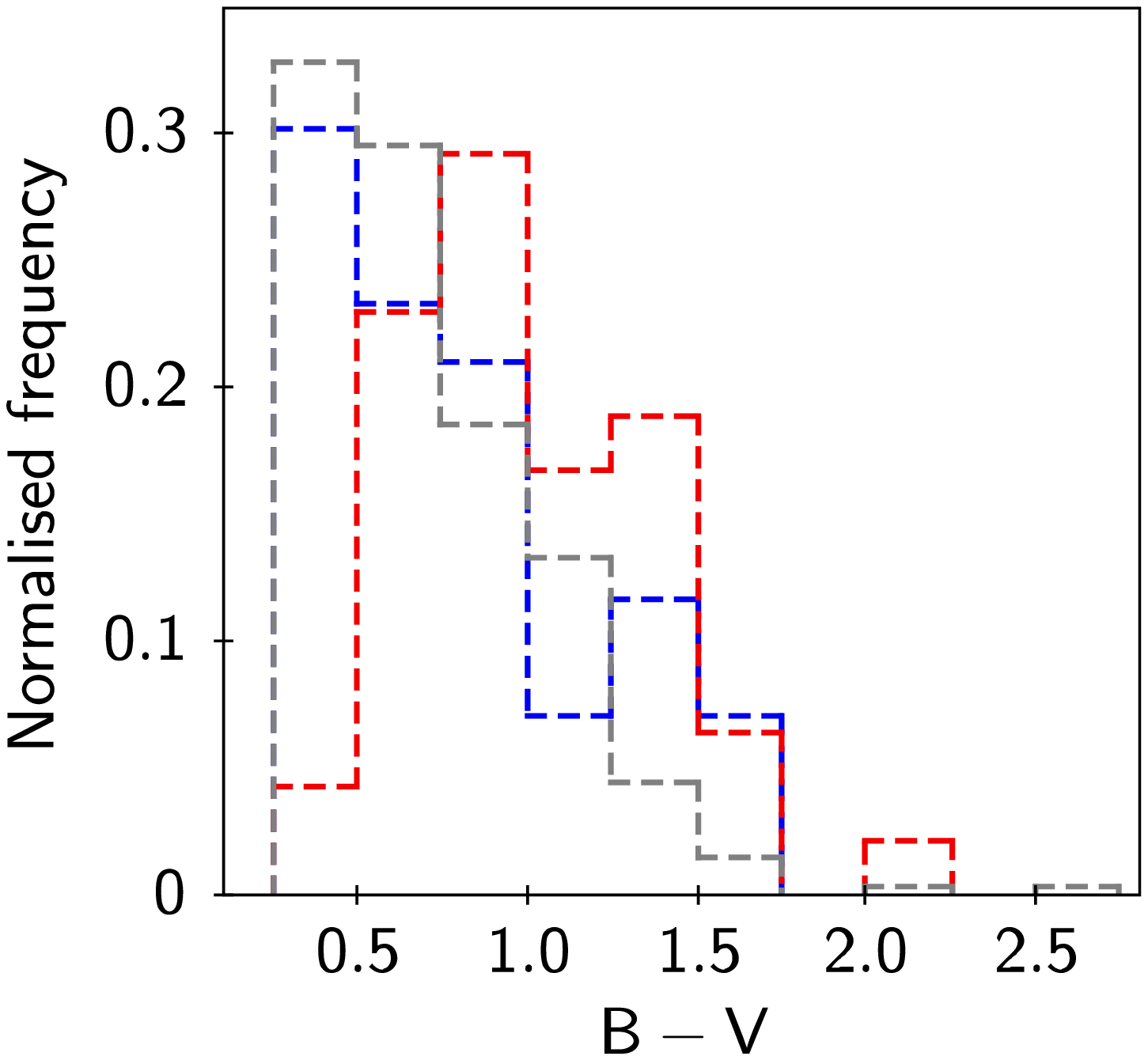}}
	\caption{Differential frequency distributions of the serendipitous variable sample (SV, blue lines), the serendipitous non-variable sample 	(SNV, grey lines) and the target variable sample (TV, red lines), for various properties of the stars and their associated observations. The distributions are normalised to unit area.
	{\it (From top left, across) (a)} Log~[X-ray to visual-band flux ratio, $f_\mathrm{X}/f_\mathrm{V}$]. 
	{\it (b)} Log~[quiescent X-ray luminosity, $L_\mathrm{X,quies}$ ($\rm  erg\,s^{-1}$)].
	{\it (c)} Distance (pc). 
	{\it (d)} On-time (s).
	{\it (e)} Visual apparent magnitude, $V$.
	{\it (f)} Colour, $B-V$.
	The distributions in (d) relate to individual detections (i.e.\ each DETID was counted once), while the others relate to individual stars (i.e. each SRCID was counted once).
	(a) and (b) use values averaged over all 2XMM detections of each source (i.e.\ derived from 2XMM $\rm SC\_$ parameters).
	}
	\label{figCompareHist}
\end{figure*}

\begin{figure*}[h]
	\resizebox{0.45 \hsize}{!}{\includegraphics{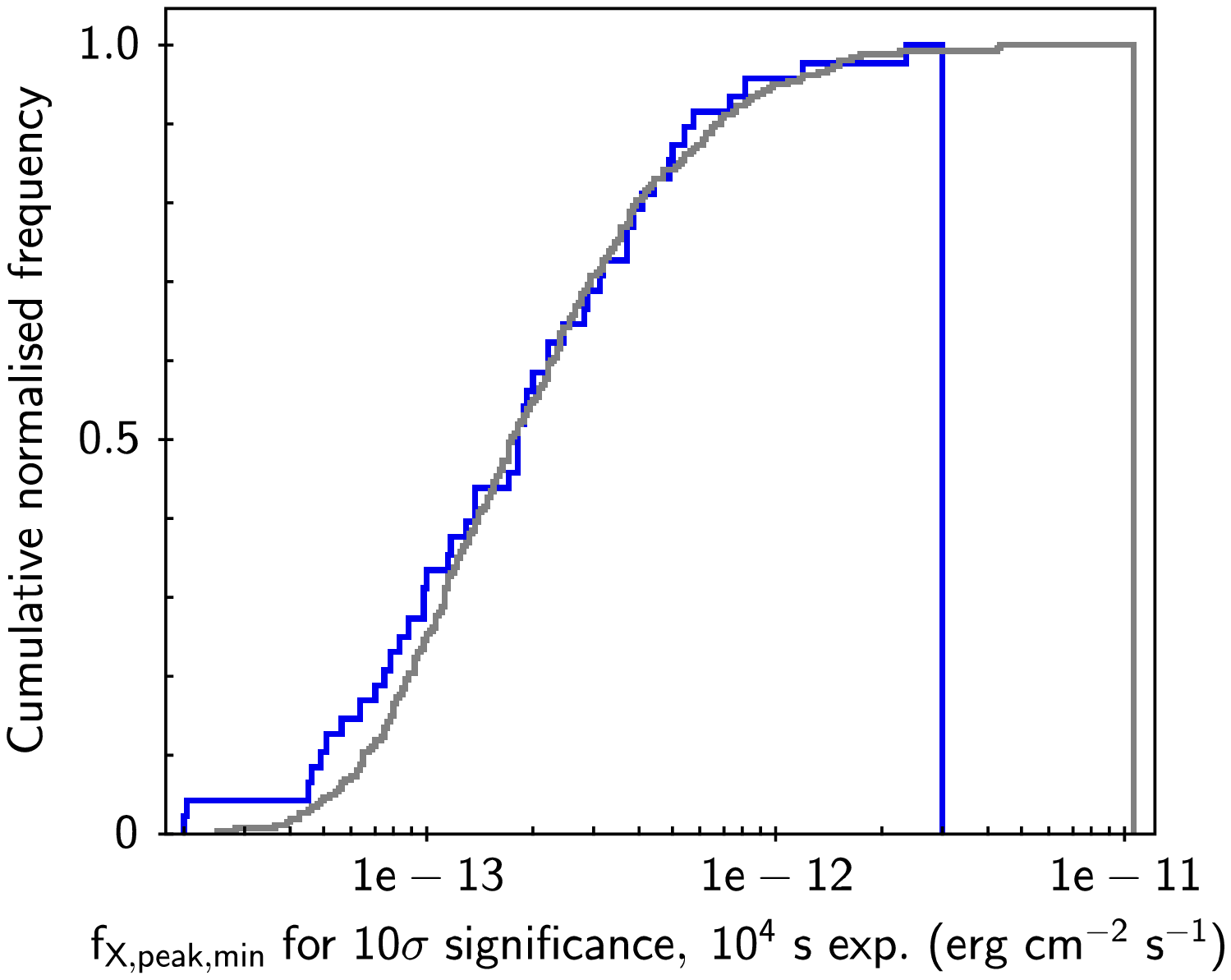}}
	\resizebox{0.45 \hsize}{!}{\includegraphics{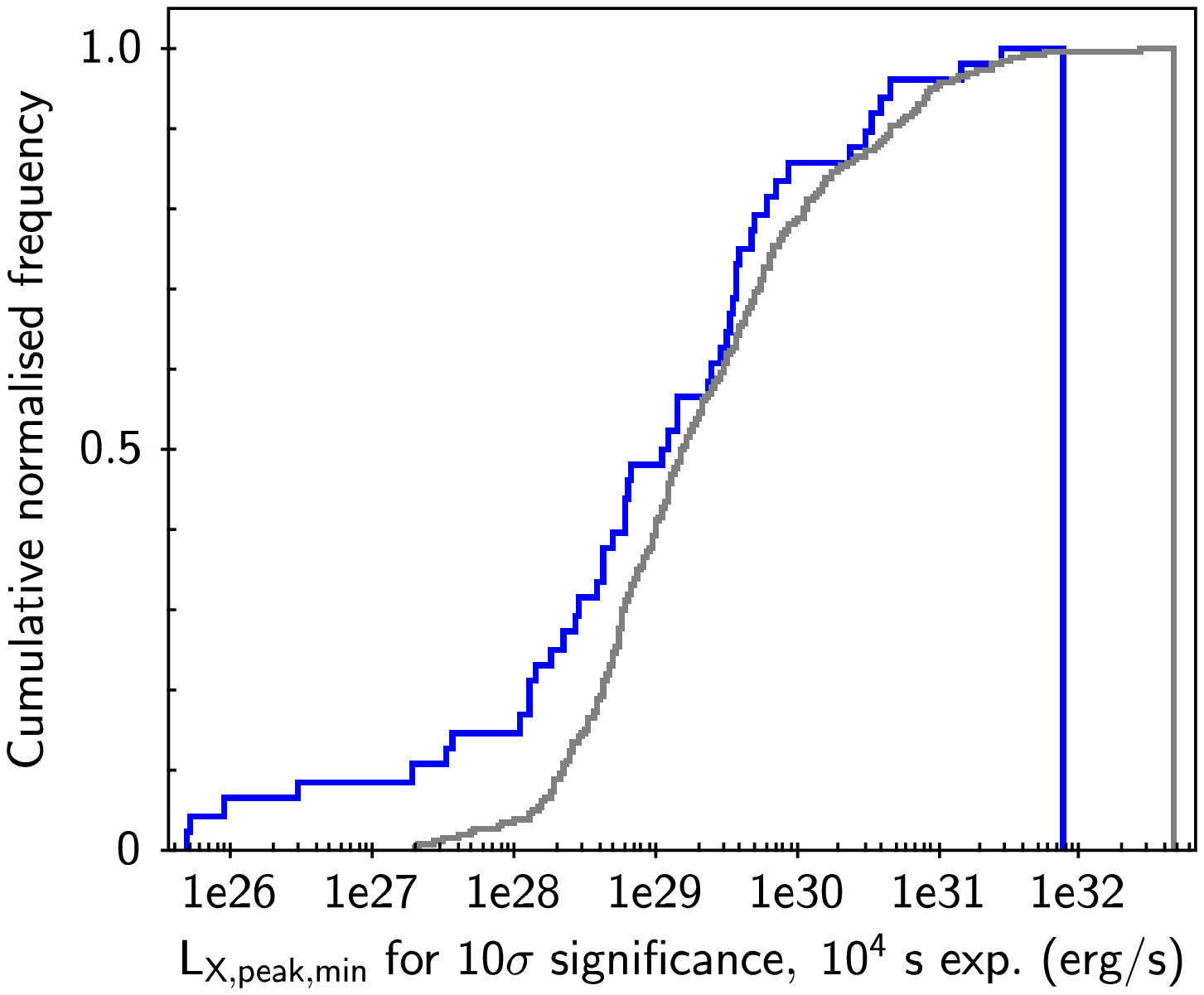}}
	\resizebox{0.45 \hsize}{!}{\includegraphics{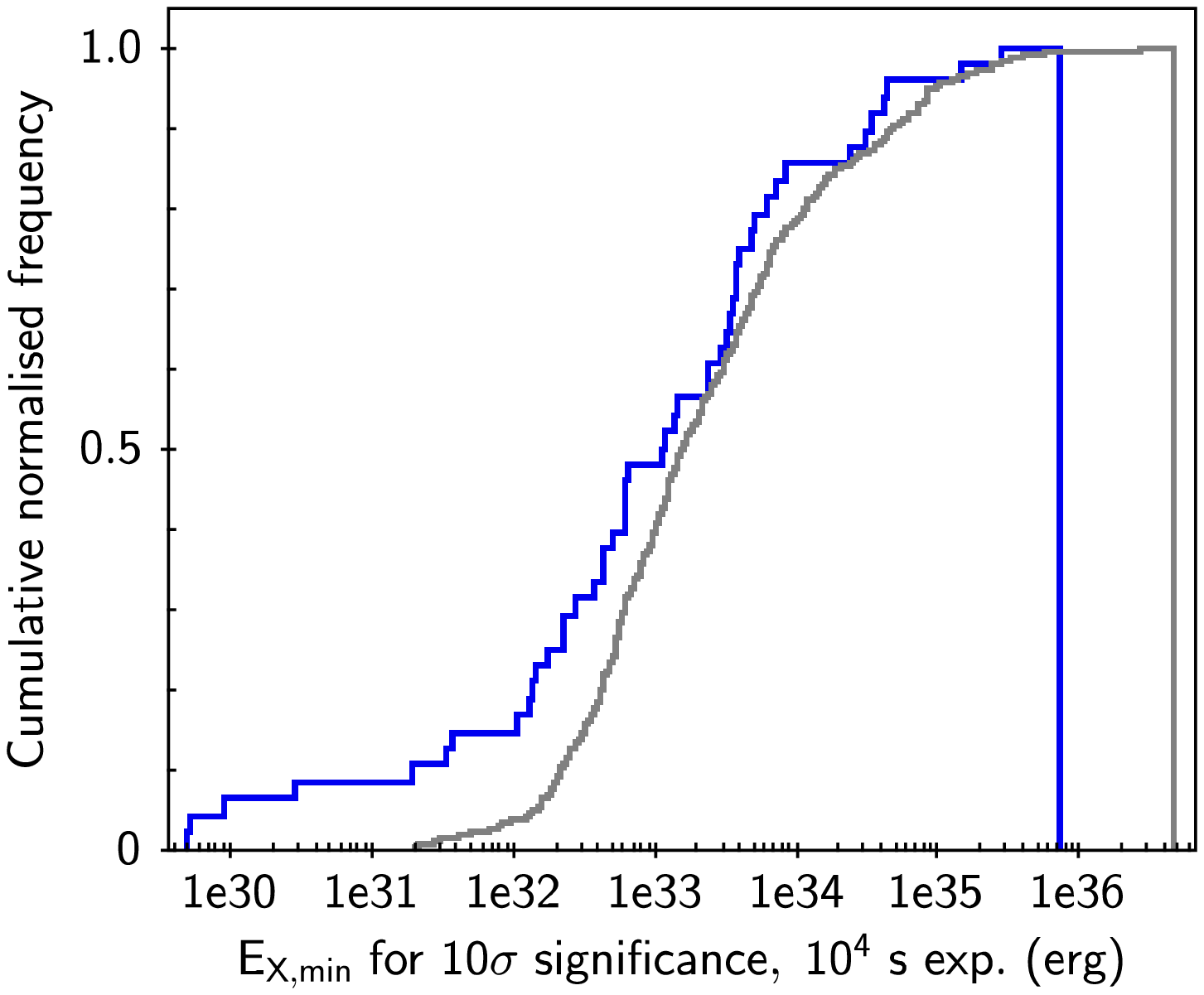}}
	\resizebox{0.45 \hsize}{!}{\includegraphics{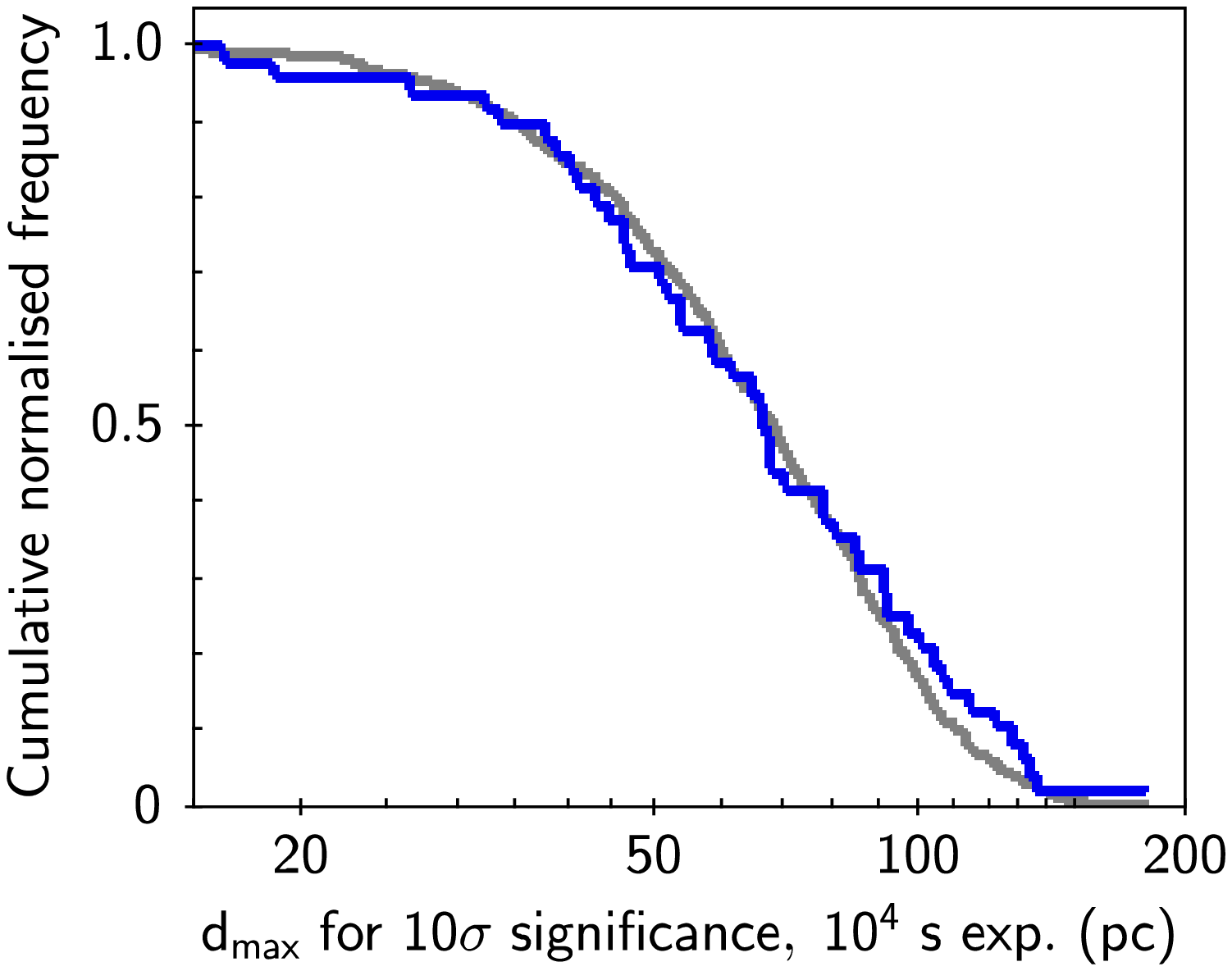}}
	\caption{
	{\it (From top left, across) (a)} Survey completeness or `coverage', C, in terms of the fraction of the survey (or specific samples) in which a flare with peak X-ray flux, $ > f_\mathrm{X,peak,min}$ $\rm (erg\,cm^{-2}\,s^{-1})$ could be detected, for a flare S:N $>10$ and a duration of $10^4$~s; $f_\mathrm{X,peak,min}$ scales as [S:N threshold] / [$\sqrt{\rm duration}$].
	Key: blue line: serendipitous variable sample (SV); grey line: serendipitous non-variable sample (SNV); the line for all serendipitous stars with time-series is indistinguishable from that for the SNV sample. 
{\it (b)} As (a), but for peak X-ray luminosity, $ > L_\mathrm{X,peak,min}$ $\rm (erg\,s^{-1})$;
$L_\mathrm{X,peak,min}$ scales as [S:N threshold] / [$\sqrt{\rm duration}$].
{\it (c)} As (a), but for X-ray emitted energy, $ > E_\mathrm{X,min}$ (erg); $E_\mathrm{X,min}$ scales as [S:N threshold] $\times$ [$\sqrt{\rm duration}$].
{\it (d)} As (a), but for maximum distance, $ < d_\mathrm{max}$ (pc) to which a flare with $ L_\mathrm{X,peak} > 10^{29}$ or $ E_\mathrm{X} > 10^{33}$ could be detected;
$d_\mathrm{max}$ scales as  ${\rm duration^{0.25}} \times \sqrt{L_\mathrm{X,peak} / {\rm [S:N\ threshold]}} $ or ${\rm duration^{-0.25}} \times \sqrt{E_\mathrm{X} / {\rm [S:N\ threshold]}} $ .
	}
	\label{figEminCover}
\end{figure*}

In Fig.~\ref{figCompareHist} we show frequency distributions of the serendipitous variable sample (SV), the serendipitous non-variable sample (SNV) and the target variable sample (TV), for various properties of the stars and their associated observations.
Visual inspection indicates:
\begin{itemize} 
	\item comparing SV and SNV: no substantial difference for any of the plotted quantities, i.e.\ $f_\mathrm{X}/f_\mathrm{V}$, $L_\mathrm{X,quies}$, distance, on-time, $V$, $B-V$, minimum detectable $E_\mathrm{X}$.
	\item comparing SV and TV: TV stars tend to have higher $f_\mathrm{X}/f_\mathrm{V}$ and $L_\mathrm{X,quies}$, smaller distance, and be optically redder (i.e. later spectral type). All these differences may plausibly be attributed to selection effects in the TV sample.
\end{itemize}
The `survey coverage' curves in Fig.~\ref{figEminCover} show that we would expect to detect at least 50\%  of all flares with $E_\mathrm{X} \ga 10^{33}$ erg and duration $\la 10^4$ s, i.e.\ the survey incompleteness within these criteria is expected to be no more than a factor two.

Why was there apparently such a low fraction of the serendipitous stars seen to flare?
We suggest two obvious, and not mutually exclusive, possibilities.
(i) The stars observed to flare are broadly representative of all the serendipitous-sample stars with 2XMM light-curves, and the non-detection of flares from $>90$\% of the sample simply reflects the true (rather low) flare rate.
(ii) The stars observed to flare have some `activity' property which manifests itself in flaring, but not in the general properties such as quiescent X-ray luminosity. Such additional sources of activity might, for example, arise in magnetic-field or tidal interactions between the components of a close binary system, or interaction between coronal magnetic field and circumstellar material in a PMS star (see e.g.\ G\"udel 2004).
We emphasise that we are not implying that the stars in the non-variable sample are intrinsically without flare activity, only that they have not produced detectable flares in the 2XMM survey observations, and we note that this result is broadly consistent with estimates based on Poisson statistics (e.g.\ Akopian 2013).
The data used in the current analysis were insufficient in terms of total observing time for each star to 
fully
resolve issue (i), or to address the related topic of degree of correlation, if any, between flares closely-spaced in time.
However, these scenarios are amenable to test via future work: both detailed spectrometry and photometry of individual stars, and examination of the growing XMM observational database, the latter now having publicly-available more than a factor three more observations than for 2XMM.
A preliminary inspection of the additional light-curves now available via the 3XMM database shows that several of the stars in the SNV sample do exhibit flares.

\section{Conclusions} \label{conclusions}

We have presented a survey of stellar flares from a sample of Tycho cool (F--M type) stars observed by XMM-Newton. The results allow a uniform visualisation and analysis of the XMM-Newton data; we have augmented the standard X-ray data products to include hardness-ratio time-series, and where available, have presented the associated XMM OM UV/visible data. The survey has enabled recognition of new, coronally-active stars, observed serendipitously by XMM-Newton. We have demonstrated the utility of such uniform, relatively large samples in the statistical investigation of the physical properties of stellar coronae and activity, and potentially as diagnostics of underlying mechanisms. 
In a wider context, stellar flares may, as in our own solar system, have significant influence on exoplanetary systems (see e.g.\ Dartnell 2011, Feigelson 2010, Horvath \& Galante 2012, Melott \& Thomas 2011), and may contribute significantly to the `hard' (photon energy $> 2$~keV) X-ray emission from the Galactic ridge (Warwick 2014).
In addition, for planning of future wide-field X-ray missions and instruments (e.g.\ Osborne et al.\ 2013), the estimates of flare rates and duty cycles provide a useful guide to the expected contribution of F--K-type stars to the overall frequency of X-ray transient events.

\begin{acknowledgements}

The authors acknowledge the financial support from the UK Science and Technology Facilities Council and the UK Space Agency during part of the time that this work was undertaken.
The authors thank the referee, Manuel G\"udel, for his comments and suggestions.
This research has made use of: the Leicester Database and Archive Service (LEDAS), and the XMM-SSC Web site, at the Department of Physics and Astronomy, University of Leicester, UK; 
the SIMBAD database, the VizieR catalogue access tool, and Aladin, operated at CDS,
Strasbourg, France; the SAO/NASA Astrophysics Data System, principally via the site hosted by the University of Nottingham, UK. This research was also greatly aided by Mark Taylor's (University of Bristol) TOPCAT (Tool for OPerations on Catalogues And Tables) software.

\end{acknowledgements}

\appendix

\section{Quality screening of the 2XMM light-curves} \label{appQual}

We list here the main features found in the data products that were likely to compromise the quality of the light-curves with respect to validity of the detection of `variability' and its characterisation. These features were identified mainly from visual inspection of the graphical products. 
Identification of any of these features in a dataset led to rejection of the light-curve for the present work.

Type of feature and number of 2XMM-Tycho variable detections rejected:
\begin{itemize}
	\item Optical loading (see 2XMM Appendix~A): 11
	\item Confusion and contamination of source flux by another (usually brighter) nearby source: 8
	\item Problems in the background determination (e.g. poor background subtraction, background contamination from a nearby source): 7
	\item Exposure correction or GTI problem (leading to zeros or NULL-values in some of the source light-curve bins): 2
	\item Spurious detection (due to a nearby bright source): 1
\end{itemize}
Thus in total, 29 out of the 157 2XMM-Tycho variable detections were rejected. 

Of these 29, 11 had 2XMM SUM\_FLAG$\ge 3$, and none had SUM\_FLAG=2 (see 2XMM Appendix D, Sect.\ D.5 for description of the SUM\_FLAG values).

\section{Calculation of the `completeness correction factor', C}  \label{appCfactor}

Here we outline the calculation of the correction factor, C, used in Section~\ref{dis-rates} to account for variations in the minimum detectable flare strength. These arose mainly from the source quiescent flux upon which the flare was superimposed. 
C is the fraction of the survey time in which the flare could have been detected.
We note that our primary aim was to demonstrate that our flare rates and related estimates were insensitive to incompleteness at the levels of accuracy needed for our analysis and warranted by the limited numbers in our samples, 
i.e.\ to estimate C to better than a factor $\sim 2$ and to utilise it at relatively modest correction values, i.e. 1/C$\la 2$.
We wished to generate an estimate of C in a simple and rapid manner, and based on quantities directly available from the 2XMM catalogue, rather than, for example, engaging in extensive simulations.

The minimum flare peak flux detectable above a signal-to-noise threshold of $SN_\mathrm{min}$ is:

$f_\mathrm{X,peak,SNmin} \approx SN_\mathrm{min} \sigma_\mathrm{f,cat} (\tau_\mathrm{on} / \tau_\mathrm{dur})^{0.5} $

where: $\sigma_\mathrm{f,cat}$ is the EPIC total-band flux error in the 2XMM catalogue (i.e.\ column ep\_8\_flux\_err), 
$\tau_\mathrm{on}$ is the observation duration (`on-time'), $\tau_\mathrm{dur}$ is the flare duration, and S:N is evaluated over the time interval $\tau_\mathrm{dur}$ (Sect.~\ref{analVar}).

The assumption here is that $\sigma_\mathrm{f,cat}$ provides a reasonable representation of the error on the source time-series. 
Comparison of the EPIC total-band count-rate error with the count-rate error derived directly from the individual camera (pn, MOS1, MOS2) light-curves indicates that the latter can be a up to a factor $\sim 2$ greater than the former (especially for MOS1, MOS2)\footnote{Though the individual camera count-rate errors from the catalogue are generally in good agreement with the corresponding light-curve values, with a mean ratio $\sim 1.0 \pm 0.2$.}.
As discussed in Sect.~\ref{dis-rates}, we have examined the effects of such errors in C on the resulting flare rates and distributions.
We have also verified that using $\sigma_\mathrm{f,cat}$, rather than an error estimate based on the quiescent count rate from the light-curve, results in $\la 20$\% change in the error value.

Strictly, the value of $\tau_\mathrm{on}$ is not precisely determined, since it varies between cameras (pn, MOS1, MOS2), and a flare (or other variability) could be flagged in any of the active cameras. Our solution for the present purpose was to use MOS1 on-time if available, else MOS2, else pn. This introduces an acceptably small `uncertainty' in $\tau_\mathrm{on}$, e.g.\ $\la 30$\% variation in derived $\tau_\mathrm{on}$ in $\sim 80$\% of cases.

C is given by the normalised, cumulative distribution of observation on-times, i.e.:

$ C(> f_\mathrm{X,peak,SNmin,j}) = \sum_{i=1}^j \tau_\mathrm{on,i} / \sum_{i=1}^n \tau_\mathrm{on,i} $

where: $n$ is the total number of observations in the sample being considered (and will, in general, include observations where no flares were detected),
and the set of [$ f_\mathrm{X,peak,SNmin,i},\ i = 1,n $] is ordered by increasing value.

C can also be expressed in terms of flare peak luminosity or emitted energy, via:

$ L_\mathrm{X,peak} = 4 \pi d^2 f_\mathrm{X,peak} $

$ E_\mathrm{X} \approx 4 \pi d^2 f_\mathrm{X,peak} \tau_\mathrm{dur} $

where: $d$ is the source distance.

Alternatively, C can be expressed in terms of maximum observable distance:

$ d_\mathrm{max} = \sqrt{L_\mathrm{X,peak} / (4 \pi f_\mathrm{X,peak}) } $

$ d_\mathrm{max} \approx \sqrt{E_\mathrm{X} / (4 \pi f_\mathrm{X,peak} \tau_\mathrm{dur} ) }  $.

Example curves are shown in Fig.~\ref{figEminCover}.

\nocite{*}
\bibliographystyle{aa} 
\bibliography{2xmm-tycho-aa-paper-bibtex} 

\newcommand{\noopsort}[1]{}
\begin{thebibliography}{109}
\expandafter\ifx\csname natexlab\endcsname\relax\def\natexlab#1{#1}\fi

\bibitem[{{Akopian}(2013)}]{2013Ap.....56..488A}
{Akopian}, A.~A. 2013, Astrophysics, 56, 488

\bibitem[{{Allen}(1973)}]{1973asqu.book.....A}
{Allen}, C.~W. 1973, {Astrophysical quantities, University of London, Athlone
  Press, London}

\bibitem[{{Ammons} {et~al.}(2006){Ammons}, {Robinson}, {Strader}, {Laughlin},
  {Fischer}, \& {Wolf}}]{2006ApJ...638.1004A}
{Ammons}, S.~M., {Robinson}, S.~E., {Strader}, J., {et~al.} 2006, \apj, 638,
  1004

\bibitem[{{Arnaud} {et~al.}(2010){Arnaud}, {Dorman}, \& {Gordon}}]{XSPEC}
{Arnaud}, K.~A., {Dorman}, B., \& {Gordon}, C. 2010, XSPEC An X-ray spectral
  fitting package, User's Guide, NASA/GSFC,
  http://heasarc.gsfc.nasa.gov/docs/xanadu/xspec/manual/manual.html

\bibitem[{{Arzner} {et~al.}(2007){Arzner}, {G{\"u}del}, {Briggs}, {Telleschi},
  {Schmidt}, {Audard}, {Scelsi}, \& {Franciosini}}]{2007A&A...468..501A}
{Arzner}, K., {G{\"u}del}, M., {Briggs}, K., {et~al.} 2007, \aap, 468, 501

\bibitem[{{Aschwanden} {et~al.}(2008){Aschwanden}, {Stern}, \&
  {G{\"u}del}}]{2008ApJ...672..659A}
{Aschwanden}, M.~J., {Stern}, R.~A., \& {G{\"u}del}, M. 2008, \apj, 672, 659

\bibitem[{{Audard} {et~al.}(2007){Audard}, {Briggs}, {Grosso}, {G{\"u}del},
  {Scelsi}, {Bouvier}, \& {Telleschi}}]{2007A&A...468..379A}
{Audard}, M., {Briggs}, K.~R., {Grosso}, N., {et~al.} 2007, \aap, 468, 379

\bibitem[{{Audard} {et~al.}(2000){Audard}, {G{\"u}del}, {Drake}, \&
  {Kashyap}}]{2000ApJ...541..396A}
{Audard}, M., {G{\"u}del}, M., {Drake}, J.~J., \& {Kashyap}, V.~L. 2000, \apj,
  541, 396

\bibitem[{{Audard} {et~al.}(1999){Audard}, {G{\"u}del}, \&
  {Guinan}}]{1999ApJ...513L..53A}
{Audard}, M., {G{\"u}del}, M., \& {Guinan}, E.~F. 1999, \apjl, 513, L53

\bibitem[{{Audard} {et~al.}(2001){Audard}, {G{\"u}del}, \&
  {Mewe}}]{2001A&A...365L.318A}
{Audard}, M., {G{\"u}del}, M., \& {Mewe}, R. 2001, \aap, 365, L318

\bibitem[{{Audard} {et~al.}(2003){Audard}, {G{\"u}del}, {Sres}, {Raassen}, \&
  {Mewe}}]{2003A&A...398.1137A}
{Audard}, M., {G{\"u}del}, M., {Sres}, A., {Raassen}, A.~J.~J., \& {Mewe}, R.
  2003, \aap, 398, 1137

\bibitem[{{Ayres}(2005)}]{2005ApJ...618..493A}
{Ayres}, T.~R. 2005, \apj, 618, 493

\bibitem[{{Ayres} {et~al.}(2007){Ayres}, {Brown}, \&
  {Harper}}]{2007ApJ...658L.107A}
{Ayres}, T.~R., {Brown}, A., \& {Harper}, G.~M. 2007, \apjl, 658, L107

\bibitem[{{Budding} {et~al.}(2004){Budding}, {Erdem}, {{\c C}i{\c c}ek},
  {Bulut}, {Soydugan}, {Soydugan}, {Baki{\c s}}, \&
  {Demircan}}]{2004A&A...417..263B}
{Budding}, E., {Erdem}, A., {{\c C}i{\c c}ek}, C., {et~al.} 2004, \aap, 417,
  263

\bibitem[{{Chen} {et~al.}(2006){Chen}, {Sanchawala}, \&
  {Chiu}}]{2006AJ....131..990C}
{Chen}, W.~P., {Sanchawala}, K., \& {Chiu}, M.~C. 2006, \aj, 131, 990

\bibitem[{{Crawford} {et~al.}(1970){Crawford}, {Jauncey}, \&
  {Murdoch}}]{1970ApJ...162..405C}
{Crawford}, D.~F., {Jauncey}, D.~L., \& {Murdoch}, H.~S. 1970, \apj, 162, 405

\bibitem[{{Crespo-Chac{\'o}n} {et~al.}(2007){Crespo-Chac{\'o}n}, {Micela},
  {Reale}, {Caramazza}, {L{\'o}pez-Santiago}, \&
  {Pillitteri}}]{2007A&A...471..929C}
{Crespo-Chac{\'o}n}, I., {Micela}, G., {Reale}, F., {et~al.} 2007, \aap, 471,
  929

\bibitem[{{Dartnell}(2011)}]{2011AsBio..11..551D}
{Dartnell}, L.~R. 2011, Astrobiology, 11, 551

\bibitem[{{Dorman} \& {Arnaud}(2001)}]{2001ASPC..238..415D}
{Dorman}, B. \& {Arnaud}, K.~A. 2001, in Astronomical Society of the Pacific
  Conference Series, Vol. 238, Astronomical Data Analysis Software and Systems
  X, ed. F.~R. {Harnden}, Jr., F.~A. {Primini}, \& H.~E. {Payne}, 415

\bibitem[{{ESA}(1997)}]{1997ESASP1200.....E}
{ESA}, ed. 1997, ESA Special Publication, Vol. 1200, {The HIPPARCOS and TYCHO
  catalogues. Astrometric and photometric star catalogues derived from the ESA
  HIPPARCOS Space Astrometry Mission}

\bibitem[{{Farrell}(2009)}]{FarrellPrivComm09}
{Farrell}, S. 2009, private communication

\bibitem[{{Favata}(2002)}]{2002ASPC..277..115F}
{Favata}, F. 2002, in Astronomical Society of the Pacific Conference Series,
  Vol. 277, Stellar Coronae in the Chandra and XMM-NEWTON Era, ed. F.~{Favata}
  \& J.~J. {Drake}, 115

\bibitem[{{Favata} {et~al.}(2003){Favata}, {Giardino}, {Micela}, {Sciortino},
  \& {Damiani}}]{2003A&A...403..187F}
{Favata}, F., {Giardino}, G., {Micela}, G., {Sciortino}, S., \& {Damiani}, F.
  2003, \aap, 403, 187

\bibitem[{{Favata} \& {Micela}(2003)}]{2003SSRv..108..577F}
{Favata}, F. \& {Micela}, G. 2003, \ssr, 108, 577

\bibitem[{{Favata} {et~al.}(1995){Favata}, {Micela}, \&
  {Sciortino}}]{1995A&A...298..482F}
{Favata}, F., {Micela}, G., \& {Sciortino}, S. 1995, \aap, 298, 482

\bibitem[{{Feigelson}(2010)}]{2010PNAS..107.7153F}
{Feigelson}, E.~D. 2010, Proceedings of the National Academy of Science, 107,
  7153

\bibitem[{{Franciosini} {et~al.}(2007{\natexlab{a}}){Franciosini},
  {Pillitteri}, {Stelzer}, {Micela}, {Briggs}, {Scelsi}, {Telleschi}, {Audard},
  {Palla}, \& {G{\"u}del}}]{2007A&A...468..485F}
{Franciosini}, E., {Pillitteri}, I., {Stelzer}, B., {et~al.}
  2007{\natexlab{a}}, \aap, 468, 485

\bibitem[{{Franciosini} {et~al.}(2003){Franciosini}, {Randich}, \&
  {Pallavicini}}]{2003A&A...405..551F}
{Franciosini}, E., {Randich}, S., \& {Pallavicini}, R. 2003, \aap, 405, 551

\bibitem[{{Franciosini} {et~al.}(2007{\natexlab{b}}){Franciosini}, {Scelsi},
  {Pallavicini}, \& {Audard}}]{2007A&A...471..951F}
{Franciosini}, E., {Scelsi}, L., {Pallavicini}, R., \& {Audard}, M.
  2007{\natexlab{b}}, \aap, 471, 951

\bibitem[{{Frasca} {et~al.}(2006){Frasca}, {Guillout}, {Marilli}, {Freire
  Ferrero}, {Biazzo}, \& {Klutsch}}]{2006A&A...454..301F}
{Frasca}, A., {Guillout}, P., {Marilli}, E., {et~al.} 2006, \aap, 454, 301

\bibitem[{{Gatewood}(1996)}]{1996AAS...188.4011G}
{Gatewood}, G. 1996, in American Astronomical Society Meeting Abstracts, Vol.
  188, American Astronomical Society Meeting Abstracts \#188, \#40.11

\bibitem[{{Gondoin}(2003)}]{2003A&A...400..249G}
{Gondoin}, P. 2003, \aap, 400, 249

\bibitem[{{Gondoin}({\noopsort{a}}2004a)}]{2004A&A...415.1113G}
{Gondoin}, P. {\noopsort{a}}2004a, \aap, 415, 1113

\bibitem[{{Gondoin}({\noopsort{b}}2004b)}]{2004A&A...426.1035G}
{Gondoin}, P. {\noopsort{b}}2004b, \aap, 426, 1035

\bibitem[{{Gondoin} {et~al.}(2002){Gondoin}, {Erd}, \&
  {Lumb}}]{2002A&A...383..919G}
{Gondoin}, P., {Erd}, C., \& {Lumb}, D. 2002, \aap, 383, 919

\bibitem[{{Griffin}(2009)}]{2009Obs...129..317G}
{Griffin}, R.~F. 2009, The Observatory, 129, 317

\bibitem[{{G{\"u}del}(2004)}]{2004A&ARv..12...71G}
{G{\"u}del}, M. 2004, \aapr, 12, 71

\bibitem[{{G{\"u}del} {et~al.}(2001){G{\"u}del}, {Audard}, {Briggs}, {Haberl},
  {Magee}, {Maggio}, {Mewe}, {Pallavicini}, \& {Pye}}]{2001A&A...365L.336G}
{G{\"u}del}, M., {Audard}, M., {Briggs}, K., {et~al.} 2001, \aap, 365, L336

\bibitem[{{G{\"u}del} {et~al.}(2007{\natexlab{a}}){G{\"u}del}, {Briggs},
  {Arzner}, {Audard}, {Bouvier}, {Feigelson}, {Franciosini}, {Glauser},
  {Grosso}, {Micela}, {Monin}, {Montmerle}, {Padgett}, {Palla}, {Pillitteri},
  {Rebull}, {Scelsi}, {Silva}, {Skinner}, {Stelzer}, \&
  {Telleschi}}]{2007A&A...468..353G}
{G{\"u}del}, M., {Briggs}, K.~R., {Arzner}, K., {et~al.} 2007{\natexlab{a}},
  \aap, 468, 353

\bibitem[{{G{\"u}del} \& {Naz{\'e}}(2009)}]{2009A&ARv..17..309G}
{G{\"u}del}, M. \& {Naz{\'e}}, Y. 2009, \aapr, 17, 309

\bibitem[{{G{\"u}del} {et~al.}(2007{\natexlab{b}}){G{\"u}del}, {Skinner},
  {Mel'Nikov}, {Audard}, {Telleschi}, \& {Briggs}}]{2007A&A...468..529G}
{G{\"u}del}, M., {Skinner}, S.~L., {Mel'Nikov}, S.~Y., {et~al.}
  2007{\natexlab{b}}, \aap, 468, 529

\bibitem[{{Guillout} {et~al.}(1999){Guillout}, {Schmitt}, {Egret}, {Voges},
  {Motch}, \& {Sterzik}}]{1999A&A...351.1003G}
{Guillout}, P., {Schmitt}, J.~H.~M.~M., {Egret}, D., {et~al.} 1999, \aap, 351,
  1003

\bibitem[{{Guillout} {et~al.}(1998{\natexlab{a}}){Guillout}, {Sterzik},
  {Schmitt}, {Motch}, {Egret}, {Voges}, \& {Neuhaeuser}}]{1998A&A...334..540G}
{Guillout}, P., {Sterzik}, M.~F., {Schmitt}, J.~H.~M.~M., {et~al.}
  1998{\natexlab{a}}, \aap, 334, 540

\bibitem[{{Guillout} {et~al.}(1998{\natexlab{b}}){Guillout}, {Sterzik},
  {Schmitt}, {Motch}, \& {Neuhaeuser}}]{1998A&A...337..113G}
{Guillout}, P., {Sterzik}, M.~F., {Schmitt}, J.~H.~M.~M., {Motch}, C., \&
  {Neuhaeuser}, R. 1998{\natexlab{b}}, \aap, 337, 113

\bibitem[{{Haisch} {et~al.}(1991){Haisch}, {Strong}, \&
  {Rodono}}]{1991ARA&A..29..275H}
{Haisch}, B., {Strong}, K.~T., \& {Rodono}, M. 1991, \araa, 29, 275

\bibitem[{{Hempelmann} {et~al.}(2006){Hempelmann}, {Robrade}, {Schmitt},
  {Favata}, {Baliunas}, \& {Hall}}]{2006A&A...460..261H}
{Hempelmann}, A., {Robrade}, J., {Schmitt}, J.~H.~M.~M., {et~al.} 2006, \aap,
  460, 261

\bibitem[{{H{\o}g} {et~al.}(2000){H{\o}g}, {Fabricius}, {Makarov}, {Urban},
  {Corbin}, {Wycoff}, {Bastian}, {Schwekendiek}, \&
  {Wicenec}}]{2000A&A...355L..27H}
{H{\o}g}, E., {Fabricius}, C., {Makarov}, V.~V., {et~al.} 2000, \aap, 355, L27

\bibitem[{{Horvath} \& {Galante}(2012)}]{2012IJAsB..11..279H}
{Horvath}, J.~E. \& {Galante}, D. 2012, International Journal of Astrobiology,
  11, 279

\bibitem[{{Jauncey}(1967)}]{1967Natur.216..877J}
{Jauncey}, D.~L. 1967, \nat, 216, 877

\bibitem[{{Klutsch} {et~al.}(2008){Klutsch}, {Frasca}, {Guillout}, {Freire
  Ferrero}, {Marilli}, {Mignemi}, \& {Biazzo}}]{2008A&A...490..737K}
{Klutsch}, A., {Frasca}, A., {Guillout}, P., {et~al.} 2008, \aap, 490, 737

\bibitem[{{Koen} \& {Eyer}(2002)}]{2002MNRAS.331...45K}
{Koen}, C. \& {Eyer}, L. 2002, \mnras, 331, 45

\bibitem[{{L{\'o}pez-Santiago} {et~al.}(2007){L{\'o}pez-Santiago}, {Micela},
  {Sciortino}, {Favata}, {Caccianiga}, {Della Ceca}, {Severgnini}, \&
  {Braito}}]{2007A&A...463..165L}
{L{\'o}pez-Santiago}, J., {Micela}, G., {Sciortino}, S., {et~al.} 2007, \aap,
  463, 165

\bibitem[{{Maggio} {et~al.}(2011){Maggio}, {Sanz-Forcada}, \&
  {Scelsi}}]{2011A&A...527A.144M}
{Maggio}, A., {Sanz-Forcada}, J., \& {Scelsi}, L. 2011, \aap, 527, A144

\bibitem[{{Malkov} {et~al.}(2006){Malkov}, {Oblak}, {Snegireva}, \&
  {Torra}}]{2006A&A...446..785M}
{Malkov}, O.~Y., {Oblak}, E., {Snegireva}, E.~A., \& {Torra}, J. 2006, \aap,
  446, 785

\bibitem[{{Marraco} \& {Rydgren}(1981)}]{1981AJ.....86...62M}
{Marraco}, H.~G. \& {Rydgren}, A.~E. 1981, \aj, 86, 62

\bibitem[{{Mason} {et~al.}(2001){Mason}, {Breeveld}, {Much}, {Carter},
  {Cordova}, {Cropper}, {Fordham}, {Huckle}, {Ho}, {Kawakami}, {Kennea},
  {Kennedy}, {Mittaz}, {Pandel}, {Priedhorsky}, {Sasseen}, {Shirey}, {Smith},
  \& {Vreux}}]{2001A&A...365L..36M}
{Mason}, K.~O., {Breeveld}, A., {Much}, R., {et~al.} 2001, \aap, 365, L36

\bibitem[{{Matranga} {et~al.}(2005){Matranga}, {Mathioudakis}, {Kay}, \&
  {Keenan}}]{2005ApJ...621L.125M}
{Matranga}, M., {Mathioudakis}, M., {Kay}, H.~R.~M., \& {Keenan}, F.~P. 2005,
  \apjl, 621, L125

\bibitem[{{Melott} \& {Thomas}(2011)}]{2011AsBio..11..343M}
{Melott}, A.~L. \& {Thomas}, B.~C. 2011, Astrobiology, 11, 343

\bibitem[{{Mitra-Kraev} {et~al.}(2005{\natexlab{a}}){Mitra-Kraev}, {Harra},
  {G{\"u}del}, {Audard}, {Branduardi-Raymont}, {Kay}, {Mewe}, {Raassen}, \&
  {van Driel-Gesztelyi}}]{2005A&A...431..679M}
{Mitra-Kraev}, U., {Harra}, L.~K., {G{\"u}del}, M., {et~al.}
  2005{\natexlab{a}}, \aap, 431, 679

\bibitem[{{Mitra-Kraev} {et~al.}(2005{\natexlab{b}}){Mitra-Kraev}, {Harra},
  {Williams}, \& {Kraev}}]{2005A&A...436.1041M}
{Mitra-Kraev}, U., {Harra}, L.~K., {Williams}, D.~R., \& {Kraev}, E.
  2005{\natexlab{b}}, \aap, 436, 1041

\bibitem[{{Morales} {et~al.}(2009){Morales}, {Torres}, {Marschall}, \&
  {Brehm}}]{2009ApJ...707..671M}
{Morales}, J.~C., {Torres}, G., {Marschall}, L.~A., \& {Brehm}, W. 2009, \apj,
  707, 671

\bibitem[{{Naz{\'e}}(2009)}]{2009A&A...506.1055N}
{Naz{\'e}}, Y. 2009, \aap, 506, 1055

\bibitem[{{Nordon} \& {Behar}(2008)}]{2008A&A...482..639N}
{Nordon}, R. \& {Behar}, E. 2008, \aap, 482, 639

\bibitem[{{Osborne} {et~al.}(2013){Osborne}, {O'Brien}, {Evans}, {Fraser},
  {Martindale}, {Atteia}, {Cordier}, \& {Mereghetti}}]{2013EAS....61..625O}
{Osborne}, J.~P., {O'Brien}, P., {Evans}, P., {et~al.} 2013, in EAS
  Publications Series, Vol.~61, EAS Publications Series, ed. A.~J.
  {Castro-Tirado}, J.~{Gorosabel}, \& I.~H. {Park}, 625--631

\bibitem[{{Osten} {et~al.}(2010){Osten}, {Godet}, {Drake}, {Tueller},
  {Cummings}, {Krimm}, {Pye}, {Pal'shin}, {Golenetskii}, {Reale}, {Oates},
  {Page}, \& {Melandri}}]{2010ApJ...721..785O}
{Osten}, R.~A., {Godet}, O., {Drake}, S., {et~al.} 2010, \apj, 721, 785

\bibitem[{{Otero} \& {Dubovsky}(2004)}]{2004IBVS.5557....1O}
{Otero}, S.~A. \& {Dubovsky}, P.~A. 2004, Information Bulletin on Variable
  Stars, 5557, 1

\bibitem[{{Pallavicini} {et~al.}(2004){Pallavicini}, {Franciosini}, \&
  {Randich}}]{2004MmSAI..75..434P}
{Pallavicini}, R., {Franciosini}, E., \& {Randich}, S. 2004, \memsai, 75, 434

\bibitem[{{Pandey} \& {Singh}(2008)}]{2008MNRAS.387.1627P}
{Pandey}, J.~C. \& {Singh}, K.~P. 2008, \mnras, 387, 1627

\bibitem[{{Pandey} \& {Singh}(2012)}]{2012MNRAS.419.1219P}
{Pandey}, J.~C. \& {Singh}, K.~P. 2012, \mnras, 419, 1219

\bibitem[{{Pojmanski}(1998)}]{1998AcA....48...35P}
{Pojmanski}, G. 1998, \actaa, 48, 35

\bibitem[{{Pollacco} {et~al.}(2006){Pollacco}, {Skillen}, {Collier Cameron},
  {Christian}, {Hellier}, {Irwin}, {Lister}, {Street}, {West}, {Anderson},
  {Clarkson}, {Deeg}, {Enoch}, {Evans}, {Fitzsimmons}, {Haswell}, {Hodgkin},
  {Horne}, {Kane}, {Keenan}, {Maxted}, {Norton}, {Osborne}, {Parley}, {Ryans},
  {Smalley}, {Wheatley}, \& {Wilson}}]{2006PASP..118.1407P}
{Pollacco}, D.~L., {Skillen}, I., {Collier Cameron}, A., {et~al.} 2006, \pasp,
  118, 1407

\bibitem[{{Pourbaix} {et~al.}(2004){Pourbaix}, {Tokovinin}, {Batten}, {Fekel},
  {Hartkopf}, {Levato}, {Morrell}, {Torres}, \& {Udry}}]{2004A&A...424..727P}
{Pourbaix}, D., {Tokovinin}, A.~A., {Batten}, A.~H., {et~al.} 2004, \aap, 424,
  727

\bibitem[{{Pribulla} {et~al.}(2001){Pribulla}, {Chochol}, {Heckert}, {Errico},
  {Vittone}, {Parimucha}, \& {Teodorani}}]{2001A&A...371..997P}
{Pribulla}, T., {Chochol}, D., {Heckert}, P.~A., {et~al.} 2001, \aap, 371, 997

\bibitem[{{Pye} \& {McHardy}(1983)}]{1983MNRAS.205..875P}
{Pye}, J.~P. \& {McHardy}, I.~M. 1983, \mnras, 205, 875

\bibitem[{{Raassen} {et~al.}(2003){Raassen}, {Mewe}, {Audard}, \&
  {G{\"u}del}}]{2003A&A...411..509R}
{Raassen}, A.~J.~J., {Mewe}, R., {Audard}, M., \& {G{\"u}del}, M. 2003, \aap,
  411, 509

\bibitem[{{Raassen} {et~al.}(2007){Raassen}, {Mitra-Kraev}, \&
  {G{\"u}del}}]{2007MNRAS.379.1075R}
{Raassen}, A.~J.~J., {Mitra-Kraev}, U., \& {G{\"u}del}, M. 2007, \mnras, 379,
  1075

\bibitem[{{Rao} \& {Vahia}(1987)}]{1987A&A...188..109R}
{Rao}, A.~R. \& {Vahia}, M.~N. 1987, \aap, 188, 109

\bibitem[{{Reale}(2007)}]{2007A&A...471..271R}
{Reale}, F. 2007, \aap, 471, 271

\bibitem[{{Reale} {et~al.}(1997){Reale}, {Betta}, {Peres}, {Serio}, \&
  {McTiernan}}]{1997A&A...325..782R}
{Reale}, F., {Betta}, R., {Peres}, G., {Serio}, S., \& {McTiernan}, J. 1997,
  \aap, 325, 782

\bibitem[{{Reid} {et~al.}(1995){Reid}, {Hawley}, \&
  {Gizis}}]{1995AJ....110.1838R}
{Reid}, I.~N., {Hawley}, S.~L., \& {Gizis}, J.~E. 1995, \aj, 110, 1838

\bibitem[{{Robichon} {et~al.}(1999){Robichon}, {Arenou}, {Mermilliod}, \&
  {Turon}}]{1999A&A...345..471R}
{Robichon}, N., {Arenou}, F., {Mermilliod}, J.-C., \& {Turon}, C. 1999, \aap,
  345, 471

\bibitem[{{Robrade} {et~al.}(2004){Robrade}, {Ness}, \&
  {Schmitt}}]{2004A&A...413..317R}
{Robrade}, J., {Ness}, J.-U., \& {Schmitt}, J.~H.~M.~M. 2004, \aap, 413, 317

\bibitem[{{Robrade} \& {Schmitt}(2005)}]{2005A&A...435.1073R}
{Robrade}, J. \& {Schmitt}, J.~H.~M.~M. 2005, \aap, 435, 1073

\bibitem[{{Robrade} \& {Schmitt}(2006)}]{2006A&A...449..737R}
{Robrade}, J. \& {Schmitt}, J.~H.~M.~M. 2006, \aap, 449, 737

\bibitem[{{Sanz-Forcada} {et~al.}(2006){Sanz-Forcada}, {Favata}, \&
  {Micela}}]{2006A&A...445..673S}
{Sanz-Forcada}, J., {Favata}, F., \& {Micela}, G. 2006, \aap, 445, 673

\bibitem[{{Scelsi} {et~al.}(2005){Scelsi}, {Maggio}, {Peres}, \&
  {Pallavicini}}]{2005A&A...432..671S}
{Scelsi}, L., {Maggio}, A., {Peres}, G., \& {Pallavicini}, R. 2005, \aap, 432,
  671

\bibitem[{{Schaefer} {et~al.}(2000){Schaefer}, {King}, \&
  {Deliyannis}}]{2000ApJ...529.1026S}
{Schaefer}, B.~E., {King}, J.~R., \& {Deliyannis}, C.~P. 2000, \apj, 529, 1026

\bibitem[{{Schneider} {et~al.}(2011){Schneider}, {Dedieu}, {Le Sidaner},
  {Savalle}, \& {Zolotukhin}}]{2011A&A...532A..79S}
{Schneider}, J., {Dedieu}, C., {Le Sidaner}, P., {Savalle}, R., \&
  {Zolotukhin}, I. 2011, \aap, 532, A79

\bibitem[{{Schrijver} {et~al.}(2012){Schrijver}, {Beer}, {Baltensperger},
  {Cliver}, {G{\"u}del}, {Hudson}, {McCracken}, {Osten}, {Peter}, {Soderblom},
  {Usoskin}, \& {Wolff}}]{2012JGRA..117.8103S}
{Schrijver}, C.~J., {Beer}, J., {Baltensperger}, U., {et~al.} 2012, Journal of
  Geophysical Research (Space Physics), 117, 8103

\bibitem[{{Smith} {et~al.}(2005){Smith}, {G{\"u}del}, \&
  {Audard}}]{2005A&A...436..241S}
{Smith}, K., {G{\"u}del}, M., \& {Audard}, M. 2005, \aap, 436, 241

\bibitem[{{Stelzer} {et~al.}(2004){Stelzer}, {Micela}, \&
  {Neuh{\"a}user}}]{2004A&A...423.1029S}
{Stelzer}, B., {Micela}, G., \& {Neuh{\"a}user}, R. 2004, \aap, 423, 1029

\bibitem[{{Stobbart} {et~al.}(2006){Stobbart}, {Roberts}, \&
  {Warwick}}]{2006MNRAS.370...25S}
{Stobbart}, A.-M., {Roberts}, T.~P., \& {Warwick}, R.~S. 2006, \mnras, 370, 25

\bibitem[{{Stocke} {et~al.}(1991){Stocke}, {Morris}, {Gioia}, {Maccacaro},
  {Schild}, {Wolter}, {Fleming}, \& {Henry}}]{1991ApJS...76..813S}
{Stocke}, J.~T., {Morris}, S.~L., {Gioia}, I.~M., {et~al.} 1991, \apjs, 76, 813

\bibitem[{{Suh} {et~al.}(2005){Suh}, {Audard}, {G{\"u}del}, \&
  {Paerels}}]{2005ApJ...630.1074S}
{Suh}, J.~A., {Audard}, M., {G{\"u}del}, M., \& {Paerels}, F.~B.~S. 2005, \apj,
  630, 1074

\bibitem[{{Telleschi} {et~al.}(2005){Telleschi}, {G{\"u}del}, {Briggs},
  {Audard}, {Ness}, \& {Skinner}}]{2005ApJ...622..653T}
{Telleschi}, A., {G{\"u}del}, M., {Briggs}, K., {et~al.} 2005, \apj, 622, 653

\bibitem[{{Thomson} {et~al.}(2010){Thomson}, {Gaunt}, {Cilliers}, {Wild},
  {Opperman}, {McKinnell}, {Kotze}, {Ngwira}, \& {Lotz}}]{2010AdSpR..45.1182T}
{Thomson}, A.~W.~P., {Gaunt}, C.~T., {Cilliers}, P., {et~al.} 2010, Advances in
  Space Research, 45, 1182

\bibitem[{{Tsujimoto} {et~al.}(2002){Tsujimoto}, {Koyama}, {Tsuboi}, {Goto}, \&
  {Kobayashi}}]{2002ApJ...566..974T}
{Tsujimoto}, M., {Koyama}, K., {Tsuboi}, Y., {Goto}, M., \& {Kobayashi}, N.
  2002, \apj, 566, 974

\bibitem[{{van Altena} {et~al.}(1995){van Altena}, {Lee}, \&
  {Hoffleit}}]{1995gcts.book.....V}
{van Altena}, W.~F., {Lee}, J.~T., \& {Hoffleit}, E.~D. 1995, {The general
  catalogue of trigonometric [stellar] parallaxes, 4th ed., Yale Univ.\ Obs.,
  New Haven, CT}

\bibitem[{{van den Besselaar} {et~al.}(2003){van den Besselaar}, {Raassen},
  {Mewe}, {van der Meer}, {G{\"u}del}, \& {Audard}}]{2003A&A...411..587V}
{van den Besselaar}, E.~J.~M., {Raassen}, A.~J.~J., {Mewe}, R., {et~al.} 2003,
  \aap, 411, 587

\bibitem[{{van Leeuwen}(2007)}]{2007A&A...474..653V}
{van Leeuwen}, F. 2007, \aap, 474, 653

\bibitem[{{Voges} {et~al.}(1999){Voges}, {Aschenbach}, {Boller},
  {Br{\"a}uninger}, {Briel}, {Burkert}, {Dennerl}, {Englhauser}, {Gruber},
  {Haberl}, {Hartner}, {Hasinger}, {K{\"u}rster}, {Pfeffermann}, {Pietsch},
  {Predehl}, {Rosso}, {Schmitt}, {Tr{\"u}mper}, \&
  {Zimmermann}}]{1999A&A...349..389V}
{Voges}, W., {Aschenbach}, B., {Boller}, T., {et~al.} 1999, \aap, 349, 389

\bibitem[{{Wall} \& {Jenkins}(2003)}]{2003psa..book.....W}
{Wall}, J.~V. \& {Jenkins}, C.~R. 2003, {Practical Statistics for Astronomers,
  Cambridge Univ.\ Press, Cambridge}

\bibitem[{{Walter} {et~al.}(1988){Walter}, {Brown}, {Mathieu}, {Myers}, \&
  {Vrba}}]{1988AJ.....96..297W}
{Walter}, F.~M., {Brown}, A., {Mathieu}, R.~D., {Myers}, P.~C., \& {Vrba},
  F.~J. 1988, \aj, 96, 297

\bibitem[{{Warwick}(2014)}]{2014MNRAS.445...66W}
{Warwick}, R.~S. 2014, \mnras, 445, 66

\bibitem[{{Watson} {et~al.}(2009){Watson}, {Schr{\"o}der}, {Fyfe}, {Page},
  {Lamer}, {Mateos}, {Pye}, {Sakano}, {Rosen}, {Ballet}, {Barcons}, {Barret},
  {Boller}, {Brunner}, {Brusa}, {Caccianiga}, {Carrera}, {Ceballos}, {Della
  Ceca}, {Denby}, {Denkinson}, {Dupuy}, {Farrell}, {Fraschetti}, {Freyberg},
  {Guillout}, {Hambaryan}, {Maccacaro}, {Mathiesen}, {McMahon}, {Michel},
  {Motch}, {Osborne}, {Page}, {Pakull}, {Pietsch}, {Saxton}, {Schwope},
  {Severgnini}, {Simpson}, {Sironi}, {Stewart}, {Stewart}, {Stobbart}, {Tedds},
  {Warwick}, {Webb}, {West}, {Worrall}, \& {Yuan}}]{2009A&A...493..339W}
{Watson}, M.~G., {Schr{\"o}der}, A.~C., {Fyfe}, D., {et~al.} 2009, \aap, 493,
  339

\bibitem[{{West}(2009)}]{WestPrivComm09}
{West}, R.~G. 2009, private communication

\bibitem[{{Wolk} {et~al.}(2005){Wolk}, {Harnden}, {Flaccomio}, {Micela},
  {Favata}, {Shang}, \& {Feigelson}}]{2005ApJS..160..423W}
{Wolk}, S.~J., {Harnden}, Jr., F.~R., {Flaccomio}, E., {et~al.} 2005, \apjs,
  160, 423

\bibitem[{{Wright} {et~al.}(2003){Wright}, {Egan}, {Kraemer}, \&
  {Price}}]{2003AJ....125..359W}
{Wright}, C.~O., {Egan}, M.~P., {Kraemer}, K.~E., \& {Price}, S.~D. 2003, \aj,
  125, 359

\bibitem[{{Zuckerman} \& {Song}(2004)}]{2004ARA&A..42..685Z}
{Zuckerman}, B. \& {Song}, I. 2004, \araa, 42, 685

\end{thebibliography}

\onecolumn		
\section{Tables and figures available in electronic form only}  \label{appEl}

\newpage
\subsection{The stars in the 2XMM-Tycho flare survey}

\begin{table*}[h]
\caption{The stars in the 2XMM-Tycho flare survey (1 row per XMM observation).}
\label{tabStars}
\centering
\begin{tabular}{l l c c c c}

\end{tabular}
\newline
\end{table*}

\newpage
\subsection{The flares in the 2XMM-Tycho flare survey}

\begin{table*}[h]
\caption{The flares and other time-variability events in the 2XMM-Tycho flare survey (1 row per event).}
\label{tabVarDetail}
\centering
\begin{tabular}{l l c c c c}

\end{tabular}
\newline
\end{table*}

\newpage
\subsection{X-ray and ultraviolet light-curves}    \label{appALLlcOM}

\begin{figure}[h]
	\caption{All 108 EPIC X-ray light-curves (top panel of each pair), and corresponding OM ultraviolet data (bottom panel of each pair) where available. Each pair of plots is labelled at the top with the 2XMM DETID, the star name, the EPIC camera and exposure number, and the X-ray time binning $\Delta t$~(s). 
The conversion factor for count rates measured in the different EPIC cameras is 1 MOS count/s $\approx 3.2$ PN count/s.
The X-ray data are for the Total energy band (0.2--12 keV); the OM waveband filters are indicated towards the right of the plot, and colour-coded.
EPIC X-ray `flux' units are Total-band count/s (in one of PN, MOS1, MOS2 cameras), while OM flux units are 
$\rm \,erg\,cm^{-2}\,s^{-1}\,\AA^{-1} $ for imaging-mode data and count/s for fast-mode data.
The conversion factors for OM count rates (count/s) to flux values are: 
$5.67 \times 10^{-15}$ (W2),
$2.20 \times 10^{-15}$ (M2),
$4.77 \times 10^{-16}$ (W1),
$1.99 \times 10^{-16}$ (U),                                    
$1.24 \times 10^{-16}$ (B),
$2.51 \times 10^{-16}$ (V)
(XMM-SOC-CAL-TN-0019 http://xmm2.esac.esa.int/docs/documents/CAL-TN-0019.pdf Table 18.)
The plots are ordered by 2XMM DETID;
within each page, DETID increases from top to bottom, then right to left, starting at top right.
}
	\label{figALLlcOM}
\end{figure}

\newpage
\subsection{X-ray count-rate and hardness-ratio light-curves}    \label{appALLlcHR}

\begin{figure}[h]
	\caption{All 108 EPIC X-ray light-curves (top panel of each pair), and corresponding hardness-ratio light-curves with approximate temperatures indicated (bottom panel of each pair). Each pair of plots is labelled at the top with the 2XMM DETID, the star name, the EPIC camera and exposure number, and the X-ray time binning $\Delta t$~(s). 
The conversion factor for count rates measured in the different EPIC cameras is 1 MOS count/s $\approx 3.2$ PN count/s.
The X-ray count rates are for the Total energy band (0.2--12 keV), in one of PN, MOS1, MOS2 cameras.
The hardness-ratios use bands 0.2-1, 1-12 keV.
The plots are ordered by 2XMM DETID;
within each page, DETID increases from top to bottom, then right to left, starting at top right.
}
	\label{figALLlcHR}
\end{figure}

\end{document}